\documentclass[11pt, A4paper]{article} 
\usepackage{jheppub}
\usepackage[utf8]{inputenc} 
\usepackage{times,relsize}
\newcommand{\north}{\text{\smaller[3]N}}
\newcommand{\south}{\text{\smaller[3]S}}
\newcommand{\atw}{\text{\smaller[1]A}}
\newcommand{\lcl}[1]{\boldsymbol #1}
\newcommand{\rch}{q}
\newcommand{\ren}{\textbf{r}}
\newcommand{\nin}{ \;\,/\hskip-11.5pt\in }
\newcommand{\susy}{{\bf Q}}
\newcommand{\brs}{{\bf Q}_\text{B}}

\newcommand{\RV}{{\text{R}_\text{V}}}
\newcommand{\RA}{{\text{R}_\text{A}}}

\title{Supersymmetric vortex defects in two dimensions}

\author[a]{Kazuo Hosomichi,}
\author[b]{Sungjay Lee}
\author[c]{and Takuya Okuda}

\affiliation[a]{
Department of Applied Physics, National Defense Academy, Kanagawa 239-8686, Japan}
\affiliation[b]{
Korea Institute for Advanced Study, 85 Hoegi-ro, Dongdaemun-gu, Seoul 02455, Korea}
\affiliation[c]{
University of Tokyo, Komaba, Meguro-ku, Tokyo 153-8902, Japan}

\emailAdd{hosomiti@nda.ac.jp}
\emailAdd{sjlee@kias.re.kr}
\emailAdd{takuya@hep1.c.u-tokyo.ac.jp}

\abstract{
We study codimension-two BPS defects in 2d $\mathcal{N}=(2,2)$ supersymmetric gauge theories, focusing especially on those characterized by vortex-like singularities in the dynamical or non-dynamical gauge field. We classify possible SUSY-preserving boundary conditions on charged matter fields around the vortex defects, and derive a formula for defect correlators on the squashed sphere. We also prove an equivalence relation between vortex defects and 0d-2d coupled systems. Our defect correlators are shown to be consistent with the mirror symmetry duality between Abelian gauged linear sigma models and Landau-Ginzburg models, as well as that between the minimal model and its orbifold.
We also study the vortex defects inserted at conical singularities.  
}

\preprint{KIAS-P17005, UT-KOMABA-17-5}
\begin{document}
\bibliographystyle{JHEP}

\maketitle

\section{Introduction and summary}

In gauge theories in different dimensions, one can introduce defects of codimension 2 in a number of ways. One standard way is the following. Let $(r,\varphi)$ be the polar coordinates for the two dimensions transverse to the defect, with $r=0$ the position of the defect. One requires the gauge field $A$ to behave near the defect like
\begin{equation}
 A \sim \eta d\varphi,
\label{ddef}
\end{equation}
where $\eta$ is a Lie algebra valued constant called vorticity. According to their dimensionality, they are called surface operators in 4d~\cite{Gukov:2014gja}, vortex loops in 3d and vortex defects in 2d. Compared to Wilson loops~\cite{Wilson:1974sk} or 't Hooft loops~\cite{tHooft:1977nqb} in four dimensions, these operators are relatively new and have much more to be explored, and yet they are expected to play equally important roles as order parameters in gauge theories~\cite{Gukov:2006jk,Gukov:2013zka}.

Another standard way to define a defect is to introduce dynamical variables or fields localized on it and let them interact with the fields in the bulk. For 4d supersymmetric gauge theories, another definition has been proposed based on embedding the theory into a larger theory and ``Higgsing'' by a position-dependent scalar vev~\cite{Gaiotto:2012xa}. Moreover, there are examples where the defects based on different definitions are believed to be equivalent~\cite{Frenkel:2015rda,Gorsky:2017hro}. Studying the relations among different definitions of defects will therefore be a key for their better understanding.

BPS defects in supersymmetric theories are especially interesting, since their protected sector can often be determined precisely. In particular, for systems with Lagrangian descriptions, SUSY localization allows us to evaluate the protected observables by explicit path integration. For example, exact path integrals have been worked out for the coupled 1d-3d systems on $S^3$ in~\cite{Assel:2015oxa}, and for the (0d-)2d-4d systems on $S^4$ in~\cite{Gomis:2014eya,Gomis:2016ljm,Pan:2016fbl}, see also~\cite{Chen:2015fta,Pan:2015hza}. On the other hand, we have not yet reached a fully satisfactory understanding for the defects defined by the boundary condition~(\ref{ddef}), though exact vortex loop observables in Abelian 3d gauge theories on $S^3$ and $S^2\times S^1$ were worked out in~\cite{Kapustin:2012iw,Drukker:2012sr}, and a proposal for surface operator vev on $S^4$ were given in~\cite{Nawata:2014nca}.  A major difficulty here is in finding the right definition of the path integral measure for the gauge fields as well as charged matter fields under the singular boundary condition.

The aim of this paper is to propose a fully precise definition of the defects of the type~(\ref{ddef}) in the path integral formalism. As the simplest setting to study this problem, we take 2d ${\cal N}=(2,2)$ supersymmetric gauge theories of vector and chiral multiplets, and focus on vortex defects preserving half of the supersymmetry. Throughout the paper we work in Euclidean signature.  In the standard ${\cal N}=(2,2)$ terminology, the defects are either in the twisted chiral or anti-twisted chiral rings. Based on our definition we study various aspects of 2d vortex defects, which include their relation to other type of defects or the transformation property under mirror symmetry.

\paragraph{Boundary conditions on charged matter. }

Defining path integration under the boundary condition~(\ref{ddef}) requires a re-examination of boundary conditions of all other fields with gauge charge. For systems with Abelian gauge symmetry, the problem simplifies considerably because all the fields in vector multiplet are neutral. In this case the fluctuations of gauge field and its superpartners around the singular classical value~(\ref{ddef}) should be regular, so the only difficulty lies in how to define path integral over charged matter fields. We will study Abelian systems in most of this paper, and discuss generalization to non-Abelian theories briefly in Section~\ref{sec:vnat}.

A common approach to defining the path integral  explicitly is via the  expansion of fields into the  eigenfunctions of the relevant Laplace or Dirac operator.  For the matter chiral multiplet  $\{\phi,\psi^\pm,F\}$ in ${\cal N}=(2,2)$ theories, the SUSY preserved by the defect introduces additional structures. For example, take a vortex defect to be twisted chiral. Then the unbroken SUSY around the defect divides the chiral multiplet into two subsets, $\{\phi,\psi^+\}$ and $\{\psi^-,F\}$, of components obeying the same boundary condition. Also, the Hilbert spaces ${\cal H},{\cal H}'$ of their wave functions are mapped to each other by $D_z,D_{\bar z}$, where $z\equiv re^{i\varphi}$ is the local complex coordinate around the defect.
\begin{equation}
 \phi,\psi^+\in~{\cal H}~
\begin{array}{c}
 \xleftarrow{~~~~D_z~~~~}\\[-3.8mm] \xrightarrow[~~~~D_{\bar z}~~~~]{}
\end{array}
~{\cal H}'~\ni \psi^-,F.
\end{equation}
The general eigenfunctions of $D_zD_{\bar z}, D_{\bar z}D_z$ on the vortex defect background{} exhibit a vorticity-dependent  non-integer  power-law behavior near the defect. Based on this, we argue that the supersymmetry is preserved only if one allows the wave functions of either ${\cal H}'$ or ${\cal H}$ to have a mild divergence $(\sim r^{\gamma}, \gamma>-1)$ at the defect. These are our definition of the {\it normal} and {\it flipped} boundary conditions on charged chiral multiplets.

This proposal can be immediately applied to the evaluation of defect correlators on squashed sphere, with a twisted chiral defect $V_{\eta_\north}$ placed at the north pole and an anti-twisted chiral defect $V_{\eta_\south}$ at the south pole. The SUSY localization~\cite{Benini:2012ui,Doroud:2012xw} allows one to reduce the path integral to an integral and sum over the constant vevs of scalars in the vector multiplet. For a  $U(1)$ gauge theory the defect correlator takes the form
\begin{equation}
 \langle V_{\eta_\north}V_{\eta_\south}\rangle =\sum_{s\in\frac12(\eta_\north-\eta_\south+\mathbb Z)}\int\frac{da}{2\pi}e^{-t(-s+\eta_\north+ia)}e^{-\bar t(s+\eta_\south+ia)}\times Z_\text{1-loop}.
\end{equation}
Here $t\equiv r+i\theta$ is the FI-theta coupling and $a,s$ are the scalar vevs, the latter taking quantized values because it is related to the integrated magnetic flux on the sphere (excluding the defect contributions).  Let us define the ceiling function $\lceil x \rceil$ (floor function $\lfloor x \rfloor$) as the smallest (largest) integer not smaller (larger) than $x$. The contribution to the one-loop determinant $Z_\text{1-loop}$ from a single chiral multiplet of electric charge 1, vector R-charge $2q$ is
\begin{equation}
 Z_\text{1-loop} = \frac{\Gamma(\kappa_\north(\eta_\north)-\eta_\north+s+q-ia)}{\Gamma(-\kappa_\south(\eta_\south)+\eta_\south+s+1-q+ia)},
\label{z1lwd}
\end{equation}
with $\kappa_\north(\eta)=\lceil\eta\rceil$ or $\lfloor\eta\rfloor$ depending on the choice of normal or flipped boundary condition at the north pole, and $\kappa_\south$ is defined in a similar way. This is the basic building block to express defect correlators of general (Abelian) GLSMs on the squashed sphere. Note the invariance of the formula under integer shifts of $\eta$'s which reflects the invariance of the matter path integral measure under large gauge transformations. 
It depends on $\eta$ discontinuously because some modes start violating, or conversely start obeying, the boundary conditions as $\eta$'s are varied.

Also, by a careful treatment of UV divergence one can show that the vortex defects have non-trivial dimension and axial R-charge. They are determined by the electric charges of matter chiral multiplets and their boundary conditions.

\paragraph{Relations among different defects.}

An interesting question is how the vortex defects~(\ref{ddef}) are related to other kinds of defects, for example those defined by 0d-2d systems. It turns out that the analysis of path integrals on smeared defect configurations gives us very useful information.

By gauge theories on smeared vortex defect configurations, we mean path integration  over the gauge and matter fields around the backgrounds not precisely of the form~(\ref{ddef}), but with the singularity replaced by a smooth vorticity distribution. On one hand, turning on  any smooth classical configuration  for a  dynamical gauge field corresponds to a trivial shift of path integration variables, which would have no effect on the value of the integrals. On the other hand, the form of a wave function for a charged matter field  does depend non-trivially on the detail of smearing.

We study the behavior of matter path integrals in the  smeared vortex defect background{} with a highly peaked vorticity distribution by defining the measure in terms of the eigenmodes of the Laplace or the Dirac operator.    We find that a certain number of matter wave functions end up having a  delta functional support and give rise to a system of 0d fields. They form 0d chiral or Fermi multiplets under the SUSY preserved by the defect. Moreover, the remaining wave functions supported in 2d bulk are shown to form the Hilbert space for a  chiral multiplet satisfying either the normal or the flipped boundary condition depending on the sign of  the vorticity multiplied by the charge. For a chiral multiplet of unit electric charge, the effect of smearing can be schematically expressed as follows.
\begin{equation}
 {\cal D}(\text{2d chiral})_{V_\eta^\text{smeared}}
=\left\{\begin{array}{ll}
\displaystyle{\cal D}(\text{2d chiral})_{V^\text{flipped}_\eta}\times
 \prod_{a=0}^{\lfloor\eta\rfloor-1}d(\text{0d chiral})_a
  \quad& (\eta>0) \\
\displaystyle{\cal D}(\text{2d chiral})_{V^\text{normal}_\eta}\times
 \prod_{\alpha=0}^{\lfloor-\eta\rfloor-1}d(\text{0d Fermi})_\alpha
  \quad& (\eta<0)
\end{array}
\right.
\label{rel001}
\end{equation}
Our initial choice of Lagrangian for the 2d system uniquely determines the 0d action with which to integrate over the 0d variables.

The triviality of the ``smeared gauge vortex defect''  allows us to turn~(\ref{rel001}) into a more useful relation between vortex defects and 0d-2d systems. As an example, consider  a $U(1)$ GLSM with chiral multiplets of $U(1)$ charge $e_i$ and twisted mass $M_i$ in  flat space. The relation~(\ref{rel001}) then implies the equivalence
\begin{equation}
 V_\eta = \prod_i (e_i\Sigma+M_i)^{\kappa_i(e_i\eta)},
\label{vtc01}
\end{equation}
where $\kappa_i(x)=\lceil x\rceil$ or $\lfloor x\rfloor$ depending on the boundary condition on   the $i$-th multiplet  at the defect, and $\Sigma$ is the twisted chiral field in the bulk $U(1)$ vector multiplet evaluated at the defect. The $i$-th factor in the product on the RHS arises from $|\kappa_i(e_i\eta)|$ copies of 0d chiral or Fermi multiplets realizing the unbroken supersymmetry algebra $\susy^2 \propto  e_i\Sigma+M_i$; the integral over each chiral or Fermi multiplet gives a factor $(e_i\Sigma+M_i)^{\mp1}$ up to equivalence in $\susy$-cohomology. 
We will see in a number of examples that an Omega-deformed version of~(\ref{vtc01}) is satisfied by the vortex defects within correlators on the squashed sphere.

Independently of the above, the invariance of matter path integrals under large gauge transformations implies
\begin{equation}
 V_{\eta+1} = e^{-t_\ren(\mu)} \mu^{\sum_ie_i}\cdot V_\eta,
\label{vdpe}
\end{equation}
where $t_\ren(\mu)$ is the renormalized FI-theta parameter at the renormalization scale $\mu$. By combining~(\ref{vtc01}) and~(\ref{vdpe}) one recovers the familiar twisted chiral ring relation for $\Sigma$ in GLSMs,
\begin{equation}
 \prod_i\left(\frac{e_i\Sigma+M_i}{\mu}\right)^{e_i} = e^{-t_\ren(\mu)}.
\end{equation}
Similar argument works also for vortex defects within correlators on the squashed sphere, but the ring relation is Omega-deformed. We point out that it can be better interpreted as a holomorphic differential equation (Picard-Fuchs equation) satisfied by the sphere partition function.

\paragraph{Mirror symmetry for vortex defects}

The Abelian GLSMs discussed above are known to have a mirror description in terms of LG models of twisted chiral multiplets~\cite{Hori:2000kt}. For example, the $U(1)$ GLSM with chiral multiplets $\Phi_i$ of charge $e_i$ and mass $M_i$ is mirror to the LG model with twisted chiral fields $(\Sigma, Y_i)$ and the superpotential
\begin{equation}
 \widetilde W = \sum_i\left\{(e_i\Sigma+M_i)Y_i+\mu e^{-Y_i}\right\}-t_\ren(\mu)\Sigma.
\end{equation}
As was shown in~\cite{Gomis:2012wy}, the comparison of exact correlators on the squashed sphere provides a non-trivial check of the mirror equivalence. By applying this argument to vortex defects we identify the corresponding operator in the mirror theory
\begin{equation}
 V_\eta~~\Longleftrightarrow ~~ \widetilde V_\eta \;\equiv\;
 e^{\eta(\sum_i e_iY_i-t_\ren)}\cdot e^{-\sum_i\kappa_i(e_i\eta)Y_i}.
\end{equation}
Here $\kappa_i$ is defined in the same way as in~(\ref{vtc01}). For vortex defects for dynamical $U(1)$ vector multiplet, the first factor in the definition of $\widetilde V_\eta$ is trivial due to the F-term condition. The above formula works also for the cases where the vector multiplet is non-dynamical; as an example, for the $\mathbb Z_{k+2}$ orbifold of the ${\cal N}=2$ minimal model at level $k$ we check that the correlators of twist fields can be reproduced rather precisely by those of appropriately chosen vortex defects.

\paragraph{Vortex defect at conical singularities}

Supersymmetric correlators can be evaluated for vortex defects inserted on conical singularities. SUSY gauge theories on spaces with conical singularities have been attracting attention even without any operator insertions mainly because of its application to R\'enyi entropy~\cite{Nishioka:2013haa,Huang:2014gca,Huang:2014pda,Crossley:2014oea,Alday:2014fsa,Hama:2014iea,Nishioka:2016guu}. Another (more technical) motivation to introduce conical singularity is that the vortex defects at $\mathbb Z_K$-fixed point with vorticity $\eta=r/K\;(r\in\mathbb Z)$ can be studied by a simple orbifold projection. In~\cite{Hosomichi:2015pia} this idea was used to derive a formula for vortex defect correlators  on the sphere. However, by comparing it with the analysis of the matter wave function we find that some of the argument in~\cite{Hosomichi:2015pia} need to be reconsidered, and the formula there needs to be corrected accordingly.

We study the BPS vortex defects at general conical singularities $(\varphi\sim\varphi+2\pi/K)$ through their correlators on the sphere. The one-loop determinant for a  charged chiral multiplet of $\RA$-charge $2q$ takes the same form as~(\ref{z1lwd}), but the integer-valued function $\kappa_{\north,\south}$ now depends also on $K$ and $q$ as  
\begin{alignat}{2}
 \kappa_\north(\eta_\north) &\equiv\lceil\eta_\north-q(1-\tfrac1K)\rceil,
&\quad&\text{(normal b.c. at NP)}
\nonumber \\
 \kappa_\north(\eta_\north) &\equiv\lfloor\eta_\north-(q-1)(1-\tfrac1K)\rfloor.
&\quad&\text{(flipped b.c. at NP)}
\end{alignat}

\vskip3mm

This paper is organized as follows. In Section~\ref{sec:preliminary} we present the  definitions of various point-like defects and also summarize the construction of 2d SUSY gauge theories. Section~\ref{sec:vcs} develops the exact formula for the path integral over charged matter fields on the squashed sphere with vortex defects with or without smearing deformation, and make comparisons. In Section~\ref{sec:relat-between-defect} we study the charged chiral multiplet on a smeared $U(1)$ vortex defect background{}, from which we derive relations between vortex defects and 0d-2d systems. Section~\ref{sec:defect-correlators} is an application of our result to some sample Abelian GLSMs. Section~\ref{sec:mirr-symm-vort} derives  the transformation property of vortex defects under mirror symmetry. The vortex defects at conical singularity are discussed in Section~\ref{sec:vcns}.  Section~\ref{sec:vnat} discusses the generalization to non-Abelian theories and summarizes the basic building blocks for expressing exact defect correlators on sphere. We conclude in Section~\ref{sec:discussion} with discussions on future directions.

Some useful materials are recorded in the Appendices. 
Appendix~\ref{sec:dominant-terms-bulk} explains some properties of charged matter wave functions around a smeared defect in  flat space,  which are  used in Section~\ref{sec:relat-between-defect}. Appendix~\ref{sec:quintic} is a brief review of mirror symmetry and Gromov-Witten invariant for the quintic Calabi-Yau.
In Appendix~\ref{sec:interpret-kappa} we present another interpretation of the integer $\kappa(\eta)\equiv \kappa_\north(\eta)$ based on the analysis of the BPS vortex solution in the Higgs phase.

\section{SUSY gauge theories and vortex defects in 2d}
\label{sec:preliminary}

Here we review some properties of supersymmetric 2d curved backgrounds and the construction of ${\cal N}=(2,2)$ SUSY theories on such backgrounds. Some detailed formulae are presented for the  squashed sphere background, which will be used for the evaluation of defect correlators in Section~\ref{sec:vcs} and later. We will be primarily interested in the gauged linear sigma models (GLSMs) of vector and chiral multiplets, for we are studying the BPS vortex defects in this class of theories. We also discuss the Landau-Ginzburg (LG) theories of twisted chiral multiplets as they give the mirror description of the Abelian GLSMs. We then introduce several inequivalent definitions of BPS defects in  flat space.  One is the vortex defect defined via the  singular gauge field configuration~(\ref{ddef})  supplemented with a choice of boundary conditions on chiral multiplets.  Another is defined  via introduction of localized degrees of freedom on the defects.  Yet another is defined by smearing the vortex singularity (\ref{ddef}).

\subsection{Construction of SUSY theories}
\label{sec:review-N22}

We begin by summarizing spinor notations. We use the $2\times 2$ matrices $\epsilon_{\alpha\beta}$ and $(\gamma^a)^\alpha_{~\beta}$,
\begin{equation}
\def\arraystretch{.85}
 \epsilon=\left(\begin{array}{cc} 0&1  \\ -1& 0 \end{array}\right),\quad
 \gamma^1=\left(\begin{array}{cc} 0&~1  \\  1&~0 \end{array}\right),\quad
 \gamma^2=\left(\begin{array}{cc} 0&-i \\  i& 0 \end{array}\right),\quad
 \gamma^3=\left(\begin{array}{cc} 1&0  \\  0& -1\end{array}\right).
\def\arraystretch{1.0}
\end{equation}
We also use $\gamma^{ab}\equiv\frac12(\gamma^{a}\gamma^{b}-\gamma^{b}\gamma^{a})$. The two components of a Dirac spinor $\psi$ for the eigenvalues $\gamma^3=\pm1$ will be denoted by $\psi^+,\psi^-$. For spinor bilinears, the contracted indices will be suppressed as follows,
\begin{equation}
 \xi\psi \equiv \xi^\alpha\epsilon_{\alpha\beta}\psi^\beta,\quad
 \xi\gamma^a\psi \equiv \xi^\alpha
 \epsilon_{\alpha\beta}(\gamma^a)^\beta_{~\gamma}\psi^\gamma,\quad
\text{etc.}
\end{equation}

\paragraph{Supersymmetric backgrounds.}

We are going to consider SUSY theories on various 2d supersymmetric, flat or curved, backgrounds. The backgrounds of our interest are characterized by the vielbein $e^a_m$, spin connection $\omega^{ab}_m$, a vector field $V_m$ and a scalar field $H$. They are chosen in such a way that the Killing spinor equations 
\begin{eqnarray}
 D_m\xi &\equiv&
 \left(\partial_m+\frac14\omega_m^{ab}\gamma^{ab}-iV_m\right)\xi
 ~=~ \frac {iH}2\gamma_m\xi\,,
 \nonumber \\
 D_m\bar\xi &\equiv&
 \left(\partial_m+\frac14\omega_m^{ab}\gamma^{ab}+iV_m\right)\bar\xi
 ~=~ \frac{iH}2\gamma_m\bar\xi
\label{KSeq}
\end{eqnarray}
have   solutions. This is a simplified version in which we kept only a part of the fields in ${\cal N}=(2,2)$ supergravity multiplet which are relevant for our purpose. See~\cite{Closset:2014pda} for the fully general Killing spinor equation. The Killing spinors $\xi,\bar\xi$ are assigned the vector R-charge $\RV=+1,-1$, so that $V_m$ is identified with the gauge field for the R-symmetry.

An obvious example of SUSY background is the flat space with $\omega^{ab}_m=V_m=H=0$, for which any constant spinors are solutions to~(\ref{KSeq}). Another important example which we will study throughout the paper is the squashed sphere. Using the standard polar coordinates $\theta,\varphi$, the vielbein and the   spin connection are given by
\begin{equation}
 e^1=f(\theta) d\theta, \quad
 e^2=\ell\sin\theta d\varphi,\quad
 \omega^{12}=-\frac{\ell}{f(\theta)}\cos\theta d\varphi,
\label{BG1}
\end{equation}
and the other background fields are chosen as 
\begin{equation}
 H = \frac1f,\quad
 V = \frac12\left(\frac{\ell}{f}-1\right)d\varphi\,.
\label{BG2}
\end{equation}
The choice $f(\theta) \equiv\sqrt{\ell^2\cos^2\theta+\tilde\ell^2\sin^2\theta}$ corresponds to an ellipsoid with axis-lengths $\ell,\ell,\tilde\ell$. In particular, $f(\theta)\equiv\ell=\tilde\ell$ gives the round sphere of radius $\ell$ with the background fields $H=1/\ell,V=0$. This family of backgrounds has the same solution to the Killing spinor equation~(\ref{KSeq}) irrespective of the choice of $f(\theta)$. For the analysis of localized path integrals in later sections, we choose the following explicit solutions.
\begin{equation}
 \xi = e^{-\frac{i\varphi}2}
 \left(\begin{array}{r}
 i\sin\frac\theta2 \\
  \cos\frac\theta2 \end{array}\right),\quad
 \bar\xi = e^{\frac{i\varphi}2}
 \left(\begin{array}{r}
  \cos\frac\theta2 \\
 i\sin\frac\theta2 \end{array}\right).
\label{KS}
\end{equation}
The regularity of the metric requires $f(\theta)=\ell$ at the north pole $\theta=0$ and the south pole $\theta=\pi$. Otherwise we have conical singularities with the deficit angle $2\pi(1-\ell/f)$ there.

\paragraph{GLSMs.}

Let us turn to the construction of SUSY theories. We begin by the gauge theory of vector and chiral multiplets called GLSMs. The vector multiplet consists of a gauge field $A_m$, scalars $\rho,\sigma$, Dirac spinors $\lambda,\bar\lambda$ and an auxiliary scalar $D$, transforming under SUSY $\susy$ as 
\begin{eqnarray}
 \susy A_m &=& \tfrac12(\xi\gamma_m\bar\lambda+\bar\xi\gamma_m\lambda),
 \nonumber \\
 \susy\rho &=& \tfrac12(\xi\gamma_3\bar\lambda+\bar\xi\gamma_3\lambda),
 \nonumber \\
 \susy\sigma &=& \tfrac i2(\xi\bar\lambda-\bar\xi\lambda),
 \nonumber \\
 \susy\lambda &=& \tfrac12\gamma^{mn}\xi F_{mn}
  -\gamma^{3m}D_m(\xi\rho)-i\gamma^mD_m(\xi\sigma)
  -\gamma^3\xi[\rho,\sigma]+\xi D,
 \nonumber \\
 \susy\bar\lambda &=& \tfrac12\gamma^{mn}\bar\xi F_{mn}
 -\gamma^{3m}D_m(\bar\xi\rho)+i\gamma^mD_m(\bar\xi\sigma)
 +\gamma^3\bar\xi[\rho,\sigma]-\bar\xi D,
 \nonumber \\
 \susy D &=& \tfrac12D_m(\xi\gamma^m\bar\lambda-\bar\xi\gamma^m\lambda)
 -\tfrac i2[\rho,\xi\gamma_3\bar\lambda-\bar\xi\gamma_3\lambda]
 +\tfrac 12[\sigma,\xi\bar\lambda+\bar\xi\lambda]\,.
\end{eqnarray}
A chiral multiplet consists of a complex scalar $\phi$, Dirac spinor $\psi$ and an auxiliary field $F$ in a complex representation of the gauge group, and their conjugates $\bar\phi,\bar\psi,\bar F$ constitute an anti-chiral multiplet. They transform under SUSY as
\begin{eqnarray}
 \susy\phi &=& \xi\psi,
 \nonumber \\
 \susy\bar\phi &=& \bar\xi\bar\psi,
 \nonumber \\
 \susy\psi &=& -\gamma^m\bar\xi D_m\phi
 +i\gamma^3\bar\xi\rho\phi-\bar\xi\sigma\phi
 -\rch\cdot\phi\gamma^mD_m\bar\xi+\xi F,
 \nonumber \\
 \susy\bar\psi &=& -\gamma^m\xi D_m\bar\phi
 -i\gamma^3\xi\bar\phi\rho-\xi\bar\phi\sigma
 -\rch\cdot\bar\phi\gamma^mD_m\xi-\bar\xi\bar F,
 \nonumber \\
 \susy F &=& -\bar\xi(\gamma^mD_m\psi
 -i\gamma^3\rho\psi-\sigma\psi-i\bar\lambda\phi)
 -\rch\cdot D_m\bar\xi\gamma^m\psi,
 \nonumber \\
 \susy\bar F &=& +\xi(\gamma^mD_m\bar\psi
 +i\gamma^3\bar\psi\rho-\bar\psi\sigma+i\bar\phi\lambda)
 +\rch\cdot D_m\xi\gamma^m\bar\psi.
\end{eqnarray}
Here $q$ is a half of the vector R-charge of $\phi$, i.e. $\RV[\phi]=2q$. We summarize the vector R-charges of fields in a GLSM in Table~\ref{tab:R}.
\begin{table}[htbp]
\def\STRUT{\rule[-0.85mm]{0mm}{5.5mm}}
{\small
\begin{center}
\begin{tabular}{c||cc|ccccc|cccccc}
\!\!field & $\xi$ & $\bar\xi$ &
 $A_m$ & $\rho\pm i\sigma$ & $\lambda$ & $\bar\lambda$ & $D$ &
 $\phi$ & $\bar\phi$ & $\psi$ & $\bar\psi$ & $F$ & $\bar F$ \STRUT\\
\hline
$\text{R}_{\text{V}}$ & $1$ & $-1$ &
 $0$ & $0$ & $1$ & $-1$ & $0$ &
 $2\rch$ \!\!\!&\!\!\! $-2\rch$ \!\!\!&\!\!\! $2\rch-1$ \!\!\!&\!\!\!
 $1-2\rch$ \!\!\!&\!\!\! $2\rch-2$ \!\!\!&\!\!\! $2-2\rch$\!\! \STRUT\\
\end{tabular}
\caption{\label{tab:R} The vector R-charges of fields in a GLSM.}
\end{center}
}
\end{table}

For the action, we take the integral of $(1/2\pi)$ times the following Lagrangians. First, the supersymmetric kinetic Lagrangians for vector and chiral multiplets are given by
\begin{eqnarray}
 {\cal L}_{\text{vec}} &=&
 \text{Tr}\Big[
 \big(F_{12}-H\rho\big)^2
 +\big(D+H\sigma\big)^2+D_m\rho D^m\rho+D_m\sigma D^m\sigma
 -[\rho,\sigma]^2
  \nonumber \\ && \hskip6mm
 -\bar\lambda\big(
  \gamma^mD_m\lambda-i\gamma^3[\rho,\lambda]-[\sigma,\lambda]\big)
  \Big],
 \nonumber \\
 {\cal L}_{\text{mat}} &=&
 D_m\bar\phi D^m\phi+\bar\phi\Big\{\rho^2+\sigma^2+2iqH\sigma
 +\tfrac12\rch R-\rch^2H^2+iD\Big\}\phi
 +\bar FF
 \nonumber \\ &&
 -\bar\psi\Big(\gamma^mD_m-i\gamma^3\rho-\sigma-i\rch H\Big)\psi
 +i\bar\psi\bar\lambda\phi-i\bar\phi\lambda\psi\,.
\label{Lvm}
\end{eqnarray}
One can also introduce interactions in the form of the F-terms of gauge invariant chiral multiplets. These Lagrangians are actually all SUSY exact. As is well known, for supersymmetric observables defined by path integrals, one can shift the integrand by SUSY-exact quantities without changing the value of the integrals. Supersymmetric observables therefore do not depend on the couplings appearing in SUSY-exact Lagrangians. For Abelian vector multiplets, we also have the FI-theta term,
\begin{equation}
 {\cal L}_{\text{FI}}~=~-irD + i\theta F_{12},
\end{equation}
which is SUSY-invariant but not exact.

\paragraph{LG models.}

Let us next summarize the construction of LG models. A twisted chiral multiplet consists of a complex scalar $Y$, fermions $\chi^+,\bar\chi^-$ and an auxiliary complex scalar $G$. Their conjugates $\bar Y,\bar\chi^+,\chi^-,\bar G$ form an anti-twisted chiral multiplet. They transform under SUSY as follows,
\begin{align}
 \susy Y &= i\xi P_-\bar\chi-i\bar\xi P_+\chi, \nonumber \\
 \susy\bar Y &= i\xi P_+\bar\chi-i\bar\xi P_-\chi, \nonumber \\
  \susy \chi &=
+ P_+(\xi G -i\gamma^m\xi D_mY)  +P_-(\xi\bar G-i\gamma^m\xi D_m\bar Y),
 \nonumber \\
 \susy\bar\chi &=
 -P_-(\bar\xi G -i\gamma^m\bar\xi D_mY)  -P_+(\bar\xi\bar G-i\gamma^m\bar\xi D_m\bar Y),
 \nonumber \\
 \susy G &=
\xi\gamma^mD_m P_-\bar\chi -\bar\xi\gamma^mD_m P_+\chi,
 \nonumber \\
 \susy\bar G &=
\xi\gamma^mD_m P_+\bar\chi - \bar\xi\gamma^mD_m P_-\chi,
\end{align}
where $P_\pm\equiv\frac12(1\pm\gamma^3)$. We summarize the vector R-charges of fields in a LG model in Table~\ref{tab:R-LG}.
\begin{table}[htbp]
\def\STRUT{\rule[-0.85mm]{0mm}{5.5mm}}
{\small
\begin{center}
\begin{tabular}{c||ccccccccccccc}
\!\!field & $Y$ & $\bar{Y}$ & $\chi$ & $\bar{\chi}$ &$G$ & $\bar{G}$ \STRUT\\
\hline
$\text{R}_{\text{V}}$ & $0$ & $0$ & $+1$& $-1$ & $0$ & $0$
\STRUT\\
\end{tabular}
\caption{\label{tab:R-LG} The vector R-charges of fields in a LG model.}
\end{center}
}
\end{table}
It is known that one can form a twisted chiral multiplet from an Abelian vector multiplets by identifying the scalar fields as
\begin{equation} \label{YG-Sigma}
Y_\Sigma =  \Sigma\equiv\sigma-i\rho,\quad
  G_\Sigma= D+iF_{12} + H \Sigma.
\end{equation}
The free kinetic Lagrangian
\begin{equation}
 {\cal L}_{\widetilde{\text{mat}}}=D^m\bar YD_mY -\bar\chi\gamma^mD_m\chi
 +\bar G G
\end{equation}
is SUSY-exact. A SUSY invariant but non-exact interaction can be introduced by choosing a holomorphic function of the twisted chiral fields $\widetilde W(Y_i)$ of $\RA=2$, called the  twisted superpotential.
\begin{eqnarray}
 {\cal L}_{\widetilde{W}} &=&
\frac i2\sum_i\left\{
 G_i \frac{\partial\widetilde W}{\partial Y_i}
+\bar G_i  \frac{\partial\overline{\widetilde W}}{\partial\bar Y_i}
\right\}
-\frac{iH}2(\widetilde W+\overline{\widetilde W})
 \nonumber \\ && 
\qquad \qquad
+\frac12\sum_{i,j}\left\{\frac{\partial^2\widetilde W}{\partial Y_i\partial Y_j}\chi^{i+}\bar\chi^{j-} +\frac{\partial^2\overline{\widetilde W}}{\partial\bar Y_i\partial\bar Y_j}\bar\chi^{i+}\chi^{j-}\right\}.
\label{LWt}
\end{eqnarray}

\subsection{BPS vortex defects}
\label{sec:BPS-defects}

Here we present definitions of BPS vortex defects in ${\cal N}=(2,2)$ SUSY theories on the flat plane. We use the coordinates $x^1,x^2$ or $z=x^1+ix^2=re^{i\varphi}$, and the defects are put at the origin unless otherwise stated. There are four supercharges corresponding to the constant components of $\xi^\pm,\bar\xi^\pm$, which we denote by $\susy_\pm,\bar\susy_\pm$.

\paragraph{Gauge versus flavor vortex defects.}

Gauge vortex defects are operators in gauge theories around which the dynamical gauge field $A$ is required to take the singular form,
\begin{equation}
 A \simeq \eta d\varphi\quad\text{or}\quad
 A \simeq \frac{\eta}{2i}\left(\frac{dz}z-\frac{d\bar z}{\bar z}\right).
\label{dfv}
\end{equation}
The vorticity $\eta$ is a Lie algebra valued constant. For non-Abelian theories, $\eta$ is assumed to be in Cartan subalgebra.

We will see in $U(1)$ examples that gauge vortex defects in fact depend on $\eta$ in a piecewise constant manner, and that they are equivalent to genuine local operators constructed from the scalars in the vector multiplet.  This is analogous to the familiar fact that for a 2d CFT with a discrete symmetry, an operator corresponding to a twisted sector is mutually non-local with the original fields, but becomes a genuine local operator when we orbifold the CFT because the operators that transform under the symmetry get projected out.

For theories with global symmetry, one can turn on,  for  its Abelian subgroup,   a flat background gauge field $A^{(\text{f})}$ with the above vortex-type singularity.   We call this a flavor vortex defect.    For example, the configuration of flavor vortex defects with vorticity $\eta_i$ at $z=z_i$ is obtained by turning on the following $A^{(\text{f})}$ in the background:
\begin{equation}
 A^{(\text{f})} = \sum_i\frac{\eta_i}{2i}\left(
 \frac{dz}{z-z_i}-\frac{d\bar z}{\bar z-\bar z_i}\right)
 =ig^{-1}dg,\qquad
 g=\prod_i\left(\frac{z-z_i}{\bar z-\bar z_i}\right)^{-\frac{\eta_i}2}.
\end{equation}
Note that, although $A^{(\text{f})}$ looks pure gauge, $g$ is generically not a single-valued function. Consider now the case with $U(1)$ flavor symmetry, and let $\phi$ be a field with a unit flavor charge. The field $\phi$ transforms as $\phi\to g\phi$ under the gauge transformation that eliminates $A^{(\text{f})}$, so that $\phi$ in the new gauge becomes generically non-periodic around each defect. For example, the monodromy of $\phi$ around a defect of vorticity $\eta$ at the origin is
\begin{equation}
 \phi(ze^{2\pi i})= e^{-2\pi i\eta}\phi(z).
\end{equation}
 This means that, unlike gauge vortex defects, a flavor vortex defect with non-integer $\eta$ does not define a genuine local operator;  it is the end point of a topological line operator \cite{Kapustin:2014gua}.  

Vortex defects without dynamical vector multiplets  have applications to orbifold theories, {\it i.e.}, theories with discrete gauge symmetries. Namely, the vortex operator with $\eta=r/K\,(r,K\in\mathbb Z_+)$ is a realization of the twist field that belongs to the $r$-th twisted sector of the $\mathbb Z_K$ orbifold theory.

The above simple discussion also implies an important property of the defects: under the assumption of charge quantization, the vorticity $\eta$ is a periodic variable as far as the behavior of charged matter fields  around the defect is concerned. This is because the defects with suitably quantized $\eta$ can be gauged away without introducing multivaluedness to the charged matter fields.

\paragraph{BPS condition.}

The SUSY transformation of the gauginos $\lambda,\bar\lambda$ on the vortex defect configuration~(\ref{dfv}) is generically non-zero, except when the auxiliary field $D$ is appropriately turned on at the same time. Half of the ${\cal N}=(2,2)$ SUSY is preserved in the following two cases.
\begin{alignat}{2}
 D&=+iF_{12}=+2\pi i\eta\delta^2(x) &\quad :  \quad \susy_-,\bar\susy_+ &\text{ are preserved.}
 \nonumber \\
 D&=-iF_{12}=-2\pi i\eta\delta^2(x) &\quad :  \quad \susy_+,\bar\susy_- &\text{ are preserved.}
\label{dfdv}
\end{alignat}
In the former (latter) case the defect is a twisted chiral (resp. anti-twisted chiral) operator.

The SUSY corresponding to the Killing spinor~(\ref{KS}) on the squashed sphere of Section~\ref{sec:review-N22} is a combination of $\susy_-$ and $\bar\susy_+$ at the north pole, while it is a combination of $\susy_+$ and $\bar\susy_-$ at the south pole. Inserting a twisted chiral defect at the north pole and an anti-twisted chiral defect at the south pole does not break the supersymmetry, so their correlators can be evaluated exactly.

\paragraph{Boundary condition on  fields at the defect.}

The vortex defects also affect the measure of path integration for charged matter fields. A standard way to define the path integral over matter fields  is to decompose the fields into the eigenfunctions of the Laplace or Dirac operators on the relevant gauge field configuration. The behavior of general matter wave functions will therefore be modified by the defects.

Take the 2d plane with a single defect of vorticity $\eta$ at the origin, and consider a (bosonic or fermionic) matter field of charge $w$ around it. (For a non-Abelian theory, $w$ is a weight vector of some representation of the gauge group, so that $w\cdot\eta$ is the standard inner product in the weight space.) The field is then expanded in the eigenfunctions  of an appropriate differential operator. Using the rotational invariance of the defect, one may assume general eigenfunctions to take the separated form $\Psi = \hat\Psi_m(r)e^{im\varphi}\,(m\in\mathbb Z)$. One can then show that, on the defect background with vorticity $\eta$, the radial part of the general solution to the differential equation takes the following form near the defect:
\begin{equation}
 \hat\Psi_m(r) \sim
 \left\{\begin{array}{ll}
  c_+ r^{m-w\cdot\eta} +c_- r^{-(m-w\cdot\eta)}
 & (m\ne w\cdot\eta) \\
  c_0 +c_1\ln r
 & (m = w\cdot\eta) \\
	\end{array}\right. .
\label{frpw}
\end{equation}
Note the appearance of  non-integer  powers of $r$. This behavior will be confirmed explicitly for the vortex defect correlators  on the  squashed sphere in Section~\ref{sec:vcs} where the spectrum becomes discrete. This implies that a solution finite at the defect may well turn into  a diverging function after being differentiated several times.  Therefore, requiring all the wave functions to be finite at the defect does not necessarily lead to a definition of supersymmetric defects, because SUSY transformation involves derivatives.

Let us take the vortex defect at the origin to be twisted chiral and denote the preserved SUSY by $\susy_\atw\equiv \varepsilon\susy_-+\bar\varepsilon\bar\susy_+$, namely we set $\xi^-=\varepsilon$, $\bar\xi^+=\bar\varepsilon$ and $\xi^+=\bar\xi^-=0$. This is the combination used for defining A-twisted topological theories. The fields in a charged chiral multiplet transform under it as 
\begin{alignat}{4}
 \susy_\atw\phi~~~ &= -\varepsilon\psi^+,\qquad &
 \susy_\atw\psi^- &= \varepsilon F-2\bar\varepsilon D_{\bar z}\phi,
\nonumber \\
 \susy_\atw\psi^+ &= -\bar\varepsilon\Sigma\phi,\qquad &
 \susy_\atw F~~ &= -\bar\varepsilon\big\{2D_{\bar z}\psi^+-\Sigma\psi^--i\bar\lambda^-\phi\big\}
\label{catw}
\end{alignat}
with $\Sigma\equiv \sigma-i\rho$.
 For  the boundary condition to  be consistent with the unbroken SUSY, we must require $\phi,\psi^+$ to satisfy the same boundary condition, and $\psi^-,F$ to satisfy the same boundary condition as $D_{\bar z}\phi$. We thus arrive at two Hilbert spaces and a map
\begin{equation}
D_{\bar z}:{\cal H}\to{\cal H}'.\quad\left(
\begin{array}{lcl}
 {\cal H} &\equiv
 \text{space of wave functions of $\phi$ and $\psi^+$}
 \nonumber \\
 {\cal H}'&\equiv
 \text{space of wave functions of $\psi^-$ and $F$}
\end{array}
\right)
\end{equation}
Applying the same argument to the anti-chiral multiplet, one finds that  $\bar\phi$ and $\bar\psi^-$ belong to $\bar{\cal H}$ (the conjugate of ${\cal H}$) while $\bar\psi^+$ and $\bar F$ belong to $\bar{\cal H}'$. Furthermore, recall that the standard Dirac Lagrangian for the fermions
\begin{equation}
 -\bar\psi\gamma^mD_m\psi = -2\bar\psi^+D_{\bar z}\psi^++2\bar\psi^-D_z\psi^-.
\end{equation}
For the Hermiticity of the Dirac operator one needs to require
\begin{equation}
D_z{\cal H}'\subset{\cal H},\qquad
D_{\bar z}{\cal H}\subset{\cal H}'.
\end{equation}
The basis of ${\cal H}$ and ${\cal H}'$ can therefore be chosen as the complete set of eigenfunctions of the operators $D_zD_{\bar z}$ and $D_{\bar z}D_z$, respectively.

The above relation between ${\cal H}$ and ${\cal H}'$ constrains the allowed boundary condition in the following way. Suppose one requires $\phi,\psi^+\in{\cal H}$ to be finite at the defect. Then their image under $D_{\bar z}$ may diverge mildly as $r^\gamma\,(\gamma>-1)$ at the defect. Therefore one must allow $\psi^-,F\in{\cal H}'$ to diverge in the same way. Note this does not cause problems with square normalizability. In addition, since $D_z{\cal H}'\subset{\cal H}$ we must require $D_z\psi^-,D_zF$ be finite at the defect. Thus we found a set of SUSY-preserving boundary conditions (on components)  summarized as
\begin{equation} \label{eq:nbc}
 \phi,\,\psi^+,\,D_z\psi^-,\,D_zF\text{ are finite at the defect }
 (\psi^-,F\text{ may diverge mildly})
\end{equation}
which we call the {\it normal boundary condition} (on the multiplet). By exchanging the role of ${\cal H}$ and ${\cal H}'$ one obtains another consistent set of supersymmetric boundary conditions
\begin{equation} \label{eq:fbc}
 D_{\bar z}\phi,\,D_{\bar z}\psi^+,\,\psi^-,\,F\text{ are finite at the defect }
 (\phi,\psi^+\text{ may diverge mildly})
\end{equation}
which we call the {\it flipped boundary condition}. The choice of normal or flipped boundary condition can be made for each irreducible set of matter chiral multiplets. The choice of matter boundary conditions should be regarded as discrete labels characterizing the defect.

For a  non-Abelian gauge theory, the fluctuations  of (dynamical) vector multiplet fields around the  vortex defect background given in (\ref{dfv}) and (\ref{dfdv}) can be treated in a way similar to the way the charged matter fields were treated. Let us again put a twisted chiral defect at the origin, and look at the action of unbroken SUSY $\susy_\atw$ on the vector multiplet fields. Using $\Sigma\equiv\sigma-i\rho$, $G\equiv D+iF_{12}$ and $\bar\Sigma\equiv\sigma +i\rho$, $\bar G\equiv D-iF_{12}$, one finds
\begin{alignat}{2}
 \susy_\atw A_z&= +\tfrac12\bar\varepsilon\lambda^+,
 \qquad &
 \susy_\atw\lambda^+& = -2i\varepsilon D_z\Sigma,
 \nonumber \\
 \susy_\atw A_{\bar z}&= -\tfrac12\varepsilon\bar\lambda^-,
 \qquad &
 \susy_\atw\bar\lambda^-& = +2i\bar\varepsilon D_{\bar z}\Sigma,
 \nonumber \\
 \susy_\atw\bar\Sigma &= -i(\varepsilon\bar\lambda^++\bar\varepsilon\lambda^-),
 \qquad &
 \susy_\atw\lambda^-& = \varepsilon\big\{\tfrac i2[\Sigma,\bar\Sigma]+\bar G\big\},
 \nonumber \\
 \susy_\atw\bar G &= \tfrac12[\Sigma,\bar\varepsilon\lambda^--\varepsilon\bar\lambda^+],
 \qquad &
 \susy_\atw\bar\lambda^+& = \bar\varepsilon\big\{\tfrac i2[\Sigma,\bar\Sigma]-\bar G\big\},
\label{vatw}
\end{alignat}
and $\susy_\atw\Sigma=\susy_\atw G=0$.
Let primed fields $A'_z,A'_{\bar z},D'$ denote the fluctuations from the given vortex defect configuration. We obtain  the groups of fields obeying the same boundary conditions
\begin{equation}
  A'_z,\lambda^+ \in\,{\cal H}, \quad
  A'_{\bar z},\bar\lambda^- \in\,\bar{\cal H}, \quad
  \sigma,\rho,D',\lambda^-,\bar\lambda^+ \in\,{\cal H}'=\bar{\cal H}',
\end{equation}
and the differential operators relating them 
\begin{equation}
 D_{\bar z}: {\cal H}\to {\cal H}'\to\bar{\cal H},\quad
 D_z: \bar{\cal H}\to {\cal H}'\to {\cal H}.
\end{equation}
Due to the insertion of the defect, the gauge symmetry at the defect is broken to a subgroup of elements which commute with $\eta$. For example, if $\eta$ is a generic element of the Cartan subalgebra, the parameter of gauge transformation has to be a Lie algebra valued function whose ladder operator part vanishes at the defect. Since (\ref{vatw}) implies $\susy_\atw^2=\varepsilon\bar\varepsilon\text{Gauge}(\Sigma)$, it follows that $\Sigma$ and all other elements of ${\cal H}'$ have to obey this boundary condition. So we impose:
\begin{equation}
 D_{\bar z}\{A'_z,\, \lambda^+\},\,
 D_z\{A'_{\bar z},\, \bar\lambda^-\},\,
 \{\sigma,\rho,D',\lambda^-,\bar\lambda^+\}\text{ are finite at the defect.}
\end{equation}
This is the only consistent boundary condition for vector multiplets.

So far we have been studying the boundary conditions on fields  near a twisted chiral vortex defect. For the boundary conditions near an anti-twisted chiral defect, the supersymmetric boundary conditions can be classified in the same way. For example, the normal boundary condition on matter requires $\phi,\psi^-,D_{\bar z}\psi^+$ and $D_{\bar z}F$ to be finite at the defect. For completeness we list the fields required to be finite around different vortex defects in Table~\ref{tab:bcs}.

\begin{table}
\begin{tabular}{|c||c|c|c|}
 \hline
 defect type & matter -- normal b.c. & matter -- flipped b.c. & vector multiplet\\
 \hline
 twisted chiral
 & $\phi,\psi^+,D_z\{\psi^-,F\}$
 & $D_{\bar z}\{\phi,\psi^+\},\psi^-,F$
 & $D_{\bar z}\{A'_z,\lambda^+\},D_z\{A'_{\bar z},\bar\lambda^-\},$ \\

 & $\bar\phi,\bar\psi^-,D_{\bar z}\{\bar\psi^+,\bar F\}$
 & $D_z\{\bar\phi,\bar\psi^-\},\bar\psi^+,\bar F$
 & $\sigma,\rho,D',\lambda^-,\bar\lambda^+$ \\
 \hline
 anti-twisted chiral
 & $\phi,\psi^-,D_{\bar z}\{\psi^+,F\}$
 & $D_z\{\phi,\psi^-\},\psi^+,F$
 & $D_z\{A'_{\bar z},\lambda^-\},D_{\bar z}\{A'_z,\bar\lambda^+\},$ \\
 & $\bar\phi,\bar\psi^+,D_z\{\bar\psi^-,\bar F\}$
 & $D_{\bar z}\{\bar\phi,\bar\psi^+\},\bar\psi^-,\bar F$
 & $\sigma,\rho,D',\lambda^+,\bar\lambda^-$ \\
 \hline
\end{tabular}
\caption{The fields required to be finite at the vortex defects preserving different supersymmetries.}
\label{tab:bcs}
\end{table}

\subsection{Smeared vortex defect configurations}
\label{sec:smear-vort-defects}

Another natural approach to studying  vortex defects is to define them as a limit of smooth gauge field configurations. This was originally done in~\cite{Kapustin:2012iw} and applied to the flavor vortex defects in 2d in~\cite{Okuda:2015yra}.

On the flat space with coordinates $x^1,x^2$ or $z=x^1+ix^2$, a smeared BPS vortex defect configuration for a $U(1)$ gauge field is defined as
\begin{equation} \label{smear-background-flat}
D= \pm 2 \pi i  \varrho  \,,\qquad
F_{z\bar z} =  \pi i \varrho \,,
\end{equation}
where the vorticity density $\varrho$ is an arbitrary real function, and  the sign $+$($-$) is for the vortex defect configuration corresponding to a twisted chiral(anti-twisted chiral) operator.%
\footnote{%
Compared with~\cite{Okuda:2015yra}, we have $\varrho = -\rho^\text{\cite{Okuda:2015yra}}$, $A_\mu = - v_\mu^\text{\cite{Okuda:2015yra}}$, $D=i D^\text{\cite{Okuda:2015yra}}$. Compared with the $\ell\rightarrow \infty$ limit of Section~\ref{sec:smear-sphere} at the north pole, we have $\varrho= (2\pi r)^{-1}\partial_r S$.}
A smeared flavor vortex defect  can be introduced by turning on  a  background vector multiplet configuration of the above form for  an  Abelian flavor symmetry of the theory. One should be able to define a local operator , or more precisely the end of a topological line operator \cite{Kapustin:2014gua},  in  the limit  $\varrho(x)\to \eta\cdot\delta^2(x)$.

For a sufficiently well-behaved $\varrho(x)$, the charged matter fields should all behave regularly at the defect. In particular, there is no such issue as the matter wave functions exhibiting  a non-integer  power law behavior or divergence. The observables associated to the smeared  defects  will therefore depend on $\eta$ in a smooth manner. Also, the gauge field configuration~(\ref{smear-background-flat}) is no longer  pure gauge, so there is no reason that the observables are periodic in $\eta$.

\subsection{0d-2d coupled systems}
\label{sec:defects-via-coupling-flat}

Another way to define  a codimension-two defect  is to introduce a set of dynamical variables localized on the defect and couple it to the 2d bulk fields,  as  in~\cite{Gomis:2016ljm}.  For the 0d degrees of freedom to realize the properties similar to those of vortex defects, they need to transform under  (a subgroup of) the gauge symmetry in the bulk. In this paper we will mostly focus on the 0d variables coupled to the bulk $U(1)$ gauge theory.

\paragraph{0d-2d couplings in flat space.}

The localized degrees of freedom also transform under unbroken SUSY. Let us here take the defect to be twisted chiral and denote the  parameters of the unbroken SUSY by $\xi^-=\varepsilon$ and $\bar\xi^+=\bar\varepsilon$.  The supercharge is $\susy_\atw=\varepsilon\susy_-+\bar\varepsilon \bar\susy_+$ as in Section~\ref{sec:BPS-defects},  and it satisfies
\begin{equation}
 \susy_\atw^2 = \varepsilon\bar\varepsilon\,\text{Gauge}(\Sigma),
\label{qa2}
\end{equation}
where $\Sigma$ here  is  the value of the 2d field $\sigma-i\rho$ at the
defect.
The bulk chiral and vector multiplet fields transform as in~(\ref{catw}) and~(\ref{vatw}), which in particular show how the 2d ${\cal N}=(2,2)$ multiplets decompose into multiplets of the smaller SUSY $\susy_\atw$. We  consider two kinds of 0d supermultiplets, called chiral and Fermi multiplets.%
\footnote{%
These multiplets are the dimensional reductions of two-dimensional $\mathcal{N}=(0,2)$ supermultiplets. We thus use the terminology familiar in that context.
}
We use bold symbols for these localized degrees of freedom.

Let us now consider a defect at the origin $x^1=x^2=0$ of the flat plane. A 0d chiral multiplet on the defect is a pair of a boson $\lcl u$ and a fermion $\lcl\zeta$ transforming in a representation of the bulk gauge group. Its conjugate anti-chiral multiplet is denoted by $(\bar{\lcl u},\bar{\lcl{\zeta}})$.  These variables  transform under $\susy_\atw$ as 
\begin{alignat}{2}
 \susy_\atw \lcl u&= \varepsilon\lcl\zeta,\qquad &
 \susy_\atw \lcl\zeta &= \bar\varepsilon\Sigma\lcl u,
 \nonumber \\
 \susy_\atw \bar{\lcl u}&= \bar\varepsilon\bar{\lcl\zeta},\qquad &
 \susy_\atw \bar{\lcl\zeta} &= -\varepsilon\bar{\lcl u}\Sigma.
\label{0dc}
\end{alignat}
It is easy to see that   $(\lcl u,\lcl\zeta)$ and $(\bar{\lcl u},\bar{\lcl{\zeta}})$ transform in the same way as the 2d chiral multiplet fields $(\phi,-\psi^+)$ and $(\bar\phi,\bar\psi^-)$ at the defect. Likewise, a Fermi multiplet consists of a fermion $\lcl\eta$ and a boson $\lcl h$, and its conjugate anti-Fermi multiplet consists of $\bar{\lcl\eta},\bar{\lcl h}$. They are simply chiral and anti-chiral multiplets with flipped statistics.%
\footnote{%
\label{holo-cont}%
More generally, the supersymmetry transformations of a Fermi multiplet can involve a holomorphic function of chiral multiplet scalars.
This is the case for the Fermi multiplet that arises as a restriction of a bulk chiral multiplet as can be seen from (\ref{catw}).
For the Fermi multiplets that originate from localized modes in Section~\ref{sec:relat-between-defect}, such terms are absent.
}
They transform under $\susy_\atw$ as
\begin{alignat}{2}
 \susy_\atw \lcl\eta &= \varepsilon\lcl h,\qquad &
 \susy_\atw \lcl h & = \bar\varepsilon\Sigma\lcl\eta,
 \nonumber \\
 \susy_\atw \bar{\lcl\eta} &= \bar\varepsilon\bar{\lcl h},\qquad &
 \susy_\atw \bar{\lcl h} & = -\varepsilon\bar{\lcl\eta}\Sigma.
\label{0df}
\end{alignat}
They correspond to the other half of the 2d chiral multiplet fields, $(\psi^-,F)$ and $(-\bar\psi^+,\bar F)$.

So far we have only discussed short multiplets. A general long multiplet $(\lcl C,\lcl\vartheta,\bar{\lcl\vartheta},\lcl M)$ coupled to the bulk vector multiplet transforms as
\begin{alignat}{2}
 \susy_\atw \lcl C
  &= \varepsilon\bar{\lcl\vartheta}+\bar\varepsilon\lcl\vartheta,
 \qquad &
 \susy_\atw\lcl\vartheta
 &= -\varepsilon\lcl M+\frac 12\varepsilon\Sigma\lcl C,
 \nonumber \\
 \susy_\atw\bar{\lcl\vartheta}
 &= \bar\varepsilon\lcl M+\frac 12\bar\varepsilon\Sigma\lcl C,
 \qquad &
 \susy_\atw \lcl M
 &= \frac 12\varepsilon\Sigma\bar{\lcl\vartheta}
  -\frac 12\bar\varepsilon\Sigma\lcl\vartheta.
\label{0dl}
\end{alignat}
An example of this is the 2d anti-twisted chiral multiplet $(\bar\Sigma,-i\lambda^-,-i\bar\lambda^+,i\bar G)$ at the defect, which is in the adjoint representation. See~(\ref{vatw}).

BPS defects are defined by the integrals over the localized degrees of freedom with a supersymmetric weight $e^{-S_\text{0d}}$. For the study of supersymmetric ($\susy_\atw$-invariant) observables, it only matters to choose  the action  $S_\text{0d}$ up to $\susy_\atw$-exact terms. The above discussions on 0d supermultiplets shows that the only $\susy_\atw$-invariants are the $\lcl M, \lcl h$ and $\bar{\lcl h}$-components of gauge-invariant supermultiplets. Moreover, they are all $\susy_\atw$-exact: for example,
\begin{equation}
\lcl M=\susy_\atw\left(\frac{\varepsilon\bar{\lcl\vartheta}-\bar\varepsilon\lcl\vartheta}{2\varepsilon\bar\varepsilon}\right).
\end{equation}
Different $S_\text{0d}$ are therefore all equivalent in the sense of $\susy_\atw$-cohomology, so the defects are classified only by the number of 0d multiplets introduced on it. The choice of $S_\text{0d}$ does matter, however, if one is interested in observables that are not protected by SUSY.

Let us now study some explicit choices of $S_\text{0d}$. First, for the Fermi multiplet $(\lcl\eta,\lcl h)$ and $(\bar{\lcl\eta},\bar{\lcl h})$, the most natural choice would be the $\lcl M$-component of the long multiplet with the lowest component~$\bar{\lcl\eta}\lcl\eta$,
\begin{equation}
 S_\text{0d}^\text{(F)} = \big(-\bar{\lcl\eta}\lcl\eta\big)_{\lcl M}
 =\bar{\lcl h}\lcl h-  \bar{\lcl\eta}\Sigma\lcl\eta.
\label{szf1}
\end{equation}
For the chiral multiplet, a similar construction leads to
\begin{equation}
\big(\bar{\lcl u}\lcl u\big)_{\lcl M}
 = \bar{\lcl u}\Sigma\lcl u+\bar{\lcl\zeta}\lcl\zeta.
\label{szc1}
\end{equation}
A better choice for which the quadratic term in $\lcl u,\bar{\lcl u}$ is positive definite would be%
\footnote{%
For the Fermi and chiral 0d multiplets obtained by restricting a bulk chiral multiplet,
the sum of~(\ref{szf1}) and~(\ref{szc2}) coincides with the bulk Lagrangian~(\ref{Lvm}) restricted to a point up to the extra contributions mentioned in footnote~\ref{holo-cont}.
}
\begin{equation}
 S^\text{(C)}_\text{0d} = \big(\bar{\lcl u}\bar\Sigma\lcl u\big)_{\lcl M}
 = \frac12\bar{\lcl u}\{\bar\Sigma,\Sigma\}\lcl u
 +i\bar{\lcl u}\bar G\lcl u +\bar{\lcl\zeta}\bar\Sigma\lcl\zeta
 -i\bar{\lcl u}\lambda^-\lcl\zeta-i\bar{\lcl\zeta}\bar\lambda^+\lcl u .
\label{szc2}
\end{equation}
The integral over the 0d Fermi or chiral multiplets with the weights~(\ref{szf1}) or~(\ref{szc1}) gives $\text{det}\Sigma$ or $(\text{det}\Sigma)^{-1}$, respectively. For the choice of the weight~(\ref{szc2}), the result of the integration is the ratio of determinants,
\begin{equation}
 \int d[\lcl u,\bar{\lcl u},\lcl\zeta,\bar{\lcl\zeta}]e^{-(\bar{\lcl u}\bar\Sigma\lcl u)_{\lcl M}} = \frac{\text{det}\bar\Sigma}{\text{det}(\frac12\{\Sigma,\bar\Sigma\}+i\bar G+\lambda^-\bar\Sigma^{-1}\bar\lambda^+)},
\label{zdit1}
\end{equation}
but it should differ from $(\text{det}\Sigma)^{-1}$ only by $\susy_\atw$-exact terms.

Let us also discuss the couplings between 0d-2d variables via $\lcl h$ and $\bar{\lcl h}$-type interactions. Let $(\phi,\psi,F)$ be a 2d chiral multiplet, and $(\phi,-\psi^+)$ its components on the defect which form a 0d chiral multiplet. Let $(\lcl\eta,\lcl h)$ be a 0d Fermi multiplet on the defect in the conjugate representation of the gauge symmetry relative to $(\phi,\psi,F)$. Then one has the $\lcl h$ and $\bar{\lcl h}$-type invariants
\begin{equation}
 \big(\lcl\eta\phi\big)_{\lcl h} = \lcl h\phi-\lcl\eta\psi^+,\qquad
 \big(\bar\phi\bar{\lcl\eta}\big)_{\bar{\lcl h}} = \bar\phi\bar{\lcl h}+\bar\psi^-\bar{\lcl\eta}.
\end{equation}
Integration over the 0d variables imposes the Dirichlet-like boundary condition on the bulk field components $(\phi,\psi^+;\bar\phi,\bar\psi^-)$. A similar boundary condition can also be imposed on the other half of 2d chiral multiplet components $(\psi^-,F;\bar\psi^+,\bar F)$ by using a suitable 0d chiral multiplet.

\paragraph{Omega-deformation of the 0d SUSY.}

In Section~\ref{sec:smear-sphere} we perform localization calculations on the squashed sphere~(\ref{BG1}) with 0d multiplets introduced at the north and south poles. There we will need the omega-deformed version of the 0d supersymmetry.

Let us consider the SUSY on the squashed sphere in Section~\ref{sec:review-N22} with the replacement of Killing spinors $\xi\to\varepsilon\xi,\bar\xi\to\bar\varepsilon\bar\xi$. The corresponding supercharge, which we could denote as $\varepsilon\susy_\xi+\bar\varepsilon\susy_{\bar\xi}$, can be shown to satisfy
\begin{equation}
 (\varepsilon\susy_\xi+\bar\varepsilon\susy_{\bar\xi})^2
 = \varepsilon\bar\varepsilon\left[
  \text{Gauge}(\hat\Sigma)
 +\frac i\ell\left(J_3+\frac12\RV\right)
 \right],\quad\hat\Sigma\equiv\sigma-i\cos\theta\rho-\frac i\ell A_\varphi
\end{equation}
on all fields on the squashed sphere. Here $J_3=-i\partial_\varphi$ is the generator of its isometry fixing the two poles. The 0d multiplets on the north pole transform under this supercharge restricted to $\theta=0$, which we denote as $\susy_{\atw,\text{\smaller[3]$\Omega$}}$. Its algebra is nearly the same as~(\ref{qa2}) but is deformed by the terms of order ${\cal O}(1/\ell)$. Let us introduce $\text{J}\equiv J_3+\frac12\RV$ and write
\begin{equation}
 \susy_{\atw,\text{\smaller[3]$\Omega$}}^2 = \varepsilon\bar\varepsilon\left[
  \text{Gauge}(\hat\Sigma)+\frac i\ell\,\text{J}\, \right]\bigg|_\text{north pole}.
\label{qao2}
\end{equation}
Near the south pole, the SUSY on the squashed sphere approaches $i\varepsilon \susy_++i\bar\varepsilon \bar\susy_-$, which is the SUSY of anti-topological A-twisted theories. Its algebra is the same as~(\ref{qao2}) above with $\hat\Sigma$ replaced by $\sigma+i\rho-i\ell^{-1}A_\varphi$ evaluated on the south pole.

For each 0d multiplet at the north or the south pole one needs to specify a representation of the bulk gauge group as well as the $\text{J}$ quantum number. With this understood, the transformation rule of 0d multiplets is the same as~(\ref{0dc}),(\ref{0df}),(\ref{0dl}) with $\text{Gauge}(\Sigma)$ replaced by $\text{Gauge}(\hat\Sigma)+i\ell^{-1}\text{J}$. For example, the 0d chiral and anti-chiral multiplets $(\lcl u,\lcl\zeta),(\bar{\lcl u},\bar{\lcl\zeta})$ with $\text{J}=J,-J$ transform as follows.
\begin{alignat}{2} \label{eq:chiral-deformed}
 \susy_{\atw,\text{\smaller[3]$\Omega$}} \lcl u&= \varepsilon\lcl\zeta, \qquad &
 \susy_{\atw,\text{\smaller[3]$\Omega$} } \lcl\zeta &= \bar\varepsilon\hat\Sigma\lcl u + i \frac{\bar\varepsilon}{\ell} J\lcl u,
 \nonumber \\
 \susy_{\atw,\text{\smaller[3]$\Omega$}}  \bar{\lcl u}&= \bar\varepsilon\bar{\lcl\zeta},\qquad &
 \susy_{\atw,\text{\smaller[3]$\Omega$}}  \bar{\lcl\zeta} &= -\varepsilon\bar{\lcl u}\hat\Sigma-i\frac{\varepsilon}{\ell} \bar{\lcl u}  J  .
\end{alignat}
The components $(\phi,-\psi^+)$ of a bulk chiral multiplet with $\RV=2q$ restricted to the north pole provide an example of a 0d chiral multiplet with $\text{J}=q$.%
\footnote{%
It is also natural to consider a pair $(\phi,\xi\psi)$ which transforms very much like $(\lcl u,\lcl\zeta)$ above {\it everywhere} on the squashed sphere. We will call them cohomological variables and use them in later sections.
}
Similarly, the components of the anti-twisted chiral multiplet $(\bar\Sigma,-i\lambda^-,-i\bar\lambda^+, i\bar G)$ restricted to the north pole transform as an adjoint 0d long multiplet with $\text{J}=0$.

The supersymmetric action for the 0d multiplets $S_\text{0d}$ can be constructed in the same way as in the previous paragraph for flat space with the replacement $\text{Gauge}(\Sigma)\to \text{Gauge}(\hat\Sigma)+i\ell^{-1}\text{J}$ mentioned above. Among different choices, we will be most interested in the physical actions for the 0d multiplets that arise as localized modes of a bulk chiral multiplet studied in Section~\ref{sec:smear-sphere}. For the 0d Fermi and chiral multiplets of $\RV=2q$ and $J_3=n$ arising in this way, the actions obtained from the curved version of~(\ref{Lvm}) are
\begin{eqnarray}
 S^\text{(F)}_{\text{0d},\Omega} &=&
 \bar{\lcl h}\lcl h-\bar{\lcl\eta}\hat\Sigma\lcl\eta-\frac i\ell(q+n)\bar{\lcl\eta\lcl\eta},
 \label{0d-deformed-action-Fermi-explicit} \\
 S^\text{(C)}_{\text{0d},\Omega} &=&
 \frac12\bar{\lcl u}\left\{\bar\Sigma+\frac{i(q-n-1)}\ell,\hat\Sigma+\frac{i(q+n)}\ell\right\}\lcl u
 +i\bar{\lcl u}\bar G\lcl u
 \nonumber \\ &&
 +\bar{\lcl\zeta}\left(\bar\Sigma+\frac{i(q-n-1)}\ell\right)\lcl\zeta
 -i\bar{\lcl u}\lambda^-\lcl\zeta-i\bar{\lcl\zeta}\bar\lambda^+\lcl u.
 \label{0d-deformed-action-chiral-explicit}
\end{eqnarray}

\section{Vortex defect correlators  on the squashed sphere}\label{sec:vcs}

In this section we study the correlation function of vortex defects on the squashed sphere introduced in Section~\ref{sec:review-N22},
\begin{equation}
 ds^2 = f(\theta)^2d\theta^2+\ell^2\sin^2\theta d\varphi^2,\quad
 H = \frac1{f(\theta)},\quad V = \frac12\left(\frac\ell{f(\theta)}-1\right)d\varphi,
\end{equation}
which has Killing spinors~(\ref{KS}). We first review the computational techniques that are necessary for evaluating partition function, and then extend it for the computation of the defect correlators.

An important building block for evaluating exact supersymmetric observables is the one-loop determinant, which is the Gaussian integral over the fluctuation of fields around given supersymmetric background. The SUSY localization ensures that this Gaussian approximation is actually exact; see~\cite{Pestun:2016zxk} for a review. For simplicity, we focus on the $U(1)$ SQED for which the contribution from  the vector multiplet to the determinant is trivial. In the following we take the system of a $U(1)$ vector multiplet coupled to a single chiral multiplet $(\phi,\psi,F)$ of electric charge $\text{e}=1$, vector R-charge $2q$.

\subsection{Computation of the partition function (review)}
\label{sec:partition-function}

According to the standard localization argument, non-zero contributions to SUSY path integrals arise only from the infinitesimal neighborhood of $\susy$-invariant field configurations called saddle points. Here $\susy$ is the specific SUSY on the squashed sphere corresponding to the choice of Killing spinor~(\ref{KS}). On saddle points, the bosonic fields can be shown to take the  form
\begin{equation}
 \sigma=\frac a\ell,\quad
 D = -\frac a{f\ell},\quad
 A= s\cdot(\cos\theta\mp1)d\varphi,\quad
 \rho = -\frac s\ell,\qquad
 \phi=F=0,
\label{sdl}
\end{equation}
with constants $a\in\mathbb R$ and $s\in\frac12\mathbb Z$, and fermions  set to zero. The value of $s$ is quantized because it is proportional to the magnetic flux through the sphere. The $\mp$ sign indicates that we work in different gauges on the northern and the southern hemispheres so that there is no Dirac string singularity. This will also help us distinguish the singularities due to vortex defects from the Dirac string in later discussions. The partition function can thus be written as
\begin{equation}
 Z
 = \sum_{s\in\frac12\mathbb Z}\int\frac{da}{2\pi}
 e^{-2ira+2is\theta} Z_\text{1-loop}
 = \sum_{s\in\frac12\mathbb Z}\int\frac{da}{2\pi}
 z^{-s+ia}\bar z^{s+ia} Z_\text{1-loop}.
\label{zs2}
\end{equation}
where $z\equiv e^{-t}=e^{-r-i\theta}$. The one-loop determinant $Z_\text{1-loop}$ arises from the integral over the fluctuations of fields around each saddle point labelled by $(a,s)$. For Abelian theories the only contribution to $Z_\text{1-loop}$ is from the charged chiral multiplet fields.

By using the cohomological variables defined by
\begin{equation}
 X\equiv\phi,
~~
 \susy X\equiv\xi\psi,
~~
 \Xi\equiv\bar\xi\psi,
~~
 \susy\Xi
 =F-\bar\xi\gamma^m\bar\xi D_m\phi+i\bar\xi\gamma^3\bar\xi\rho\phi,
\label{dfch}
\end{equation}
one can show that $Z_\text{1-loop}$ is given by the ratio of determinants of $\susy^2$ (see~\cite{Hosomichi:2015pia} for more detail),
\begin{equation}
 Z_\text{1-loop} =
 \frac{\text{det}_{{\cal H}'}(\susy^2)}
      {\text{det}_{{\cal H} }(\susy^2)},
\label{zol}
\end{equation}
acting on the Hilbert spaces ${\cal H},{\cal H}'$ of wave functions of $X$ and $\Xi$, respectively. The elements of ${\cal H}$ are therefore scalars with electric charge $\text{e}=1$ and the R-charge $\RV=2q$, while the elements of ${\cal H}'$ are scalars with $\text{e}=1, \RV=2q-2$. The operator $\susy^2$ acts on such charged scalar fields in general as
\begin{eqnarray}
 \susy^2 &=&
 -\bar\xi\gamma^m\xi\partial_m
 +(\frac i{2f}\bar\xi\xi-iv^mV_m)\RV
 +(\bar\xi\xi\sigma+i\bar\xi\gamma^m\xi A_m+i\bar\xi\gamma^3\xi\rho)\,\text{e}
 \nonumber \\ &=&
 \frac1\ell\left\{\partial_\varphi +\frac i2\RV
 +(a\pm is)\hskip0.5mm\text{e}\right\}.
\label{q2gn}
\end{eqnarray}
The $(+/-)$ sign here means $\susy^2$ takes different form on the north/south patches. Note also that~(\ref{sdl}) implies that the wave function $\Psi$ of a charged scalar on the two patches are related as
\begin{equation}
 \Psi_\south=\Psi_\north\,e^{2is\varphi},
 \quad
 (\susy^2)_\south = e^{2is\varphi}(\susy^2)_\north e^{-2is\varphi}.
\end{equation}
The way the bose-fermi pairs $(X,\susy X)$ and $(\Xi,\susy\Xi)$ belong to the Hilbert spaces ${\cal H}$ and ${\cal H}'$ is similar to what we observed in Section~\ref{sec:BPS-defects}. This structure becomes important in what follows.

The ratio of determinants can be further simplified by noticing that there is a mutually conjugate pair of differential operators $J^\pm$ with R-charges $\mp2$,
\begin{equation}
 J^+ \equiv \ell(\bar\xi\gamma^m\bar\xi D_m-i\bar\xi\gamma^3\bar\xi\rho),\quad
 J^- \equiv \ell(\xi\gamma^m\xi D_m-i\xi\gamma^3\xi\rho),
\end{equation}
both commuting with $\susy^2$. The operator $J^+$ maps the elements of ${\cal H}$ to ${\cal H}'$  while $J^-$ is a map in the opposite direction, both  preserving the eigenvalue of $\susy^2$. Therefore, when taking the ratio of determinants of $\susy^2$ in~(\ref{zol}), one can restrict to the kernels of $J^\pm$. Their explicit form on the two patches is
\begin{eqnarray}
(\text{north})~~~ J^\pm &=& e^{\pm i\varphi}\left\{
  \pm\frac\ell f\partial_\theta + i\cot\theta\partial_\varphi
 +\frac12\left(\frac\ell f-1\right)\cot\theta\RV
 +s\tan\frac \theta2
 \right\},
 \nonumber \\
(\text{south})~~~ J^\pm &=& e^{\pm i\varphi}\left\{
  \pm\frac\ell f\partial_\theta + i\cot\theta\partial_\varphi
 +\frac12\left(\frac\ell f-1\right)\cot\theta\RV
 +s\cot\frac \theta2
 \right\}.
\label{Jwod}
\end{eqnarray}
The differential operators $J^\pm$ with the abovementioned nice properties can be found by inspection, or one can also read them from the fact that the curved space version of the Lagrangian for the chiral multiplet in~(\ref{Lvm}) can be expressed as ${\cal L}_\text{mat}=\susy{\cal V}_\text{mat}$ with
\begin{equation}
 {\cal V}_\text{mat} =
 \bar\Xi\,\susy\Xi + \frac1\ell(\bar\Xi J^+X-\bar X J^-\Xi)
 +\susy\bar X\Big(\susy^2-2\sigma-\frac{i(2q-1)}f\Big)X.
\label{Lch}
\end{equation}

The determinants of $\susy^2$ restricted to the kernels of $J^\pm$ can be obtained by finding out all the eigenvalues explicitly. Suppose that a scalar wave function $\Psi\in{\cal H}$ takes the separated form,
\begin{equation}
\Psi=\hat\Psi(\theta)e^{im_\north\varphi}~~\text{(north patch)},\quad
\Psi=\hat\Psi(\theta)e^{im_\south\varphi}~~\text{(south patch)},
\label{sep}
\end{equation}
with the two integers $m_\north,m_\south$ related as $m_\south=m_\north+2s$. It is an eigenfunction of $\susy^2$ with
\begin{equation}
 \susy^2 = \frac i\ell(m_\north+q-ia+s) = \frac i\ell(m_\south+q-ia-s).
\end{equation}
If $\Psi\in\text{ker}J^+$, then one can show that $\hat\Psi(\theta)$ satisfies a certain first-order differential equation, and behaves near $\theta=0,\pi$ as
\begin{equation}
 \hat\Psi(\theta)\sim \sin^{m_\north}\theta~~(\theta\sim 0),\quad
 \hat\Psi(\theta)\sim \sin^{m_\south}\theta~~(\theta\sim \pi).
\end{equation}
The regularity at the poles requires that $m_\north,m_\south\ge0.$ Therefore
\begin{equation}
 \text{det}(\susy^2)|_{\text{ker}J^+}
 = \prod_{n\ge0}\frac i\ell(n+q-ia+|s|).
\label{dkj1}
\end{equation}
Similarly, if a wave function $\Psi\in{\cal H}'$ takes the separated form~(\ref{sep}), it has the eigenvalue
\begin{equation}
 \susy^2 = \frac i\ell(m_\north+q-1-ia+s) = \frac i\ell(m_\south+q-1-ia-s).
\end{equation}
If $\Psi\in\text{ker}J^-$, then $\hat\Psi(\theta)$ can be shown to behave near the poles as
\begin{equation}
 \hat\Psi(\theta)\sim \sin^{-m_\north}\theta~~(\theta\sim 0),\quad
 \hat\Psi(\theta)\sim \sin^{-m_\south}\theta~~(\theta\sim \pi).
\end{equation}
Therefore the determinant restricted on $\text{ker}J^-\subset{\cal H}'$ is given by
\begin{equation}
 \text{det}(\susy^2)|_{\text{ker}J^-}
 = \prod_{n\ge0}\frac i\ell(-n+q-1-ia-|s|).
\label{dkj2}
\end{equation}

The infinite products of eigenvalues thus obtained are usually rewritten in terms of gamma function assuming zeta function regularization. The final result for the one-loop determinant is~\cite{Doroud:2012xw,Benini:2012ui}
\begin{equation}
 Z_\text{1-loop}
 = \frac{\Gamma(s+q-ia)}{\Gamma(s+1-q+ia)}.
\label{1l0}
\end{equation}
Note that the absolute value signs in~(\ref{dkj1}) and~(\ref{dkj2}) have disappeared.

More honest renormalization allows us to see why the replacement $|s|\to s$ gives the correct formula, and also to understand how it is related to the renormalization of the FI-theta coupling. Let us here  review  a renormalization procedure given in \cite{bulk-renormalization} based on the Pauli-Villars regularization and supergravity counterterms. We introduce a set of ghost chiral multiplets $\{\Phi_1,\Phi_2,\cdots\}$, which have the same charges $\text{e}=1, \RV=2q$ as the original chiral multiplet $\Phi=(\phi,\psi,F)$ but with the same or opposite statistics depending on the label $\epsilon_j=+1$ or $-1$. To make these ghosts massive, we also introduce a background vector multiplet and turn on a constant value $\Lambda$ for its $\sigma$ component, $-\Lambda/f$ for the $D$ component. It is coupled to the ghost chiral fields so that $\Phi_j$ has twisted mass $\alpha_j\Lambda$. We choose $\Lambda,\alpha_j$ to be positive, and also require that $\epsilon_j,\alpha_j$ satisfy
\begin{equation}
\sum_j\epsilon_j=-1,\quad
\sum_j\epsilon_j\alpha_j=0.
\label{epal}
\end{equation}
The one-loop determinant including the ghosts is given by a convergent infinite product, which can be safely rewritten in terms of gamma functions.
\begin{eqnarray}
Z_\text{1-loop}^\text{reg} &=& \prod_{n\in\mathbb Z_{\ge0}}
\left[
 \frac{\frac {-i}\ell(n+1-q+ia+|s|)}{\frac i\ell(n+q-ia+|s|)}
 \prod_{j}
 \left(
 \frac{\frac {-i}\ell(n+1-q+ia+|s|+i\alpha_j\ell\Lambda)}
      {\frac i\ell(n+q-ia+|s|-i\alpha_j\ell\Lambda)}
 \right)^{\epsilon_j}
\right]
 \nonumber \\ &=&
 \frac{\Gamma(|s|+q-ia)}{\Gamma(|s|+1-q+ia)}
 \prod_j\left[\frac{\Gamma(|s|+q-ia-i\alpha_j\ell\Lambda)}
                 {\Gamma(|s|+1-q+ia+i\alpha_j\ell\Lambda)}\right]^{\epsilon_j}.
\label{zolpv}
\end{eqnarray}
At this point one can replace $|s|$ in the last expression by $s$ without changing its value thanks to~(\ref{epal}) and quantization of the flux $s$. Next we take the limit $\Lambda\to\infty$ using
\begin{equation}
  \ln\Gamma(\Lambda+x) = \Lambda\ln\Lambda+(x-\frac12)\ln\Lambda-\Lambda+\frac12\ln(2\pi)+\cdots.
\end{equation}
$Z_\text{1-loop}^\text{reg}$ then becomes
\begin{eqnarray}
 Z_\text{1-loop}^\text{reg} &= &
 \frac{\Gamma(s+q-ia)}{\Gamma(s+1-q+ia)}\cdot
 \exp\Big({-}2i\ell\Lambda\sum_j\epsilon_j\alpha_j\ln\alpha_j\Big)
 \nonumber \\ &&
\quad
\times
 (\ell\tilde\Lambda)^{1-2q}
 (-i\ell\tilde\Lambda)^{-s+ia}
  (i\ell\tilde\Lambda)^{s+ia} \left(1+\mathcal{O}(\Lambda^{-1})\right)
\,,\quad
  \tilde\Lambda\equiv \Lambda\prod_j\alpha_j^{-\epsilon_j} .
\end{eqnarray}
The dependence on $\Lambda$ is removed by adding to the Lagrangian a supergravity counterterm $\mathcal{L}_\text{ct}$ constructed from the twisted superpotential 
\begin{equation} \label{Wtct}
  \widetilde{W}_\text{ct} = \frac{\mathcal{H} }{2} \log\frac {\tilde\Lambda}{ i \mu} - \sum_j \epsilon_j (\alpha_j \Lambda +\Sigma + \mathcal{H})\log \frac{\alpha_j \Lambda +\Sigma + \mathcal{H}}{\mu e}
\end{equation}
by applying the formula~(\ref{LWt}).
Here $\mathcal{H}$ is the lowest component of the twisted chiral multiplet constructed from the gravity multiplet, and is equated to $-iH$ in the supersymmetric background under consideration.%
\footnote{%
We equate
$  (a_j, b_j, c_j, B,\mathcal{H},\overline{\mathcal{H}},\sigma=\sigma_1+i\sigma_2,D, \widetilde{W}, t=r-i\theta)$
 in~\cite{bulk-renormalization} with
$  (-\alpha_j,1, 1,2s,-iH,-iH, -\Sigma=-\sigma+i\rho,-i D,\tilde{W}/4\pi,t_\ren=r+i\theta) $ in this paper.
}
For ease of presentation we set the renormalization scale $\mu$ to $1/\ell$.
Then one can show that the renormalized one-loop determinant
\begin{equation} \label{one-loop-det-ren}
\lim_{\Lambda\rightarrow +\infty}  e^{-\frac{1}{2\pi}\int \cal{L}_\text{ct}}   Z_\text{1-loop}^\text{reg}
\end{equation}
is precisely~(\ref{1l0})~\cite{bulk-renormalization}.
The bare FI-theta parameter $t_0 \equiv r_0+i\theta_0$ is related to the renormalized one $t_\ren$ as
\begin{equation}
 t_0(\Lambda) = \ln(\ell\tilde\Lambda)-\frac{i\pi}2+t_\ren.
\label{tren}
\end{equation}
Note that the counterterms are not unique and suffer from ambiguities from finite counterterms. Effectively, we chose~(\ref{1l0}) as the renormalization condition.

\subsection{Introduction of vortex defects}

Let us next introduce vortex defects with vorticities $\eta_\north$ and $\eta_\south$ at the north and the south poles. We require that
\begin{equation}
 A \sim \eta_\north d\varphi~~(\text{at }\theta\sim0),\quad
 A \sim \eta_\south d\varphi~~(\text{at }\theta\sim\pi).
\end{equation}
In order to preserve the SUSY on the squashed sphere~(\ref{KS}), one also needs to turn on the auxiliary field $D$ so that the defect at the north (south) pole is twisted chiral (resp. anti-twisted chiral). The saddle point configuration~(\ref{sdl}) for vector multiplet fields is modified as follows.
\begin{alignat}{2}
 \sigma&=\frac a\ell, &\qquad
 D &= -\frac a{f\ell}+2\pi i\eta_\north\delta^2_{(\text{NP})}
 +2\pi i\eta_\south\delta^2_{(\text{SP})},
 \nonumber \\
 \rho &= -\frac s\ell,&\qquad
 A &= \left\{\begin{array}{ll}
      s(\cos\theta-1)d\varphi+\eta_\north d\varphi & (\text{north}) \\
      s(\cos\theta+1)d\varphi+\eta_\south d\varphi & (\text{south})
	    \end{array} \right. .
\label{spv}
\end{alignat}
Note that the quantization condition on $s$ gets modified to
\begin{equation}
 s\in\frac12(\eta_\north-\eta_\south+\mathbb Z)
\label{quant-cond-modified}
\end{equation}
due to the magnetic flux carried by the defect. The correlator of vortex defects thus takes the following form,
\begin{equation}
 \langle V_{\eta_\north}V_{\eta_\south}\rangle
 = \sum_{s\in\frac12(\eta_\north-\eta_\south+\mathbb Z)}
 \int\frac{da}{2\pi}\,
 z^{-s+\eta_\north+ia}\bar z^{s+\eta_\south+ia}
 \cdot Z_\text{1-loop}.
\label{defco}
\end{equation}

We would like to compute $Z_\text{1-loop}$ on the vortex defect background in the same way as before by reformulating it in terms of the operators $\susy^2$ and $J^\pm$. This time, $\susy^2$ on the north and the south patches are given by
\begin{equation}
 \susy^2 = \frac1\ell\left\{\partial_\varphi+\frac i2\RV+a\pm is-i\eta_{\north/\south}\right\},
\label{q2wd}
\end{equation}
and $J^\pm$ become vorticity-dependent.
\begin{eqnarray}
(\text{north})~~
 J^\pm &=& e^{\pm i\varphi}\left\{
  \pm\frac\ell f\partial_\theta + \cot\theta(i\partial_\varphi+\eta_\north)
 +\frac12\left(\frac\ell f-1\right)\cot\theta\RV
 +s\tan\frac \theta2
 \right\},
 \nonumber \\
(\text{south})~~
 J^\pm &=& e^{\pm i\varphi}\left\{
  \pm\frac\ell f\partial_\theta + \cot\theta(i\partial_\varphi+\eta_\south)
 +\frac12\left(\frac\ell f-1\right)\cot\theta\RV
 +s\cot\frac \theta2
 \right\}.
\end{eqnarray}
 In this modified setting we again use the symbols ${\cal H}$ and ${\cal H}'$ to denote the spaces of normalizable wave functions that $J^\pm$ map in opposite directions.
We require that
\begin{equation}
J^+{\cal H}\subset {\cal H}',\qquad
J^-{\cal H}'\subset {\cal H},
\label{maps}
\end{equation}
so that the Lagrangian~(\ref{Lch}), and in particular its second term, make sense.
Though we did not pay special attention when deriving~(\ref{1l0}) in the previous subsection, this property is crucial for the pairing of the modes which are not in $\text{ker}J^\pm$. On vortex defect backgrounds,~(\ref{maps}) is guaranteed only if an appropriate set of boundary conditions is imposed on matter wave functions.

Given the relation~(\ref{maps}) between the Hilbert spaces ${\cal H}$ and ${\cal H}'$, it is natural to choose the complete sets of eigenfunctions of $J^-J^+$ and $J^+J^-$ as their basis. Assuming the separated form~(\ref{sep}) for the wave functions, one can reduce the eigenvalue equation to a second-order ODE in $\theta$. By analyzing the characteristic exponents of the ODE, one finds the general solution to the eigenvalue equation behave near the north pole as
\begin{alignat}{2}
 \hat\Psi(\theta) &\sim
 c_+(\sin\theta)^{(m_\north-\eta_\north)} +  c_-(\sin\theta)^{-(m_\north-\eta_\north)}&~(m_\north\ne\eta_\north),
 \nonumber \\
 \hat\Psi(\theta) &\sim
 c_0+c_1\ln(\sin\theta) &~(m_\north=\eta_\north)~
\end{alignat}
with constants $c_\pm,c_0,c_1$. The behavior near the south pole is the same with the obvious replacements $(m_\north,\eta_\north)\to(m_\south,\eta_\south)$. The wave functions thus exhibit  non-integer  power-law behaviors.   By the same argument as in Section~\ref{sec:BPS-defects}, finite values of the wave functions at the defects are not compatible with~(\ref{maps}).   As consistent sets of boundary conditions, we  again  consider the normal and the flipped boundary conditions. In terms of the cohomological variables $X\in{\cal H}$ and $\Xi\in{\cal H}'$ they are expressed as follows:
\begin{alignat}{2}
 \text{normal:}&\quad &
 X\,\text{ and }\, J^-\Xi\,\text{ are finite.}
 \quad& (\Xi\,\text{ may diverge mildly.})
 \nonumber \\
 \text{flipped:}&\quad &
 J^+X\,\text{ and }\, \Xi\,\text{ are finite.}
 \quad& (X\,\text{ may diverge mildly.})
\end{alignat}
For either choice of boundary conditions, the operators $J^-J^+$ and $J^+J^-$ become Hermitian on ${\cal H}$ and ${\cal H}'$ respectively, and their spectrum is discrete due to the compactness of the sphere. Note also that, under the assumption that the wave functions are at most mildly divergent, the substitution of saddle-point configuration~(\ref{spv}) into $J^\pm$ gives
\begin{alignat}{3}
 (\text{north}) &\quad &
 J^+ &\simeq +2\ell D_{\bar z},\quad &
 J^- &\simeq -2\ell D_z,
 \nonumber \\
 (\text{south}) &\quad &
 J^+ &\simeq -2\ell D_w,\quad &
 J^- &\simeq +2\ell D_{\bar w}.
\end{alignat}
where $z\equiv 2\ell e^{i\varphi}\tan\tfrac\theta2,\,w\equiv 2\ell e^{-i\varphi}\cot\tfrac\theta2$ are local complex coordinates near the poles.

The one-loop determinant on the defect background can be obtained in the same way as before, by multiplying the eigenvalues of $\susy^2$ for all the zeromodes of $J^\pm$. As an example, let us choose the normal boundary condition at both defects and evaluate the determinant explicitly. First, suppose a wave function $\Psi\in\text{ker}J^+\subset {\cal H}$ takes the separated form~(\ref{sep}). Then by inspection one finds that $\hat\Psi(\theta)$ behaves near the poles as
\begin{equation}
 \hat\Psi(\theta)\sim \sin^{m_\north-\eta_\north}\theta~~(\theta\sim 0),\quad
 \hat\Psi(\theta)\sim \sin^{m_\south-\eta_\south}\theta~~(\theta\sim \pi).
\label{kerjp}
\end{equation}
Since $\Psi$ has to be finite near the poles, the integers $m_\north,m_\south$ satisfy
\begin{equation}
m_\north-\eta_\north\ge0,\quad
m_\south-\eta_\south\ge0\quad
(m_\south-\eta_\south = m_\north-\eta_\north+2s).
\end{equation}
The determinant of $\susy^2$~(\ref{q2wd}) restricted to $\text{ker}J^+\subset{\cal H}$ is thus given by
\begin{eqnarray}
 \text{det}(\susy^2)|_{\text{ker}J^+}
 &=& \prod_{m_\north\ge\eta_\north}
 \frac i\ell(m_\north-\eta_\north+q-ia+s)~~(s\ge0),
\nonumber \\
 \text{det}(\susy^2)|_{\text{ker}J^+}
 &=& \prod_{m_\south\ge\eta_\south}
 \frac i\ell(m_\south-\eta_\south+q-ia-s)~~~(s\le0).
\label{detq1}
\end{eqnarray}
Similarly, consider a zeromode wave function $\Psi\in\text{ker}J^-\subset{\cal H}'$ of the separated form~(\ref{sep}). The behavior of $\hat\Psi(\theta)$ near the poles are given by
\begin{equation}
 \hat\Psi(\theta)\sim \sin^{-m_\north+\eta_\north}\theta~~(\theta\sim 0),\quad
 \hat\Psi(\theta)\sim \sin^{-m_\south+\eta_\south}\theta~~(\theta\sim \pi).
\label{kerjm}
\end{equation}
Allowing mild divergence at the poles, the determinant of $\susy^2$ restricted to $\text{ker}J^-\subset{\cal H}'$ is given by
\begin{eqnarray}
 \text{det}(\susy^2)|_{\text{ker}J^-}
 &=& \prod_{m_\south<1+\eta_\south}\frac i\ell(m_\south-\eta_\south+q-1-ia-s)
~~~(s\ge0),
\nonumber \\
 \text{det}(\susy^2)|_{\text{ker}J^-}
 &=& \prod_{m_\north<1+\eta_\north}\frac i\ell(m_\north-\eta_\north+q-1-ia+s)
~~(s\le0).
\label{detq2}
\end{eqnarray}

The total one-loop determinant on the vortex defect background{} is given by the ratio of~(\ref{detq2}) and~(\ref{detq1}). Up to renormalization of UV divergences it is given by
\begin{equation}
 Z_\text{1-loop} =
 \frac{\Gamma(\lceil\eta_\north\rceil-\eta_\north+s+q-ia)}
      {\Gamma(-\lceil\eta_\south\rceil+\eta_\south+s+1-q+ia)}.
\label{zold1}
\end{equation}
If we regularize the infinite product explicitly by Pauli-Villars, it becomes
\begin{eqnarray}
 Z_\text{1-loop}^\text{reg} &=&
 \frac{\Gamma(\lceil\eta_\north\rceil-\eta_\north+s+q-ia)}
      {\Gamma(-\lceil\eta_\south\rceil+\eta_\south+s+1-q+ia)}
 \exp\Big({-}2i\ell\Lambda\sum_j\epsilon_j\alpha_j\ln\alpha_j\Big)
 \nonumber \\ && \times
 (\ell\tilde\Lambda)^{1-2q}
 (-i\ell\tilde\Lambda)^{\eta_\north-\lceil\eta_\north\rceil-s+ia}
  (i\ell\tilde\Lambda)^{\eta_\south-\lceil\eta_\south\rceil+s+ia},\qquad
 \tilde\Lambda\equiv\Lambda\prod_j\alpha_j^{-\epsilon_j}.
\label{rZd}
\end{eqnarray}
Several important facts can be read from the regularized expression. First, some of the $\Lambda$-dependence gets absorbed by the renormalization~(\ref{tren}) of the FI-theta parameter, or equivalently canceled by the defect contributions to the counterterm $\mathcal{L}_\text{ct}$. Second, the remaining $\Lambda$-dependence ($\eta$-dependent part) needs to be absorbed by the wave function renormalization of the defect operators in the following way\footnote{%
At this point the renormalization of the operators $V_\eta$ has ambiguities of finite renormalization. We fixed it so that the relations among different defects (discussed in Section~\ref{sec:relat-between-defect}) become the  simplest.
}
\begin{equation}
 V_{\eta_\north}^\ren =
 \tilde\Lambda^{\lceil\eta_\north\rceil}\cdot V_{\eta_\north},
 \quad
 V_{\eta_\south}^\ren =
 \tilde\Lambda^{\lceil\eta_\south\rceil}\cdot V_{\eta_\south}.
\label{Vrn}
\end{equation}
This implies that $V_\eta$ acquires an anomalous dimension.

The evaluation of $Z_\text{1-loop}$ on the defect background for other choices of boundary conditions proceeds in the same way. For example, if the flipped boundary condition is chosen at the two poles, the one-loop determinant is given by
\begin{equation}
 Z_\text{1-loop} =
 \frac{\Gamma(\lfloor\eta_\north\rfloor-\eta_\north+s+q-ia)}
      {\Gamma(-\lfloor\eta_\south\rfloor+\eta_\south+s+1-q+ia)},
\label{zold2}
\end{equation}
namely the ceiling functions are replaced by the floor functions. Similar replacement of ceiling functions by floor functions is understood in other formulae such as~(\ref{Vrn}). Note that the choice of boundary condition on charged chiral multiplets should be regarded as part of the definition of the defect. It is therefore possible that a chiral multiplet obeys the normal boundary condition at $V_{\eta_\north}$ and the flipped boundary condition at $V_{\eta_\south}$. The one-loop determinant depends on the choices of boundary conditions in an obvious manner.

\paragraph{Vortex defects for flavor symmetry (or without dynamical gauge fields).}\label{pg:vfl}

The formulae we have developed so far are applicable also to the vortex defects for flavor $U(1)$ symmetry, which can be introduced by turning on a flat gauge field $A^\text{(f)} = \eta d\varphi$ for the flavor symmetry in the background. It can be gauged away at the cost of making the charged matter fields all multi-valued (or introducing a branch cut between the two poles).  This procedure also makes sense in an orbifold theory, where the gauge symmetry is discrete.  Thus our results have applications to the study of twist operators in orbifold theories. As an example, we will demonstrate in Section~\ref{sec:mirr-symm-vort} the mirror symmetry of the Landau-Ginzburg model realization of the minimal model, where the twist operators in the orbifold are mapped to the original chiral operators.

One subtle difference in the behaviors of gauge and flavor vortex defects is their renormalization property. 
A part of the cutoff dependence, which was previously absorbed into the renormalization of FI-theta couplings for the correlators of gauge vortex defects, now turns into an extra wave function renormalization for the flavor vortex defects. Thus the renormalization property of the flavor vortex defects, coupled to a single chiral matter of unit flavor charge obeying the normal boundary condition, is given by
\begin{equation} \label{flavor-renormalization}
 (V^{\text{(f)}}_{\eta_\north})^\ren = \tilde\Lambda^{\lceil\eta_\north\rceil-\eta_\north}\cdot V^{\text{(f)}}_{\eta_\north},\quad
 (V^{\text{(f)}}_{\eta_\south})^\ren = \tilde\Lambda^{\lceil\eta_\south\rceil-\eta_\south}\cdot V^{\text{(f)}}_{\eta_\south}.
\end{equation}
For the choice of flipped boundary condition, ceiling functions are replaced by floor functions.
Mass dimensions of flavor vortex defects and defects without dynamical vector multiplets can be read off from (\ref{flavor-renormalization}).

\paragraph{Mass deformation by a superpotential.}\label{pg:mads}

In ${\cal N}=(2,2)$ SUSY gauge theories, the chiral multiplets $\Phi,\tilde\Phi$ of opposite charges satisfying $\RV[\Phi]+\RV[\tilde\Phi]=2$ can be simultaneously made massive by the superpotential
  \begin{equation}
W \propto \tilde\Phi \Phi.
  \end{equation}
Since F-terms on the squashed sphere are SUSY exact, the massive pair of chiral multiplets should not contribute to the sphere partition function at all. In the presence of vortex defects, however, the decoupling of these fields is subtle.

Let us couple the chiral multiplets $\Phi,\tilde\Phi$ of electric charges $(1,-1)$ and $\RV$-charges $(2q,2-2q)$ to a $U(1)$ vector multiplet. The localized path integral in the presence of the vortex defects takes the form
\begin{equation}
 \langle V_{\eta_\north}V_{\eta_\south}\rangle
 = \sum_{s\in\frac12(\eta_\north-\eta_\south+\mathbb Z)}
 \int\frac{da}{2\pi}\,
 z^{-s+\eta_\north+ia}\bar z^{s+\eta_\south+ia}\times Z_\text{1-loop}.
\end{equation}
The one-loop determinant depends on the boundary condition. If $\Phi$ obeys the normal boundary condition and $\tilde\Phi$ obeys the flipped boundary condition at the two poles, the determinant becomes
\begin{eqnarray}
Z_\text{1-loop} &=&
 \frac{\Gamma(\lceil\eta_\north\rceil-\eta_\north+s+q-ia)}
      {\Gamma(-\lceil\eta_\south\rceil+\eta_\south+s+1-q+ia)}
 \frac{\Gamma(\lfloor-\eta_\north\rfloor+\eta_\north-s+1-q+ia)}
      {\Gamma(-\lfloor-\eta_\south\rfloor-\eta_\south-s+q-ia)}
 \nonumber \\ &=&
(-1)^{2s+\eta_\south-\lceil\eta_\south\rceil-\eta_\north+\lceil\eta_\north\rceil},
\end{eqnarray}
which implies that all the matter degrees of freedom are lifted, with the only remnant being a shift of the theta angle by $\pi$.

If $\Phi$ and $\tilde\Phi$ both obey the normal boundary condition at the two poles,%
\footnote{%
For a potential quantum inconsistency of this assumption, see Section~\ref{sec:discussion}.
} 
it is given by
\begin{eqnarray}
Z_\text{1-loop} &=&
 \frac{\Gamma(\lceil\eta_\north\rceil-\eta_\north+s+q-ia)}
      {\Gamma(-\lceil\eta_\south\rceil+\eta_\south+s+1-q+ia)}
 \frac{\Gamma(\lceil-\eta_\north\rceil+\eta_\north-s+1-q+ia)}
      {\Gamma(-\lceil-\eta_\south\rceil-\eta_\south-s+q-ia)}
 \nonumber \\ && \hskip-12mm =
(-1)^{2s+\eta_\south-\lceil\eta_\south\rceil-\eta_\north+\lceil\eta_\north\rceil+1}(\eta_\north-\lfloor\eta_\north\rfloor-q-s+ia)
(\eta_\south-\lfloor\eta_\south\rfloor-q+s+ia),
\label{dppt}
\end{eqnarray}
where we assumed that $\eta_\north,\eta_\south\nin\mathbb Z$.
This would imply that, after the pair of massive chiral multiplets is integrated out, there remains a fermionic degree of freedom localized on each defect with non-integer vorticity.

The difference can be understood by looking at the mass term written in terms of cohomological variables,
\begin{equation}
 {\cal L}_\text{mass} \,=\, \phi\tilde F+F\tilde\phi-\psi\tilde\psi+\text{c.c.}
 \,\simeq\, \susy(X\tilde\Xi+\Xi\tilde X)+\text{c.c.}.
\end{equation}
Here $\simeq$ means the equality up to total derivatives. Now let us denote by ${\cal H}$ and ${\cal H}'$ the Hilbert spaces of the wave functions for $X$ and $\Xi$, and by $\tilde{\cal H}$ and $\tilde{\cal H}'$ the analogous spaces for $\tilde X$ and $\tilde\Xi$.  The charge assignment to the multiplets $\Phi,\tilde\Phi$ is such that
\begin{equation}
 \tilde{\cal H} = \bar{\cal H}',\qquad
 \tilde{\cal H}' = \bar{\cal H}
\end{equation}
in the absence of the defects. This equality holds also in the presence of the defects only if the boundary condition on $\tilde\Phi$ is flipped relative to that of $\Phi$. In the above example~(\ref{dppt}) where $\Phi,\tilde\Phi$ both obey the normal boundary condition at both defects, one finds
\begin{equation}
 \text{dim}\bar{\cal H}'=\text{dim}\tilde{\cal H}+1,\qquad
 \text{dim}\tilde{\cal H}'=\text{dim}\bar{\cal H}+1.
\end{equation}
The mismatch precisely corresponds to the modes having a mild divergence at either defect. These modes are responsible for the uncanceled fermionic eigenvalues in~(\ref{dppt}).

\subsection{Comparison with smearing}
\label{sec:smear-sphere}

Here we study the supersymmetric path integral over the fluctuations around the smeared defect configuration, of which the flat space version was introduced in Section~\ref{sec:smear-vort-defects}. We would like to check that the result is trivially equal to partition function, and also study how the smearing changes the Hilbert spaces of wave functions ${\cal H},{\cal H}'$ of the previous subsection. As before, we focus on the $U(1)$ gauge theory with a single chiral multiplet $(\phi,\psi,F)$ of electric charge 1 and  vector R-charge $2q$.

\paragraph{Correlator of smeared vortex defects.}
\label{pg:two-point-smeared}

The smeared defect configuration with vorticity $\eta_\north,\eta_\south$ at the two poles is defined by the following modification of~(\ref{spv}),
\begin{alignat}{2}
 \sigma&=\frac a\ell, &\quad
 D &= -\frac a{\ell f(\theta)}
 +i\frac{S'(\theta)}{\ell f(\theta)\sin\theta},\quad
 \rho = -\frac{s  +S(\theta)  }{\ell},
 \nonumber \\
 A_\theta &=0, &\quad
 A_\varphi &= \left\{\begin{array}{ll}
      s(\cos\theta-1)+\eta_\north+S(\theta)\cos\theta & (\text{north}) \\
      s(\cos\theta+1)+\eta_\south+S(\theta)\cos\theta & (\text{south})
	    \end{array} \right. ,
\label{spsm}
\end{alignat}
which solves the saddle point condition for arbitrary $S(\theta)$.\footnote{%
This form can be found by allowing $D$ to have an imaginary part while requiring other fields to be real. It is also possible to turn on ${\rm Im}\hspace{.3mm}\sigma$ and solve the SUSY conditions, but the interpretation is not clear. We turn it off for simplicity.} 
The function $S(\theta)$ here specifies the smearing: for now we only require $S(0)=-\eta_\north$, $S(\pi)=\eta_\south$ and proceed. The classical FI-theta action evaluated on this background takes the same value as that on the saddle point~(\ref{spv}). Also, the quantization condition on the $s$ above,
\begin{equation}
 2s-\eta_\north+\eta_\south\in\mathbb Z,
\end{equation}
is the same as~(\ref{quant-cond-modified}).

As in previous subsections, we evaluate $Z_\text{1-loop}$ by moving to the cohomological variables and studying the spectrum of $\susy^2$ on $\text{ker}J^\pm$.
It turns out that $\susy^2$ is the same as~(\ref{q2wd}), but $J^\pm$ now depend also on $S(\theta)$ as
\begin{eqnarray}
(\text{north})~
 J^\pm &=& e^{\pm i\varphi}\left\{
  \pm\frac\ell f\partial_\theta + \cot\theta(i\partial_\varphi+\eta_\north)
 +\frac12\left(\frac\ell f-1\right)\cot\theta\RV
 +s\tan\frac \theta2
 +\frac{S(\theta)}{\sin\theta}
 \right\},
 \nonumber \\
(\text{south})~
 J^\pm &=& e^{\pm i\varphi}\left\{
  \pm\frac\ell f\partial_\theta + \cot\theta(i\partial_\varphi+\eta_\south)
 +\frac12\left(\frac\ell f-1\right)\cot\theta\RV
 +s\cot\frac \theta2
 +\frac{S(\theta)}{\sin\theta}
 \right\}.~~~~~~~
\end{eqnarray}
Due to the effect of smearing, the behaviors of these operators near the poles are the same as those in the {\it absence} of the defects~(\ref{Jwod}). Therefore, the smearing eliminates the issue of wave functions with non-integer power law behaviors.

Let $\Psi$ be a scalar wave function in the separated form~(\ref{sep}).
If $\Psi \in {\rm ker}J^+$ the local behaviors near the poles are
\begin{equation}
 \hat\Psi(\theta)\sim \sin^{m_\north}\theta~~(\theta\sim 0),\quad
 \hat\Psi(\theta)\sim \sin^{m_\south}\theta~~(\theta\sim \pi),
\end{equation}
As a consequence we find
\begin{eqnarray}
 \text{det}(\susy^2)|_{\text{ker}J^+}
 &=& \prod_{m_\north\ge 0}
 \frac i\ell(m_\north-\eta_\north+q-ia+s)
 \quad(2s-\eta_\north+\eta_\south\ge0),
\nonumber \\
 \text{det}(\susy^2)|_{\text{ker}J^+}
 &=& \prod_{m_\south\ge 0}
 \frac i\ell(m_\south-\eta_\south+q-ia-s)
 \quad~(2s-\eta_\north+\eta_\south\le0).
\label{detq1-smeared}
\end{eqnarray}
Similarly, if $\Psi \in {\rm ker}J^-$ the local behaviors are
\begin{equation}
 \hat\Psi(\theta)\sim \sin^{-m_\north}\theta~~(\theta\sim 0),\quad
 \hat\Psi(\theta)\sim \sin^{-m_\south}\theta~~(\theta\sim \pi).
\end{equation}
The determinant is then
\begin{eqnarray}
 \text{det}(\susy^2)|_{\text{ker}J^-}
 &=& \prod_{m_\south \leq 0} \frac i\ell(m_\south-\eta_\south+q-1-ia-s)
~~~(2s-\eta_\north+\eta_\south\ge0),
\nonumber \\
 \text{det}(\susy^2)|_{\text{ker}J^-}
 &=& \prod_{m_\north \leq 0}\frac i\ell(m_\north-\eta_\north+q-1-ia+s)
~~(2s-\eta_\north+\eta_\south \le0).
\label{detq2-smeared}
\end{eqnarray}
The renormalized one-loop determinant $Z_\text{1-loop}^\text{smeared} $, defined by the same expression (\ref{one-loop-det-ren}) but now for the smeared defect background, is given for either sign of $2s-\eta_\north+\eta_\south$ as{} 
\begin{equation} \label{one-loop-smear}
Z_\text{1-loop}^\text{smeared} =  (\ell \mu)^{1+\eta_\north+ \eta_\south + 2ia -2q} 
\frac{\Gamma(s-\eta_\north-ia +q)}{\Gamma(s+\eta_\south+ia-q+1)} \,.
\end{equation}
To avoid clutter we will set $\mu$ to $1/\ell$ in the following.%
\footnote{%
For the smeared flavor defect, one can read off the mass dimension $[(V_{\eta_{\north/\south}}^\text{(f,smeared)})^{\bf r}   ]=-\eta_{\north/\south}$ of each operator from the $\ell$-dependence of the normalized correlator
\begin{equation}
  \langle 
(V_{\eta_\north}^\text{(f,smeared)})^{\bf r}  
(V_{\eta_\south}^\text{(f,smeared)})^{\bf r}  
\rangle_\text{normalized}
= Z_\text{1-loop}^\text{smeared}/Z_\text{1-loop} \propto (\ell\mu)^{\eta_\north+\eta_\south}.
\end{equation}
Expression (\ref{one-loop-smear}) is obtained by evaluating the supergravity counterterm \cite{bulk-renormalization} in the smeared defect background.
} 
Note that it depends only on $\eta_\north,\eta_\south$ and not on the detail of the smearing function $S(\theta)$. Also, the determinant is analytic but non-periodic in $\eta$'s in contrast to the earlier results~(\ref{zold1}) and~(\ref{zold2}).

One may want to define the correlation function of ``the smeared gauge vortex defects'' by
\begin{equation}
 \langle V_{\eta_\north}^\text{smeared} V_{\eta_\south}^\text{smeared}\rangle
 = \sum_{s\in\frac12(\eta_\north-\eta_\south+\mathbb Z)}
 \int\frac{da}{2\pi}\,
 z^{-s+\eta_\north+ia}\bar z^{s+\eta_\south+ia}
 \cdot Z_\text{1-loop}^\text{smeared}.
\label{defsco}
\end{equation}
As discussed in Section~\ref{sec:smear-vort-defects} we expect it to be trivial, i.e., equal to the partition function. Indeed, by rewriting it using
\begin{equation}
 \tilde a\equiv a-\frac i2(\eta_\north+\eta_\south),\quad
 \tilde s\equiv s+\frac12(\eta_\south-\eta_\north),
\end{equation}
one finds~(\ref{defsco}) looks almost identical to~(\ref{zs2}). However, there is an important difference that the integration contour for $\tilde a$ is shifted relative to that for $a$. Shifting the contour across the poles of the integrand changes the value of the integral, so the triviality of the smeared vortex defects for general $\eta_\north,\eta_\south$ is in fact subtle. Suffice it to say for now that, for general $\eta_\north,\eta_\south$, there should be a choice of the $a$-integration contour in~(\ref{defsco}) such that the defect correlator coincides with the partition function. In Section~\ref{sec:defect-correlators} we will discuss the issue of contour shifts in more detail in slightly different problems.

\paragraph{Localized modes and frozen bulk modes.}
\label{pg:localized-modes-zero}

We now take a specific form of the function $S(\theta)$. Let $g_\epsilon(u)$ be a smooth function such that $g_\epsilon(0)=0$ and $g_\epsilon(u)\simeq 1$ for $u\gg \epsilon>0$, and set
\begin{equation} \label{S-g-epsilon}
 S(\theta) = -\eta_\north\, \left\{1-g_\epsilon\left(\frac\theta\pi\right)\right\}
 +\eta_\south\,\left\{1-g_\epsilon\left(1-\frac\theta\pi\right)\right\}.
\end{equation}
For small $\epsilon$, the modification of the saddle-point configuration by $S(\theta)$ takes place only in the $\epsilon$-neighborhood of the two poles. How does this smearing affect the Hilbert spaces ${\cal H},{\cal H}'$ which were previously defined by the (normal or flipped) boundary conditions? There are two possible effects:
\begin{itemize}
 \item Some wave functions which did not satisfy the boundary condition at either pole become regular due to the smearing. Such wave functions necessarily have profiles peaked around that pole, so we call them the localized modes. For illustration we plot a sample localized mode for $\Psi\in\text{ker}J^+$ on the round sphere with a smooth choice of $g_\epsilon$ in Figure~\ref{figure:localized-mode-smooth}.

 \item Some wave functions which satisfied the boundary conditions become singular at either pole due to the smearing. We call such modes the frozen bulk modes.

\begin{figure}[htbp]
\centering
\includegraphics[bb=0 0 288 195,width=10cm]{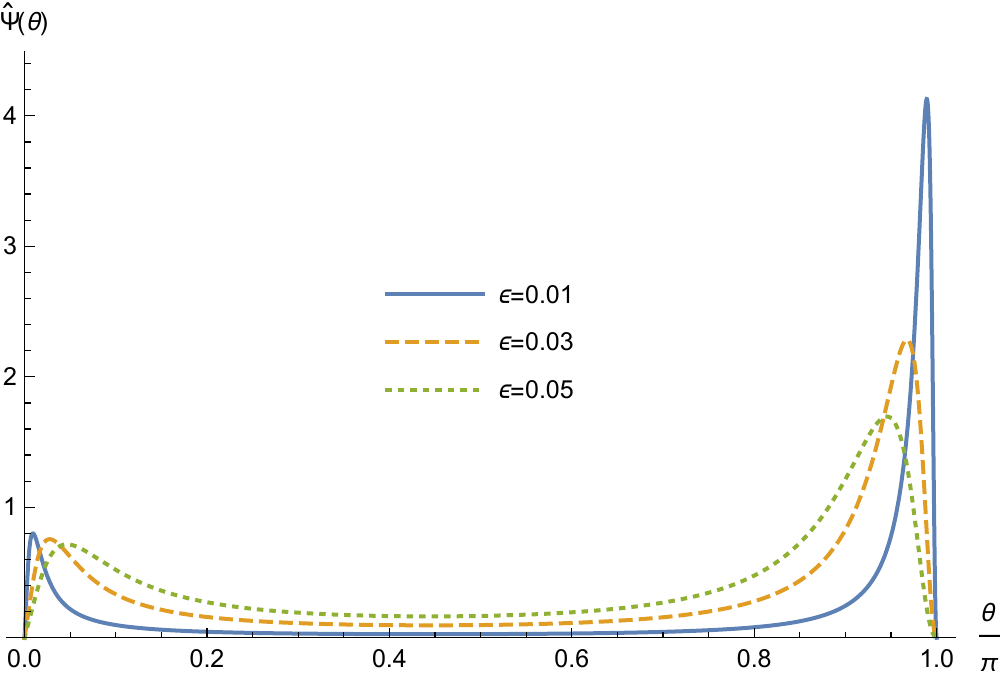}
\caption{\label{figure:localized-mode-smooth}%
An example of a localized mode $\Psi =\hat{\Psi}(\theta) e^{i m_{\north/\south}\varphi} \in {\rm ker}J^+$ on the round sphere for $m_\north=1$, $m_\south=2$, $\eta_\north=2.2$, $\eta_\south=3.7$.
The $\theta$-dependent part $\hat{\Psi}$ of the separated wave function is plotted with the choice $g_\epsilon=\beta\circ \alpha_\epsilon$, where $\alpha_\epsilon$ is the M\"obius transformation that maps the triple $\{0,\epsilon, 1\}$ to $\{0,\frac 12, 1\}$, and $\beta$ is a smooth function $\displaystyle \beta(u)=\left[1+\exp\left( \frac{1}{u}-\frac{1}{1-u} \right)\right]^{-1} $.
}
\end{figure}

\end{itemize}

Before studying the change in the Hilbert spaces due to smearing, we would like to introduce one convenient trick that simplifies our discussion, which we call $t$-deformation. See \cite{MR792703} for a motivation. The $t$-deformation is simply the similarity transformation of $J^\pm$,
\begin{equation}
 J^+ \to e^{tU(\theta)}J^+ e^{-tU(\theta)},\quad
 J^- \to e^{-tU(\theta)}J^- e^{tU(\theta)}
\label{jpmsm}
\end{equation}
with $t\in\mathbb R$ and $U(\theta)$ a monotonic function satisfying $U(0)>U(\pi)$. The relations $(J^\pm)^\dagger = J^\mp$ and $[\susy^2,J^\pm]=0$ are preserved under this transformation. Moreover, in the limit $t\to\infty$ all the $J^+$ ($J^-$)-zeromodes are automatically vanishingly small at the south (resp. the north) pole, so we only need to examine their behavior at the other pole. With this deformation of $J^\pm$, we have a rather simple interpretation of the one-loop determinants~(\ref{zold1}),~(\ref{zold2}),~(\ref{one-loop-smear}). Namely the gamma function in the numerator accounts for the contribution of $J^+$-zeromodes supported on the north hemisphere, and the one in the denominator arises from $J^-$-zeromodes supported on the south hemisphere. This simplifies the spectrum analysis significantly.

With the help of this deformation, let us study the effect of smearing on the wave functions of the charged chiral multiplet in $\text{ker}J^\pm$. We concentrate on the effect of smearing at the north pole since the other pole can be studied in the same way. In the limit $t\to\infty$, only $J^+$-zeromodes are supported around the north pole. Before the smearing, the $J^+$-zeromodes satisfying the boundary condition takes the form $X = \hat\Psi(\theta)e^{im_\north\varphi}$ on the north hemisphere, with
\begin{equation}
 m_\north\ge\kappa_\north(\eta_\north),\qquad
 \kappa_\north(\eta_\north)=\left\{\begin{array}{ll}
 \lceil\eta_\north\rceil &(\text{normal b.c. at the north pole})\\
 \lfloor\eta_\north\rfloor &(\text{flipped b.c. at the north pole})
 \end{array}
\right..
\end{equation}
After the smearing, the normalizable $J^+$-zeromodes have $m_\north\ge0$. Therefore,
\begin{itemize}
 \item If $\eta_\north>0$, the smearing brings in localized $J^+$-zeromodes with $0\le m_\north < \kappa_\north(\eta_\north)$.
 \item If $\eta_\north<0$, the smearing freezes the bulk $J^+$-zeromodes with $\kappa_\north(\eta_\north)\le m_\north<0$.
\end{itemize}
This effect can be easily read off from the ratio of $Z_\text{1-loop}$ before and after the smearing; for example,
\begin{equation}
 \frac{Z_\text{1-loop}^\text{smeared}}
      {Z_\text{1-loop}} =
 \frac{\Gamma(-\eta_\north+s+q-ia)}{\Gamma(\eta_\south+s+1-q+ia)}
 \bigg/
 \frac{\Gamma(\kappa_\north(\eta_\north)-\eta_\north+s+q-ia)}
      {\Gamma(-\kappa_\south(\eta_\south)+\eta_\south+s+1-q+ia)}.
\label{raold}
\end{equation}

In the opposite limit $t\to-\infty$ of $t$-deformation, only $J^-$-zeromode are supported around the north patch. The wave functions take the form $\Xi = \hat\Psi(\theta)e^{im_\north\varphi}$ there, and the boundary condition at the north pole requires $m_\north\le\kappa_\north(\eta_\north)$. After the smearing, the normalizability requires $m_\north\le0$. Therefore
\begin{itemize}
 \item If $\eta_\north>0$, the smearing freezes the bulk $J^-$-zeromodes with $0< m_\north \le \kappa_\north(\eta_\north)$.
 \item If $\eta_\north<0$, the smearing brings in localized $J^-$-zeromodes with $\kappa_\north(\eta_\north)< m_\north\le0$.
\end{itemize}
This effect can again be read off from the ratio of $Z_\text{1-loop}$~(\ref{raold}), but one needs to rewrite them using gamma function identities so that their denominator is interpreted as the contribution of $J^-$-zeromodes supported on the north patch.
\begin{eqnarray}
 \frac{Z_\text{1-loop}^\text{smeared}}
      {Z_\text{1-loop}} &=&
 (-1)^{\kappa_\north(\eta_\north)+\kappa_\south(\eta_\south)}
 \nonumber \\ &&\times
 \frac{\Gamma(-\eta_\south-s+q-ia)}{\Gamma(\eta_\north-s+1-q+ia)}
 \bigg/
 \frac{\Gamma(\kappa_\south(\eta_\south)-\eta_\south-s+q-ia)}
      {\Gamma(-\kappa_\north(\eta_\north)+\eta_\north-s+1-q+ia)}.
\label{raold2}
\end{eqnarray}
The sign factor on the RHS plays an important role later.

\paragraph{0d multiplets describing the smearing effect.}

Now let us identify the 0d multiplets on the defect at the north pole which reproduce the effects described in the previous paragraph. As explained in Section~\ref{sec:defects-via-coupling-flat}, those 0d multiplets represent the simple SUSY algebra
\begin{eqnarray}
 \susy^2 &=& \left[\Big(\sigma-i\cos\theta\rho-\frac i\ell A_\varphi\Big)\text{e}+\frac i\ell\Big(J_3+\frac12\RV\Big)\right]_\text{north pole}
 \nonumber \\ &=& \frac i\ell\big\{(-ia+s-\eta_\north)\text{e}+J_3+\frac12\RV\big\}\qquad\text{on the saddle point}.
\end{eqnarray}
In the first line, the terms with $1/\ell$ are due to Omega deformation. The operator $J_3$ is the angular momentum, which acts simply as $-i\partial_\varphi$ on scalar wave functions.

Let us take $\eta_\north>0$. First, in the limit $t\to\infty$ of $t$-deformation, the smearing at the north pole gives rise to some localized $J^+$-zeromodes, which turns into 0d chiral multiplets $\{\lcl u_n,\lcl\zeta_n\}_{n=0}^{\kappa_\north(\eta_\north)-1}$ with the charges
\begin{equation}
 \text{e}=1, \quad \RV=2q,\quad
 J_3 =n= 0,1,\cdots,\kappa_\north(\eta_\north)-1.
\label{chod1}
\end{equation}
Integrating over them with an appropriate weight $S_\text{0d}$ such as the one in~(\ref{0d-deformed-action-Fermi-explicit}) gives
\begin{equation}
 \int d[\lcl u_n,\lcl\zeta_n,\bar{\lcl u}_n,\bar{\lcl\zeta}_n]e^{-S_{\text{0d}}} = \prod_{n=0}^{\kappa_\north(\eta_\north)-1}\left(\frac{-i\ell}{-ia+s-\eta_\north+n+q}\right),
\label{lzss}
\end{equation}
which explains a part of the change of $Z_\text{1-loop}$~(\ref{raold}) due to smearing. In the opposite limit $t\to-\infty$, the smearing freezes the modes of $\Xi$ in $\text{ker}J^-$ with $J_3 = 1\,,\,\cdots\,,\,\kappa_\north(\eta_\north)$, each of which gives rise to a Fermi multiplet. To freeze them we need the 0d chiral multiplets $\{\tilde{\lcl u}_n,\tilde{\lcl\zeta}_n\}_{n=1}^{\kappa_\north(\eta_\north)}$ with the charges
\begin{equation}
 \text{e}=-1, \quad
 J_3 +\frac12\RV = -n-q+1,\qquad n=1,2,\cdots,\kappa_\north(\eta_\north).
\label{chod2}
\end{equation}
Note that the charges of $\tilde{\lcl u}_n,\tilde{\lcl\zeta}_n$ are precisely the opposite of those (\ref{chod1}) of $\lcl u_n,\lcl\zeta_n$. In addition, the sign factor in~(\ref{raold2}) implies that the integral measure over the 0d multiplets gets multiplied by $(-1)^{\kappa_\north(\eta_\north)}$. The integral over $\tilde{\lcl u}_n,\tilde{\lcl\zeta}_n$ which freezes the $J^-$-zeromodes of $\Xi$ and its superpartner $\susy\Xi$ is given by,
\begin{equation}
 (-1)^{\kappa_\north(\eta_\north)}
 \int d[\tilde{\lcl u}_n,\tilde{\lcl\zeta}_n,\bar{\tilde{\lcl u}}_n,\bar{\tilde{\lcl\zeta}}_n]\exp\bigg(
  -\sum_{n=1}^{\kappa_\north(\eta_\north)}\left[
 \tilde{\lcl\zeta}_n\,J^{+n}\Xi+\tilde{\lcl u}_n\,J^{+n}(\susy\Xi)+\text{c.c.}
 \right]_\text{north pole}
 \bigg).
\end{equation}
If one chose a different weight for $\tilde{\lcl u}_n,\tilde{\lcl\zeta}_n$ such as~(\ref{0d-deformed-action-Fermi-explicit}) which does not involve 0d-2d couplings, the integral would not freeze the bulk modes but one would instead obtain precisely the same result as~(\ref{lzss}). This implies that the two systems of 0d multiplets $\{\lcl u_n,\lcl\zeta_n\}$ and $\{\tilde{\lcl u}_n,\tilde{\lcl\zeta}_n\}$ are equivalent in $\susy$-cohomology although they are introduced to explain different effects of the smearing.

In the same way, for $\eta_\north<0$ we obtain two different 0d systems depending on the limit $t\to \pm\infty$. One is the system of $-\kappa_\north(\eta_\north)$ units of Fermi multiplets with charges
\begin{equation}
 \text{e}=-1, \quad
 J_3+\frac12\RV \;=\; 1-q\,,\,2-q\,,\,\cdots\,,\,-\kappa_\north(\eta_\north)-q.
\label{chlf1}
\end{equation}
The other is the same number of Fermi multiplets with the charges opposite to~(\ref{chlf1}), and additional sign factor $(-1)^{\kappa_\north(\eta_\north)}$ multiplied onto the measure. Again, these two 0d systems are equivalent in the sense of $\susy$-cohomology.

\section{Relations between defect operators}
\label{sec:relat-between-defect}

In the last section we have seen that the difference between the singular and smeared vortex defect background{}s is accounted for by a suitable set of 0d variables on the defect. Here we analyze this relation in more detail. We also combine this relation with the triviality of smeared gauge vortex defects to derive an equivalence relation between vortex defects and 0d-2d systems. We choose to use physically relevant actions and avoid $\susy$-exact deformations or cohomological arguments, so that the arguments and the results presented here could potentially be applied to non-SUSY settings. We will study the defects in flat space; once we understand them, the generalization to those in curved backgrounds is dictated by the symmetries of supergravity.

For simplicity, we assume $\eta\nin\mathbb Z$ throughout this section. We continue to focus on a single chiral multiplet $(\phi,\psi,F)$ of electric charge $\text{e}=1$ and vector R-charge $2q$, coupled to a $U(1)$ vector multiplet.

\subsection{Relations between wave functions}
\label{sec:eigenv-probl-flat}

Consider a smeared $U(1)$ vortex defect background{} centered at the origin of the flat space. We use the polar coordinate $(r,\varphi)$, and set the vector multiplet fields $A_\mu,D$ as follows,
\begin{equation} \label{smearing-background-flat}
 A_r=0,\qquad
 A_\varphi(r)=\eta\cdot g(r/\epsilon),\qquad
 D = \frac ir\partial_rA_\varphi.
\end{equation}
The smearing function $g(u)$ in this section satisfies $g(u)\sim0$ for $u\ll1$ and $g(u)\sim1$ for $u\gg1$. For this smeared vortex defect background{}, the vorticity density~(\ref{smear-background-flat}) is $\varrho = (2\pi r \epsilon)^{-1}\eta\cdot g'(r/\epsilon)$. As in the previous section, we couple a chiral multiplet with unit electric charge to this vector multiplet. Both the vector and chiral multiplets are dynamical, but the path integral measure for the chiral multiplet is defined by the mode expansion on the above vortex defect background{}.

To avoid clutter, in the following we will suppress the fluctuations of vector multiplet fields around the background~(\ref{smear-background-flat}) except $\sigma$ and $\rho$ which we write explicitly. The matter Lagrangian in~(\ref{Lvm}) gives the action
\begin{equation}
  S = \int\frac{d^2x}{2\pi} \left[ \bar\phi (-4 D_z D_{\bar z} +    \bar\Sigma \Sigma)\phi +  i \bar\psi
  \begin{pmatrix}
-i  { \bar \Sigma }   & 2i D_z\\
2i D_{\bar z}    & -i {  \Sigma}
  \end{pmatrix}
\psi \right]   \,.
\end{equation}
The Laplacian and $D$ combined to give $-4D_z D_{\bar z}$, and we set $\Sigma = \sigma -i \rho$, $\bar\Sigma = \sigma + i \rho$.

We will expand $\phi$ and $\psi^+$ in the eigenmodes of $-4 D_z D_{\bar z} $, while $\psi^-$ and $F$ will be expanded in the eigenmodes of $-4 D_{\bar z} D_z$. Since $ i D_z$ and $i D_{\bar z}$ are conjugate to each other, the eigenvalues are all non-negative. In flat space, these modes are either delta function normalizable or normalizable. As we will see, normalizable modes are all supported on the $\epsilon$-neighborhood of the defect, very much like the localized zeromodes of $J^\pm$ on the squashed sphere which we studied in Section~\ref{sec:smear-sphere}. For a non-zero eigenvalue $\lambda>0$ of $-4 D_z D_{\bar z} $, the differential equation for the radial part $\hat\Psi_m$ of the separated wave function $\Psi = \hat\Psi_m(r)e^{im\varphi}\,(m\in\mathbb Z)$ reduces to the Bessel equation for $r\gg\epsilon$.
$\hat\Psi_m(r)$ behaves asymptotically as a linear combination of $ r^{-1/2} \exp(\pm i \lambda^{1/2} r)$ for $r\gg \text{max}\{\epsilon,\lambda^{-1/2}\}$. Thus a non-zero mode of $-4 D_z D_{\bar z} $ is necessarily a delta function normalizable mode and does not correspond to a localized mode.

A normalizable zeromode of $-4D_z D_{\bar z}$ is annihilated by $D_{\bar z}$.
The radial wave function $\hat\Psi_m$ must satisfy
\begin{equation}\label{eq:D-bar-z-radial}
( r\partial_r - m + A_\varphi)\hat\Psi_m=0   \qquad  \Longrightarrow \qquad \hat\Psi_m(r) =\exp \left[ - \int^r \frac{ds}{s} (A_\varphi(s) -m) \right]  \,.
\end{equation}
The behaviors for large and small values of $r$ are
\begin{equation} \label{holo-behavior-flat}
\hat\Psi_m(r)
\sim
\left\{
\begin{array}{lcl}
r^{+m}  & \text{ for } &r \ll \epsilon \,, \\
r^{+m-\eta} & \text{ for }& r \gg \epsilon \,.
\end{array}
\right.
\end{equation}
For $\Psi$ to be smooth at $r=0$ we need $m\geq 0$, and for it to be normalizable we need $m-\eta <-1$. Therefore, for $\eta>1$ and non-integer, there are normalizable zeromodes labeled by $m=0,\cdots,\lfloor\eta\rfloor-1$. We denote them by $\{\phi^{(0)}_a\}_{a=0}^{\lfloor\eta\rfloor-1}$. Since there were no zeromodes on the singular vortex defect background{} before smearing, these normalizable zeromodes have arisen as a consequence of the smearing.

A small remark is in order regarding the limit of sending $\eta$ to a positive integer from above. In this limit, one of the $\lfloor\eta\rfloor$ normalizable zeromodes becomes non-normalizable since it behaves as $\hat\Psi_m(r)\sim 1/r$ at large $r$. The non-normalizability of this mode is due to the infinite volume of flat space. It seems more appropriate to regard it as ``marginally'' normalizable; for example one can make the index theorem hold on $\mathbb R^2$ with this prescription. See~\cite{Bilal:2008qx}.

Similar argument applies also to the mode $\Psi = \hat\Psi_m(r)e^{im\varphi}$ annihilated by $D_z$. It is smooth and normalizable if and only if $\eta+1\le m\leq 0$. We denote such normalizable modes $\Psi$ by $\tilde\phi^{(0)}_\alpha$ with $\alpha \equiv-m=0,1,\ldots,-\lceil \eta \rceil -1$.
There exists at least one such mode if and only if $\eta\le-1$.

Let us next study non-zero modes in detail. We are interested in the limit $\epsilon\to 0$ with a non-zero eigenvalue kept finite. If $\Psi$ is a delta function normalizable eigenmode of $-4 D_z D_{\bar z}$ with eigenvalue $\lambda>0$,
\begin{equation} \label{eigenvalue-equation}
  -4 D_z D_{\bar z} \Psi = \lambda \Psi \,,
\end{equation}
then
\begin{equation}
    \tilde\Psi:=\lambda^{-1/2} \cdot 2i D_{\bar z} \Psi
\end{equation}
is a delta function normalizable eigenmode of $-4D_{\bar z} D_z $ with eigenvalue $\lambda$. We will mostly focus on $\Psi$. For the eigenfunction of separated form $\Psi= \hat{\Psi}_m(r) e^{i m \varphi}$ the eigenvalue equation~(\ref{eigenvalue-equation}) reads
\begin{equation} \label{eigenvalue-equation-in-r}
\left(
 \partial_r^2 + \frac{1}{r} \partial_r + \frac{\partial_r A_\varphi}{r} - \frac{(m-A_\varphi)^2}{r^2} + \lambda \right)  \hat{\Psi}_m =0  \,.
\end{equation}
Since $A_\varphi$ approaches a constant $\eta$ for $r\gg \epsilon$, the solution there is a linear combination of two Bessel functions $J_{\pm|m-\eta|}(\lambda^{1/2} r)$. For a given $\lambda>0$, we assume $\epsilon$ was chosen small enough so that $\epsilon\ll \lambda^{-1/2}$ holds. Then the form of $\hat\Psi_m$ is determined from the regularity at $r=0$ as
\begin{equation} \label{Psi-hat-J}
 \hat{\Psi}_m=
\left\{
  \begin{array}{clc}
\text{(non-zero)}\times    r^{|m|} &  \text{ for }& 0\leq r\ll \epsilon \,, \\
   \alpha_+ J_{+|m-\eta|}(\lambda^{1/2}r) + \alpha_-  J_{-|m-\eta|}(\lambda^{1/2}r) \qquad
& \text{ for } & r \gg \epsilon
  \end{array}
\right.
\end{equation}
with some ($\epsilon$-dependent) constants $\alpha_\pm$. To compare this with the wave function obeying the normal or the flipped boundary condition, we need to find which of the two terms dominates in the bulk region $r\gg \epsilon$. This amounts to comparing the coefficients $\alpha_\pm$ as functions of $\epsilon$ in the limit of small $\epsilon$.

In Appendix~\ref{sec:dominant-terms-bulk} we show that
\begin{equation}\label{alpha-inequalities}
 \frac{\alpha_+}{\alpha_-} ~\sim~ \left\{
 \begin{array}{ll}
  \epsilon^{-2|m-\eta|+\text{non-positive}} \quad & (m<0) \\
  \epsilon^{-2|m-\eta|+2} \quad & (0\le m<\eta) \\
  \epsilon^{-2|m-\eta|-2} \quad & (\text{max}\{0,\eta\}<m)
 \end{array}
 \right.\,.
\end{equation}
This implies that the first term with coefficient $\alpha_+$ is always dominant except when $\eta>0$ and $m=\lfloor\eta\rfloor$. For this exceptional case, the corresponding wave function behaves like $\Psi\sim r^{\lfloor\eta\rfloor-\eta}e^{i\lfloor\eta\rfloor\varphi}$ for $\epsilon\ll r \ll \lambda^{-1/2}$; it diverges mildly but $D_{\bar z}\Psi$ is finite. According to~(\ref{eq:fbc}), this is the mode in ${\cal H}$ satisfying the flipped boundary condition. Note that on the smeared vortex defect background{} this mode receives a correction in the small region $r\ll \epsilon$ so that it is continued to a smooth solution.%
\footnote{%
For a vanishing eigenvalue, the local solution allowed by the flipped boundary condition near the origin is non-normalizable at infinity. Thus there is no frozen mode in flat space.}

The above result implies that, depending on whether $\eta>0$ or $\eta<0$, the set of eigenfunctions of $D_zD_{\bar z}$ with non-zero eigenvalues on the smeared background reproduces those satisfying the flipped or the normal boundary condition in the limit $\epsilon\to0$. The correspondence is one to one and, in particular, no modes are frozen by the smearing. Including the results on the zero modes, one can express the relations between the Hilbert spaces as
\begin{equation}
  \lim_{\epsilon \rightarrow 0} \mathcal{H}_\text{smeared} =
\left\{
  \begin{array}{cll}
\mathcal{H}_\text{flipped} \oplus \mathbb{C}^{\lfloor \eta \rfloor}    \qquad  & \text{ if } & \eta>0 \,,     \\
\mathcal{H}_\text{normal} & \text{ if } & \eta<0 \,.
  \end{array}
\right.
\end{equation}
The power of $\mathbb{C}$ represents localized zeromodes $\{\phi^{(0)}_a\}_{a=0}^{\lfloor\eta\rfloor-1}$.

For wave functions in $\mathcal{H}'$ the analysis above goes through if we replace $(m,\eta) \rightarrow (-m,-\eta)$ on the right hand side of~(\ref{Psi-hat-J}).
We find the relation
\begin{equation}
  \lim_{\epsilon \rightarrow 0} \mathcal{H}'_\text{smeared} =
\left\{
  \begin{array}{cll}
\mathcal{H}'_\text{flipped}    & \text{ if } & \eta>0 \,,     \\
\mathcal{H}'_\text{normal} \oplus \mathbb{C}^{\lfloor -\eta \rfloor} \qquad & \text{ if } & \eta<0 \,.
  \end{array}
\right.
\end{equation}

It is possible to cancel the effects of the localized modes by inserting extra 0d multiplets with the opposite statistics.
We may regard this construction as a way to regularize the vortex singularities defined in Section~\ref{sec:BPS-defects} by the smearing and 0d multiplets.

\subsection{0d multiplets from localized modes}
\label{sec:0d-localized}

Once we have a complete set of basis wave functions, the path integral can be formulated as an infinite dimensional integral over the coefficients in the mode expansion of the fields. For the smeared vortex defect background{} with small $\epsilon$, the modes $\phi^{(0)}_a$ and $\tilde\phi^{(0)}_\alpha$ are strongly localized near the defect, so they couple with the vector multiplet fields only at the defect. In the limit $\epsilon\to 0$ the coefficients of these modes become the 0d multiplets on the defect. The remaining variables describe the bulk dynamics of matter fields obeying the normal or the flipped boundary condition.

Let us write down the action for the 0d multiplets explicitly. First, the coefficients $({\boldsymbol u}_a,{\boldsymbol \zeta}_a)$ of $\phi^{(0)}_a$ in the expansions of $\phi$ and $\psi^+$ form a 0d chiral multiplet for each $a$. The conjugate fields give 0d anti-chiral multiplets. Their physical action is obtained by substituting the corresponding localized eigenmodes into the 2d physical action~(\ref{Lvm}):
\begin{equation}\label{0dspc}
  S^\text{(C)}_\text{0d} = \bar{\boldsymbol u}_a \bar{\Sigma}\Sigma {\boldsymbol u}_a + \bar{\boldsymbol \zeta}_a \bar{\Sigma} {\boldsymbol \zeta}_a\,.
\end{equation}
If we include other fields in the vector multiplet or put the system at the north pole of the squashed sphere, the 0d action that follows from the bulk action has more couplings. It is given explicitly in~(\ref{0d-deformed-action-chiral-explicit}), where the eigenvalue $J$ of ${\rm J}=\frac12 \RV+J_3$ is identified with $q+a$.

Similarly, the coefficients $({\boldsymbol \eta}_\alpha , {\boldsymbol h}_\alpha)$ of $\tilde\phi^{(0)}_\alpha$ in $\psi^+$ and $F$ form a Fermi multiplet, and their conjugates $(\bar{\boldsymbol \eta}_\alpha, \bar{\boldsymbol h}_\alpha)$ form an anti-Fermi multiplet. The action is
\begin{equation}
  S^\text{(F)}_\text{0d} \label{0dspf}
= \bar{\boldsymbol h}_\alpha{\boldsymbol h}_\alpha - \bar{\boldsymbol \eta}_\alpha  \Sigma {\boldsymbol \eta}_\alpha \,.
\end{equation}
The path integral over each Fermi multiplet yields an insertion of $\Sigma$ at the origin. For the Fermi multiplet at the north pole of the sphere, the complete action is~(\ref{0d-deformed-action-Fermi-explicit}). If we started with the 2d matter with $\RV[\phi]=2q$, the $J$-quantum number of $({\boldsymbol \eta}_\alpha , {\boldsymbol h}_\alpha)$ is $q-1-\alpha$.

\subsection{The relations}
\label{sec:relations}

From the detailed comparison of matter path integrals on smeared and singular vortex defect configurations, we found that they are different by a certain number of 0d multiplets with actions~(\ref{0dspc}) and~(\ref{0dspf}). Our result can be summarized as the following operator relation (up to renormalization):
\begin{equation} \label{smear-relations}
\text{``smeared gauge vortex defect''}=
\left\{
  \begin{array}{ll}
\displaystyle
 V_\eta^\text{flipped} \times
\int \prod_{a=0}^{\lfloor \eta\rfloor -1}
d[\lcl u_a,\bar{\lcl u}_a,\lcl\zeta_a,\bar{\lcl\zeta}_a] e^{-   S^\text{(C)}_\text{0d}  }
   &  \text{ for } \eta > 0 \,,\\
\displaystyle
 V_\eta^\text{normal} \times
\int \prod_{\alpha=0}^{\lfloor -\eta\rfloor -1}
d[\lcl \eta_\alpha,\bar{\lcl \eta}_\alpha,\lcl h_\alpha,\bar{\lcl h}_\alpha] e^{-   S^\text{(F)}_\text{0d}  }
&  \text{ for } \eta <0 \,.\\
  \end{array}
\right.
\end{equation}
We note that renormalization can be applied and works out as expected. On the left hand side of~(\ref{smear-relations}) we obtain the ``smeared gauge vortex defect'' renormalized by Pauli-Villars and counterterms. On the right hand side we get Pauli-Villars ghosts for each localized mode labeled by $a$ or $\alpha$. In the large $\Lambda> 0$ limit a 0d ghost multiplet contributes $(\alpha_j \Lambda)^{\pm 1}$ to the leading order. Taking their product, the effect is simply to replace $ V_\eta^\text{normal/flipped} $ by $V_\eta^{\ren,\text{normal/flipped}} = \tilde{\Lambda}^{\kappa(\eta)}\cdot V_\eta^{\text{normal/flipped}} $.

Since the left hand side of (\ref{smear-relations}) is trivial, {\it i.e.}, equal to the identity operator, it is useful to rewrite the relations by inverting the contributions of the 0d multiplets. The superdeterminants inserted by the 0d path integrals can be inverted by flipping the statistics while keeping the same actions, up to standard sign corrections.
\begin{alignat}{2}
  V_\eta^{\ren,\text{flipped} }&= \int \prod_{a=0}^{\lfloor \eta\rfloor -1} d[\lcl \eta_a,\bar{\lcl \eta}_a,\lcl h_a,\bar{\lcl h}_a] e^{-   S^\text{(C)}_\text{0d}  } \qquad& \text{ for }& \eta > 0 \,,
 \\
  V_\eta^{\ren,\text{normal}} &= \int \prod_{\alpha=0}^{\lfloor -\eta\rfloor -1}   d[\lcl u_\alpha,\bar{\lcl u}_\alpha,\lcl \zeta_\alpha,\bar{\lcl \zeta}_\alpha] e^{-   S^\text{(F)}_\text{0d}  }  \qquad& \text{ for }& \eta <0\,.
\end{alignat}

By integrating out the 0d multiplets one obtains the relation between vortex defects and local functionals of vector multiplet fields. Up to $\susy$-cohomology equivalence, the integral over a single Fermi or chiral multiplet gives $\Sigma^{\pm1}$ in  flat space, or $(\hat\Sigma+\frac i\ell\text{J})^{\pm1}$ at the north pole of the sphere. Let us write out the relation explicitly. For the flat space we have
\begin{equation}
 V_\eta^{\ren,\text{flipped}} = \Sigma^{\lfloor\eta\rfloor}\quad(\eta>0),\qquad
 V_\eta^{\ren,\text{normal}} = \Sigma^{\lceil\eta\rceil}\quad(\eta<0).
\end{equation}
For the defect at the north pole of the sphere we have
\begin{alignat}{2}
 V_\eta^{\ren,\text{flipped}} &= \prod_{a=0}^{\lfloor\eta\rfloor-1}
 \left(\hat\Sigma+\frac i\ell(q+a)\right) = \left(-i\ell\right)^{-\lfloor\eta\rfloor}\frac{\Gamma(q+\lfloor\eta\rfloor-i\ell\hat\Sigma)}{\Gamma(q-i\ell\hat\Sigma)},
 \quad &(\eta>0)
 \nonumber \\
 V_\eta^{\ren, \text{normal}} &= \prod_{\alpha=0}^{\lfloor-\eta\rfloor-1}
 \left(\hat\Sigma+\frac i\ell(q-1-\alpha)\right)^{-1} = \left(-i\ell\right)^{-\lceil\eta\rceil}\frac{\Gamma(q+\lceil\eta\rceil-i\ell\hat\Sigma)}{\Gamma(q-i\ell\hat\Sigma)}.
 \quad &(\eta<0)
\label{vd0d}
\end{alignat}

\section{Defect correlators in Abelian GLSMs}
\label{sec:defect-correlators}

Here we apply the formula developed in the previous sections to study the defect correlators on the squashed sphere for some GLSMs with $U(1)$ gauge group. Let $(e_i,2q_i, M_i)$ be the electric charge, R-charge and twisted mass of the $i$-th chiral multiplet, and denote $m_i\equiv \ell M_i+iq_i$. Under our renormalization prescriptions, the renormalized vortex defect correlators  are given by
\begin{eqnarray}
 \langle V_{\eta_\north}^\ren V_{\eta_\south}^\ren\rangle
 &=& \sum_{s\in\frac12(\eta_\north-\eta_\south+\mathbb Z)}
 \int_{\mathbb R}\frac{da}{2\pi}
 \,e^{-it_\ren(a+is)-\eta_\north t_\ren-i\bar t_\ren(a-is)-\eta_\south\bar t_\ren}
 \nonumber \\ && \hskip-9mm\cdot\prod_i
 \left(-i\ell\right)^{-\kappa_{i,\north}(e_i\eta_\north)}
 \left(i\ell\right)^{-\kappa_{i,\south}(e_i\eta_\south)}
\frac{\Gamma( \kappa_{i,\north}(e_i\eta_\north)-e_i\eta_\north+e_is-im_i-ie_ia)}
     {\Gamma(-\kappa_{i,\south}(e_i\eta_\south)+e_i\eta_\south+e_is+1+im_i+ie_ia)}.
\label{dc1}
\end{eqnarray}
The integer-valued function $\kappa_{i,\north}(x)$ equals $\lceil x\rceil$ or $\lfloor x\rfloor$ depending on whether the $i$-th matter satisfies the normal or the flipped boundary condition at the north pole, and the definition of $\kappa_{i,\south}$ is similar.

The first thing we notice is that, under the assumption of charge integrality $e_i\in\mathbb Z$, the one-loop determinant (ratio of products of gamma functions) has unit periodicity in $\eta_\north$ and $\eta_\south$. This implies that the defect operators satisfy shift relations
\begin{equation}
 V_{\eta_\north+1}^\ren = e^{-t_\ren}
 \left(-i\ell\right)^{-\sum_ie_i} V_{\eta_\north}^\ren,\quad
 V_{\eta_\south+1}^\ren = e^{-\bar t_\ren}
 \left(i\ell\right)^{-\sum_ie_i} V_{\eta_\south}^\ren.
\label{etpe2}
\end{equation}
This is a consequence of the invariance of matter path  integral measures  under large gauge transformations. We expect that this shift relation holds generally, not only inside supersymmetric correlators.

We will see below that, as a function of vorticities, the correlator is locally constant and  varies discontinuously only when $e_i\eta_\north$ or $e_i\eta_\south$ crosses an integer value for some $i$.    In particular $V_{\eta_\north}$ is trivial within the range of $\eta_\north$ such that $\kappa_{i,\north}(e_i\eta_\north)=0$ for all $i$.  Another aim of this section is to check the relation~(\ref{vd0d}) between vortex defects and local operators made of $\Sigma$ inside SUSY correlators. For (twisted chiral) vortex defects in GLSMs in  flat space, we expect
\begin{eqnarray}
 V^\ren_{\eta}
 &=& \prod_i(e_i\Sigma+M_i)^{\kappa_i(e_i\eta)}
 \nonumber \\ &=&
  \prod_{e_i\eta>0}(e_i\Sigma+M_i)^{\lfloor e_i\eta\rfloor}
  \prod_{e_i\eta<0}(e_i\Sigma+M_i)^{\lceil e_i\eta\rceil},
\label{vtcf}
\end{eqnarray}
for a vortex defect which requires flipped (or normal) boundary conditions on all the chiral multiplets such that $e_i\eta>0$ (or $e_i\eta<0$, respectively). Similarly, at the north pole of the squashed sphere we expect
\begin{equation}
 V^\ren_\eta
 = \prod_i(-i\ell)^{-\kappa_i(e_i\eta)}
 \frac{\Gamma(-im_i+\kappa_i(e_i\eta)-i\ell\hat\Sigma)}
      {\Gamma(-im_i-i\ell\hat\Sigma)}
\label{vtcs}
\end{equation}
for the defect characterized by the same boundary condition as above.

A natural question is whether these relations hold for any other choice of boundary conditions on matter fields. At this point we notice that by combining the relation~(\ref{vtcf}) in  flat space with the shift relation~(\ref{etpe2}) one finds
\begin{equation}
 e^{-t_\ren(\mu)}\cdot\mu^{\sum_ie_i} = \prod_i(e_i\Sigma+M_i)^{e_i},
\label{rre1}
\end{equation}
where we inserted the renormalization scale $\mu$ to match the dimension of the two sides. This is nothing but the twisted chiral ring relation for $\Sigma$ which follows from the effective twisted superpotential
\begin{equation}
 \widetilde W_\text{eff}(\Sigma) = -t_\ren(\mu)\Sigma-\sum_i(e_i\Sigma+M_i)\left\{\ln\frac{e_i\Sigma+M_i}\mu-1\right\}.
\label{twef}
\end{equation}
The ring relation suggests that the first line of the formula~(\ref{vtcf}) holds for more general choices of boundary conditions than those specified above. In the following we will derive, by a simple manipulation of the integration contour, that the corresponding equation on the squashed sphere~(\ref{vtcs}) holds for somewhat different choice of boundary conditions and range of $\eta$.
We will also see below that the ring relation is {\it quantized} (in the sense that $t_\ren$ and $\Sigma$ should be treated as non-commuting variables) and interpreted as a differential operator that annihilates the sphere partition function.%
\footnote{%
It has been known for some while that the sphere partition function and the related partition functions are annihilated by differential operators.
See~\cite{Closset:2015rna} for a reference that studied such differential operators in ways similar to ours.
The reduction of differential operators to ring relations in the classical limit (i.e., $\ell\to\infty$)  was originally studied by Givental.
See, for example, \cite{MR1354600}.
}

Let us now turn to the formula~(\ref{dc1}) and study the location of poles in the complex $a$-plane. The $i$-th chiral matter contributes a factor $\Gamma(-k_\north)/\Gamma(1+k_\south)$, where
\begin{align}
 k_\north &\equiv ie_ia+im_i-e_is+e_i\eta_\north-\kappa_{i,\north}(e_i\eta_\north),
 \nonumber \\
 k_\south &\equiv ie_ia+im_i+e_is+e_i\eta_\south-\kappa_{i,\south}(e_i\eta_\south).
\end{align}
Since $k_\north-k_\south\in\mathbb Z$ due to flux quantization, this factor diverges when $k_\north,k_\south$ are both non-negative integers. So the poles arising from the $i$-th matter are labelled by some $k_\north,k_\south\in\mathbb Z_{\ge0}$.
\begin{equation}
 a = -\frac i{2e_i}\big(
 \kappa_{i,\north}(e_i\eta_\north)-e_i\eta_\north
+\kappa_{i,\south}(e_i\eta_\south)-e_i\eta_\south
 -2im_i+k_\north+k_\south\big).
\end{equation}
If the $i$-th matter satisfies the normal boundary condition at both poles, then with the additional assumption $q_i>0$ one can show all these poles are in the lower half plane if $e_i>0$, or all in the upper half-plane if $e_i<0$. If we choose the flipped boundary condition at either pole, we need to put more stringent condition on $q_i$ to ensure all the poles lie on one side of the real axis.

Below we study the $\eta$-dependence of the defect correlators in three well-known examples. The vortex defects become increasingly more non-trivial in later examples. 

\paragraph{Example 1.}

\begin{figure}[t]
\begin{center}
\includegraphics[scale=1.2,bb=0 0 201 156]{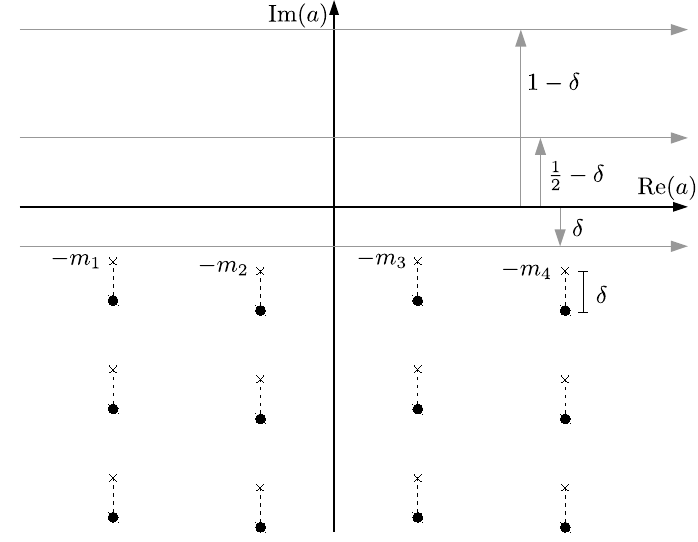}
\caption{The insertion of vortex defects shifts the poles of the integrand in the negative imaginary direction by $\delta$. The triviality of defect correlators in this theory can be shown by shifting the contour of $a$-integration below the real axis by $\delta$. The shifts of contour in the other direction are also shown.}
\label{fig:poles1}
\end{center}
\end{figure}

Let us consider the $U(1)$ gauge theory with $N$ chiral multiplets of charge $+1$, which is the GLSM for $\mathbb C\mathbb P^{N-1}$. We insert the defects  $V_{\eta_\north}$ and $V_{\eta_\south}$ on the two poles, and put the normal boundary condition on all the chiral multiplets. Their correlation function is given by
\begin{eqnarray}
 \langle V^\ren_{\eta_\north}V^\ren_{\eta_\south}\rangle
 &=& (-i\ell)^{-N\lceil\eta_\north\rceil}(i\ell)^{-N\lceil\eta_\south\rceil}
 \nonumber \\ &&\cdot
 \hskip-3mm\sum_{s\in\frac12(\eta_\north-\eta_\south+\mathbb Z)}
 \int\frac{da}{2\pi}z_\ren^{ia-s+\eta_\north}\bar z_\ren^{ia+s+\eta_\south}
 \prod_{i=1}^N\frac{\Gamma(\lceil\eta_\north\rceil-\eta_\north+s-ia-im_i)}
 {\Gamma(-\lceil\eta_\south\rceil+\eta_\south+s+1+ia+im_i)}.
\label{vcp1}
\end{eqnarray}
As shown in the Figure~\ref{fig:poles1}, the poles of the integrand are shifted in the negative imaginary direction by $\delta=\frac12(\lceil\eta_\north\rceil-\eta_\north+\lceil\eta_\south\rceil-\eta_\south)$ because of the defects.

The integrand can be simplified by replacing the variables $a,s$ by
\begin{equation}
a'\equiv  a+\frac i2(
  \lceil\eta_\north\rceil-\eta_\north
 +\lceil\eta_\south\rceil-\eta_\south),\qquad
s'\equiv  s+\frac 12(
  \lceil\eta_\north\rceil-\eta_\north
 -\lceil\eta_\south\rceil+\eta_\south).
\end{equation}
At the same time, we shift the contour of the original $a$-integration below the real axis by $\delta$ so that the integration contour for $a'$ is along the real axis. The contour does not cross any of the poles during the shift provided $\text{Im}(m_i)=q_i>0$. The defect correlator then becomes
\begin{equation}
 \langle V_{\eta_\north}^\ren V_{\eta_\south}^\ren\rangle
 = (-i\ell)^{-N\lceil\eta_\north\rceil}e^{-t_\ren\lceil\eta_\north\rceil}\cdot
    (i\ell)^{-N\lceil\eta_\south\rceil}e^{-\bar t_\ren\lceil\eta_\south\rceil}
\langle 1\rangle.
\label{vtriv}
\end{equation}
So the vortex defects are proportional to identity operators for any $\eta_\north,\eta_\south$. Actually this could have been expected from~(\ref{vtcf}) and~(\ref{rre1}).

Let us explore other ways of shifting the integration contour. For simplicity we turn off the vorticity at the south pole, so that the poles of the integrand are at
\begin{equation}
 a = -m_j-\frac i2(\lceil\eta_\north\rceil-\eta_\north+k_\north+k_\south),
 \quad j\in\{1,\cdots,N\},~~k_\north,k_\south\in\mathbb Z_{\ge0}.
\label{poles12}
\end{equation}
Rewriting~(\ref{vcp1}) in terms of $s'\equiv s-\frac12\eta_\north$ and $a'=a-\frac i2\eta_\north$ we find
\begin{equation}
 \langle V_{\eta_\north}^\ren\rangle =
 (-i\ell)^{-N\lceil\eta_\north\rceil}
 \sum_{s'\in\frac12\mathbb Z}
 \int\frac{da'}{2\pi}e^{-it_\ren(a'+is')-i\bar t_\ren(a'-is')}
 \prod_{i=1}^N\frac{\Gamma(\lceil\eta_\north\rceil+s'-ia'-im_i)}{\Gamma(s'+1+ia'+im_i)}.
\label{vcp2}
\end{equation}
In order that the contour of $a'$-integration be along the real line, the contour of the original $a$-integration has to be shifted to $\mathbb R+\frac i2\eta_\north$. This does not pass any of the poles~(\ref{poles12}) as long as
\begin{equation}
 2q_j+\lceil\eta_\north\rceil>0.
\end{equation}
If we assume $q_j>0$, we need to restrict to $\eta_\north>-1$. For $\eta\in(-1,0]$ this is the same shift of contour as the previous one, but for positive $\eta$ the shift is in the opposite direction. Some examples of the shifted contours are shown in the Figure~\ref{fig:poles1}. A comparison of~(\ref{vcp2}) with the partition function leads to the relation
\begin{equation}
 V^\ren_{\eta_\north} = (-i\ell)^{-N\lceil\eta_\north\rceil}\prod_{i=1}^N
 \frac{\Gamma(\lceil\eta_\north\rceil-i\ell\Sigma-im_i)}
      {\Gamma(-i\ell\Sigma-im_i)}.
\end{equation}
This is the same as the relation~(\ref{vtcs}), but $\eta_\north>-1$ includes outside of the range where it was originally proposed. Writing more explicitly, we have
\begin{alignat}{2}
\eta_\north&\in(-1,0] &\qquad
 V_{\eta_\north}^\ren &= 1,
\nonumber \\[3mm]
\eta_\north&\in(0,1]  &\qquad
 (-i\ell)^NV_{\eta_\north}^\ren
 &= \prod_{j=1}^N(-i\ell\Sigma-im_j),
\nonumber \\
\eta_\north&\in(1,2]  &\qquad
 (-i\ell)^{2N}V_{\eta_\north}^\ren
 &= \prod_{j=1}^N(-i\ell\Sigma-im_j)(-i\ell\Sigma-im_j+1),
~~\cdots
\label{vosi}
\end{alignat}

By comparing the second line with~(\ref{vtriv}) one finds a relation corresponding to~(\ref{rre1}) in  flat space,
\begin{equation}
 \prod_{j=1}^N(-i\ell\Sigma-im_j) = e^{-t_\ren}.
\label{tcr}
\end{equation}
The comparison of the third of~(\ref{vosi}) with~(\ref{vtriv}) gives another relation,
\begin{equation}
 \prod_{j=1}^N(-i\ell\Sigma-im_j)(1-i\ell\Sigma-im_j) = e^{-2t_\ren}.
\label{nss}
\end{equation}
This is not the simple square of the previous relation, but reduce to it in the limit $\ell\to\infty$ with $m_j/\ell$ fixed. The failure of the simple ring relation on the squashed sphere can be interpreted as an effect of Omega deformation: namely the SUSY near the north pole of the sphere is that of the topological A-twisted theory in an Omega background with $\epsilon=1/\ell$.

The above failure can be fixed by regarding $\Sigma$ as a differential operator,
\begin{equation}
 -i\ell\Sigma
 = \frac{\partial}{\partial t_\ren}
 = -z_\ren\frac{\partial}{\partial z_\ren},\qquad
 (z_\ren\equiv e^{-t_\ren})
\end{equation}
and regarding the relation~(\ref{tcr}) as the differential equation satisfied by the sphere partition function (here we omit the suffix $\ren$),
\begin{equation}
 \bigg\{\prod_{j=1}^N(-z\partial_z-im_j)-z\bigg\}Z_{S^2}=0.
\label{pf0}
\end{equation}
The differential operator corresponding to the third line of~(\ref{vosi}) then acts consistently,
\begin{eqnarray}
 \bigg\{\prod_{j=1}^N(1-z\partial_z-im_j)(-z\partial_z-im_j)\bigg\}Z_{S^2}
 &=& \prod_{j=1}^N(1-z\partial_z-im_j)\cdot zZ_{S^2}
 \nonumber \\
  &=&
 z\cdot\prod_{j=1}^N(-z\partial_z-im_j)\, Z_{S^2} = z^2Z_{S^2},
\end{eqnarray}
where we used~(\ref{pf0}) repeatedly. The translation of the operator relation into a differential equation might look somewhat ad hoc. However, the above differential equation is actually satisfied by the holomorphic blocks that appear in the Higgs branch localization formula for the sphere partition function. It is known that $Z_{S^2}$ for this model factorizes into a sum of product of a holomorphic and an anti-holomorphic blocks if the $a$-integral is rewritten as a sum over pole residues. The holomorphic blocks are given by the following $N$ independent solutions to~(\ref{pf0}).
\begin{equation}
 F_i(z) = \sum_{n\ge0}\frac{(-1)^{Nn}z^{n+\rho}}{\prod_{j=1}^N\Gamma(n+\rho+1+im_j)}\bigg|_{\rho=-im_i}.\qquad(i=1,\cdots,N)
\end{equation}
These agree precisely with the vortex partition functions at $N$ distinct Higgs branch vacua.

\paragraph{Example 2.}

Let us next couple $N$ chirals with charge $+1$ (electrons) and $\tilde N$ chirals with charge $-1$ (positrons). For positive FI parameter $r$, the theory is the GLSM for the total space of the vector bundle ${\cal O}(-1)^{\oplus\tilde N}$ over $\mathbb C\mathbb P^{N-1}$. The $N$ electrons have the mass and R-charge $m_i=\ell M_i+iq_i$, while the $\tilde N$ positrons have $\tilde m_i=\ell\tilde M_i+i\tilde q_i$. The R-charges $q_i,\tilde q_i$ are assumed all positive. We put a vortex defect on the north pole and require the normal boundary condition on all the chiral multiplets.

\begin{figure}[t]
\begin{center}
\includegraphics[bb=0 0 198 142,scale=1.2]{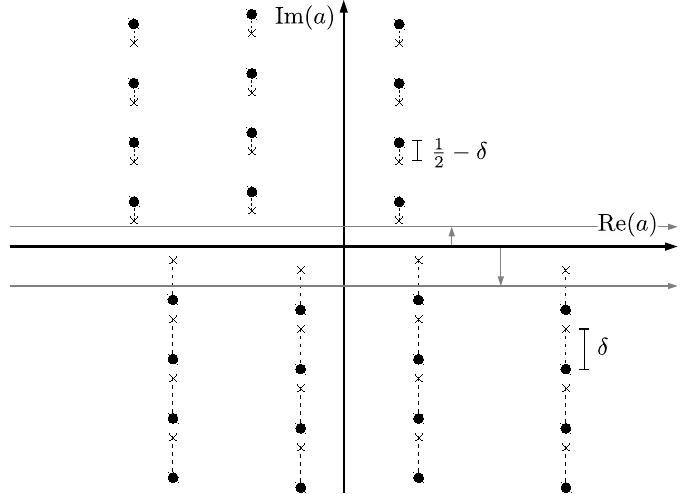}
\caption{The insertion of vortex defects moves down the poles in the lower half-plane by $\delta=\frac12(\lceil\eta_\north\rceil-\eta_\north)$, and moves up the poles in the upper half-plane by $\frac12-\delta$. We shift the integration contour by $\frac i2\eta_\north$, the direction depending on the sign of $\eta_\north$.}
\label{fig:poles2}
\end{center}
\end{figure}

The defect expectation value is given by
\begin{eqnarray}
 \langle V_{\eta_\north}^\ren\rangle
 &=& (-i\ell)^{-N\lceil\eta_\north\rceil-\tilde N\lceil-\eta_\north\rceil}
 \hskip-3mm\sum_{s\in\frac12(\eta_\north+\mathbb Z)}
 \int\frac{da}{2\pi}z_\ren^{ia-s+\eta_\north}\bar z_\ren^{ia+s}
 \nonumber \\ && \hskip1.5mm\times
 \prod_{i=1}^N\frac{\Gamma(\lceil\eta_\north\rceil-\eta_\north+s-ia-im_i)}{\Gamma(s+1+ia+im_i)}
 \prod_{i=1}^{\tilde N}\frac{\Gamma(\lceil-\eta_\north\rceil+\eta_\north-s+ia-i\tilde m_i)}
 {\Gamma(-s+1-ia+i\tilde m_i)}.
\label{vcp3}
\end{eqnarray}
This time the integrand has sequences of poles both in the lower and upper half planes. For non-integer vorticity, the effect of $\eta_\north$ is to move the poles in the lower half-plane downwards by $\delta=\frac12(\lceil\eta_\north\rceil-\eta_\north)$ and those in the upper half-plane upwards by $\frac12-\delta$, as shown in Figure~\ref{fig:poles2}.

The vortex defect is a non-trivial operator in this case for any non-zero $\eta_\north$, as one cannot eliminate the $\eta_\north$-dependence of the integrand by any change of variables $(s,a)$. We study this by rewriting~(\ref{vcp3}) in terms of $s'=s-\frac12\eta_\north$, $a'=a-\frac i2\eta_\north$ and shifting the $a$-integration contour by $\frac i2\eta_\north$. Note that the allowed shift of contour is bounded from both sides since there are poles in both the lower and the upper half-planes.

The contour does not pass any poles as long as
\begin{equation}
 2q_j+\lceil\eta_\north\rceil>0,\quad
 2\tilde q_j+\lceil-\eta_\north\rceil>0.
\end{equation}
The allowed values of $\eta_\north$ is thus restricted to $-1<\eta_\north<1$. For $\eta_\north=0$ the defect is trivial, and for other values of $\eta_\north$ one finds the relations
\begin{alignat}{2}
\eta_\north&\in(-1,0) &\qquad
 (-i\ell)^{\tilde N}V_{\eta_\north}^\ren
 &=  \prod_{j=1}^{\tilde N}(i\ell\Sigma-i\tilde m_j),
\nonumber \\
\eta_\north&\in(0,1)  &\qquad
 (-i\ell)^N V_{\eta_\north}^\ren
 &= \prod_{j=1}^N(-i\ell\Sigma-im_j).
\label{vosi2}
\end{alignat}
Again these agree with~(\ref{vtcs}), but the agreement is for the choice of $\eta_\north$ and boundary conditions which were not covered by the original proposal. Finally, combining this with the shift relation~(\ref{etpe2}) one recovers~(\ref{rre1})
\begin{equation}
\prod_{j=1}^N(-i\ell\Sigma-im_j)
= e^{-t_\ren}
 \prod_{j=1}^{\tilde N}(i\ell\Sigma-i\tilde m_j).
\end{equation}

\paragraph{Example 3.}

Let us next consider the GLSM which flows to a conformally invariant NLSM on quintic Calabi-Yau, that is the $U(1)$ gauge theory with $5$ chiral multiplets $\{X_i\}_{i=1}^5$ with charges $\text{e}=1$, $\RV=2q$ and a chiral multiplet $P$ with $\text{e}=-5, \RV=2-10q$. The theory has a superpotential $W = PG(X_i)$ with $G$ a generic quintic polynomial of $X_i$, and therefore has no continuous flavor symmetry. The free parameter $q$ corresponds to the shift of R-charge by the electric charge, so physical observables are supposed to be independent of $q$ as long as the R-charges of all chiral multiplets are positive. As in the previous case, we introduce a vortex defect $V_{\eta_\north}$ only at the north pole and derive the relation between $V_{\eta_\north}$ and polynomials of $\Sigma$.

The specific form of the superpotential is chosen so that one has $P=G(X)=0$ at low energy in the geometric phase ($r>0$). In order that we have this lifting in the presence of the defect, it is most natural to flip the boundary condition of $P$, though the other choices of boundary conditions are also allowed.

The poles of the integrand which are closest to the real axis are at
\begin{equation}
 a= -\frac i2(\lceil\eta_\north\rceil-\eta_\north+2q)\in\text{LHP},\qquad
 a= \frac i{10}(\lfloor-5\eta_\north\rfloor+5\eta_\north+2-10q)\in\text{UHP}.
\end{equation}
We are going to shift the contour of $a$-integration later by $i\eta_\north/2$. The contour does not cross any poles as long as
\begin{equation}
 \lceil\eta_\north\rceil+2q>0,\quad
 \lfloor-5\eta_\north\rfloor+2-10q>0.
\end{equation}
For a small positive $q$, one can vary $\eta_\north$ within $-1<\eta_\north<1/5$.

The expectation value of the defect can then be written as
\begin{eqnarray}
 \langle V_{\eta_\north}^\ren\rangle
 &=& (-i\ell)^{-5\lceil\eta_\north\rceil-\lfloor-5\eta_\north\rfloor}\sum_{s\in\frac12\mathbb Z}
 \int_{\mathbb R}\frac{da}{2\pi}e^{-it_\ren(a+is)-i\bar t_\ren(a-is)}
 \nonumber \\ && \hskip15mm\times
 \left[\frac{\Gamma( \lceil\eta_\north\rceil+s+q-ia)}{\Gamma(s+1-q+ia)}\right]^5
 \frac{\Gamma( \lfloor-5\eta_\north\rfloor-5s+1-5q+5ia)}
      {\Gamma(-5s+5q-5ia)}\,.
\end{eqnarray}
From this one can identify $V_{\eta_\north}$ with the local polynomial of $\Sigma$,
\begin{equation}
 V_{\eta_\north}^\ren
 =(-i\ell)^{\lceil5\eta_\north\rceil-5\lceil\eta_\north\rceil}
 \left[
 \frac{\Gamma(\lceil\eta_\north\rceil-\Theta)}
      {\Gamma(-\Theta)}\right]^5
 \frac{\Gamma( \lfloor-5\eta_\north\rfloor+1+5\Theta)}
      {\Gamma(1+5\Theta)},\qquad
 \Theta\equiv i\ell\Sigma-q.
\end{equation}
More explicitly, one finds
\begin{alignat}{4}
& 5\eta_\north\in(&0,&\,&1]&\qquad&
 (-i\ell)^4V_{\eta_\north}^\ren &= -\Theta^4/5,
 \nonumber \\
& 5\eta_\north\in(&-1,&\,&0]&\qquad&
 V_{\eta_\north}^\ren &= 1,
 \nonumber \\
& 5\eta_\north\in(&-2,&\,&-1]&\qquad&
 (-i\ell) V_{\eta_\north}^\ren &= 1+5\Theta,
 \nonumber \\
& 5\eta_\north\in(&-3,&\,&-2]&\qquad&
 (-i\ell)^2 V_{\eta_\north}^\ren &= (1+5\Theta)(2+5\Theta),
 \nonumber \\
& 5\eta_\north\in(&-4,&\,&-3]&\qquad&
 (-i\ell)^3 V_{\eta_\north}^\ren &= (1+5\Theta)(2+5\Theta)(3+5\Theta),
 \nonumber \\
& 5\eta_\north\in(&-5,&\,&-4]&\qquad&
 (-i\ell)^4 V_{\eta_\north}^\ren &= (1+5\Theta)(2+5\Theta)(3+5\Theta)(4+5\Theta).
\end{alignat}
Combining this with the shift relation~(\ref{etpe2}) one obtains a formal relation
\begin{equation}
 \Theta^4 = -5^5e^{-t}(\Theta+\frac45)(\Theta+\frac35)(\Theta+\frac25)(\Theta+\frac15)
\label{rlq}
\end{equation}
that holds inside the sum and the integral.

Although the FI-theta parameter $t$ is not renormalized in the sense of~(\ref{tren}), there is a finite renormalization so that the singularity in the K\"ahler moduli space at which a non-compact Coulomb branch emerges is shifted from the origin to \cite{Witten:1993yc, Morrison:1994fr} 
\begin{equation}
 t~=~ -\sum_{\Phi=X_{1,\cdots,5},P}\text{e}_\Phi\log(\text{e}_\Phi)
 ~ =~ 5\ln 5 + i\pi \quad \text{ mod } 2\pi i \,.
\end{equation}
This can be understood from the effective twisted superpotential~(\ref{twef}). Taking this into account we define the worldsheet instanton expansion parameter $z\equiv -5^5e^{-t}$ so that the singularity is at $z=1$, and translate $\Theta$ into a differential operator accordingly.
\begin{equation}
 \Theta = -\partial_t-q = z^q(z\partial_z)z^{-q}.
\end{equation}
One then finds that the relation~(\ref{rlq}) turns into a fourth-order differential equation satisfied by the sphere partition function,
\begin{equation}
 \left\{(z\partial_z)^4
 -z(z\partial_z+\frac45)(z\partial_z+\frac35)
   (z\partial_z+\frac25)(z\partial_z+\frac15)\right\}z^{-q} Z_{S^2}=0.
\label{pfd}
\end{equation}
This is the familiar Picard-Fuchs differential equation for period integrals of the mirror quintic. We summarize some basic facts about the mirror symmetry of quintic Calabi-Yau in Appendix~\ref{sec:quintic}.

\section{Mirror symmetry for vortex defects}
\label{sec:mirr-symm-vort}

\subsection{Hori-Vafa mirror symmetry}
\label{sec:hori-vafa-mirror}

It is known that the Abelian ${\cal N}=(2,2)$ GLSMs discussed in the previous section are  dual  to LG theories of twisted chiral multiplets.
 This duality is called mirror symmetry~\cite{Hori:2000kt}. 
Here we identify the local operators in the LG model which are  the mirror dual  of the vortex defects.

Let us begin by recalling mirror symmetry in the simplest setting. Take a $U(1)$ SQED with a single chiral multiplet of charge $1$. Its mirror is a LG model of two twisted chiral multiplets $\Sigma$ and $Y$, with the twisted superpotential
\begin{equation}
 \widetilde W = \Sigma(Y-t_\ren(\mu))+\mu e^{-Y}.
\label{tWf}
\end{equation}
As in previous sections, $t_\ren(\mu)$ here is the renormalized FI coupling at scale $\mu$, and the multiplet of $\Sigma$ is made of the vector multiplet fields of the original SQED. The imaginary part of $Y$ is of period $2\pi$, and it is T-dual to the phase of the charged scalar in the original theory.

The action of mirror symmetry on vortex defects was studied in~\cite{Okuda:2015yra} for the system of a chiral multiplet coupled to a {\it background} $U(1)$ vector multiplet on flat plane. There it was shown that the mirror operator is an exponential of the twisted chiral field $Y$. Let us reproduce this result using a different argument. Consider a defect with vorticity density $\varrho(x)$.
SUSY requires that the background gauge field and the auxiliary field are given as
\begin{equation}
 F_{12} = 2\pi \varrho,  \qquad  D = 2\pi i \varrho.
\label{gfbg}
\end{equation}
See (\ref{smear-background-flat}) and (\ref{dfdv}).
When $\varrho=\eta \cdot \delta^2(x)$, the defect defines a local operator (or the end point of a topological line) and is twisted chiral.
The twisted F-term  (\ref{LWt})  in the mirror LG theory has a $\varrho$-dependent contribution
\begin{equation}
 {\cal L} \sim
 i\text{Re}(Y-t_\ren)D-i\text{Im}(Y-t_\ren)F_{12}
 ~=~ 2\pi\varrho (t_\ren-Y),
\label{tfl}
\end{equation}
which can be regarded as an insertion of $\exp[\int d^2 x \varrho(x)(Y-t_\ren)]$, which becomes $e^{\eta (Y-t_\ren)}$ in the local limit. Note that the vorticity turned into the momentum of $Y$. For generic $\eta$ the mirror operator does not meet the quantization law of momentum, which implies that the phase of the original chiral multiplet scalar has  a non-integer  winding number around the defect. As explained in a paragraph  in page \pageref{pg:vfl}, vortex defects without dynamical vector multiplets can be used to describe twist fields in orbifold theories because of this property.

In the above and in~\cite{Okuda:2015yra} the defect was defined with smearing regularization.  As a result  the corresponding mirror operator depends analytically on $\eta$.  
For vortex defects defined by boundary conditions and  without smearing, this gets  modified and the mirror operator  depends discontinuously on  $\eta$.

\paragraph{Correlators on the squashed sphere}
\label{pg:mirr-symm-squash}

We would like to find the mirror of vortex defects via comparison of exact correlators on the squashed sphere. For this purpose, we need to recall first the localization formula for the general ${\cal N}=(2,2)$ LG models of twisted chiral multiplets on the squashed sphere~\cite{Gomis:2012wy}.

The saddle point condition for a general twisted chiral multiplet reads
\begin{equation}
 D_mY=D_m\bar Y=0\,,\qquad
 G=0~~\text{except at NP},\qquad
 \bar G=0~~\text{except at SP}.
\label{spc}
\end{equation}
The saddle point configuration for the vector multiplet~(\ref{spv}) corresponds to
\begin{equation}
 \Sigma=\frac1\ell(a+is),\quad
 \bar\Sigma=\frac1\ell(a-is),\quad
 G_\Sigma = 4\pi i\eta_\north\delta^2_\text{(NP)},
 \quad
 \bar G_\Sigma = 4\pi i\eta_\south\delta^2_\text{(SP)},
\label{sdv}
\end{equation}
which satisfies the above condition. For partition functions of general LG models, one only takes account of the saddle points with $G=\bar G=0$ everywhere. Moreover, the one-loop determinant is independent of the choice of saddle points, so the path integral reduces to an ordinary integral over constant modes of the twisted chiral fields
\begin{equation}
 Z \sim \int [dYd\bar Y] e^{i\ell \widetilde W(Y)+i\ell\overline{\widetilde W}(\bar Y)}.
\label{lgtpf}
\end{equation}
The localization works in the same way if there are insertions of a twisted chiral operator ${\cal O}(Y)$ at the north pole and/or an anti twisted chiral operator $\bar{\cal O}(\bar Y)$ at the south pole. Their correlators are given by the same formula~(\ref{lgtpf}) with additional factors ${\cal O}(Y)\bar{\cal O}(\bar Y)$ in the integrand.

As was shown in~\cite{Gomis:2012wy}, the sphere partition functions for general Abelian GLSMs can be brought into the above form~(\ref{lgtpf}). As the simplest example, let us take the $U(1)$ SQED with a single charged chiral multiplet.
\begin{equation}
 Z_\text{SQED} = \sum_{s\in\frac12\mathbb Z}
 \int_{\mathbb R}\frac{da}{2\pi}\,
 e^{-it_\ren(a+is)-i\bar t_\ren(a-is)}
 \frac{\Gamma(s+q-ia)}
      {\Gamma(s+1-q+ia)}.
\end{equation}
Using the formula
\begin{equation}
 \frac{\Gamma(q+s-ia)}{\Gamma(1-q+s+ia)} =
 \frac1\pi\int_\mathbb{R}d\text{Re}Y\int_{-\pi}^\pi d\text{Im}Y
 e^{-q(Y+\bar Y)+i(a+is)Y+i(a-is)\bar Y+e^{-\bar Y}-e^{-Y}},
\label{ggr}
\end{equation}
which holds for any $q,a\in\mathbb R$ and $s\in\frac12\mathbb Z$, the partition function can be rewritten as
\begin{equation}
 Z_\text{SQED} =
 \sum_{s\in\frac12\mathbb Z}
 \int_{\mathbb R}\frac{da}{2\pi}
 \int\frac{d^2Y}{\pi}
 e^{-q(Y+\bar Y)}
  e^{i\ell\widetilde W(\Sigma,Y)+i\ell\overline{\widetilde W}(\bar\Sigma,\bar Y)}.
\end{equation}
Here $\frac1\ell(a+is)\equiv\Sigma$ and the twisted superpotential is given by
\begin{equation}
 \widetilde W=\Sigma(Y-t_\ren)  +\frac i\ell e^{-Y},\quad
 \overline{\widetilde W} =\bar\Sigma(\bar Y-\bar t_\ren)  -\frac i\ell e^{-\bar Y},
\end{equation}
which agrees with~(\ref{tWf}).

Note that for complex $q$, the factor $e^{-i {\rm Im}(q)(Y+\bar Y)}$ can be absorbed into $\widetilde W$ and $\overline{\widetilde W}$ by regarding ${\rm Im}(q)$ as the twisted mass for the chiral multiplet in the original system.
On the other hand one can show that the factor $e^{-{\rm Re}(q) (Y + \bar Y)}$ plays the role of a measure factor in related models \cite{Hori:2000kt}.

Let us next introduce the vortex defects $V_{\eta_\north},V_{\eta_\south}$ in the same SQED and identify the local operators $\widetilde V_{\eta_\north},\widetilde V_{\eta_\south}$ in the mirror LG model they correspond to.  The auxiliary super partners of $\Sigma$ and $\bar\Sigma$ take non-zero values localized at the poles~(\ref{sdv}), leading to a modification of the formula~(\ref{lgtpf}). We determine the mirror operators by the following ansatz:
\begin{equation}
\langle V_{\eta_\north}V_{\eta_\south}\rangle
 = \sum_{s\in\frac12(\eta_\north-\eta_\south+\mathbb Z)}
 \int\frac{da}{2\pi}\int\frac{d^2Y}\pi 
e^{-q(Y+\bar Y)}
 e^{i\ell\widetilde W +i\ell\overline{\widetilde W}}
 e^{\eta_\north(Y-t_\ren)    +\eta_\south(\bar Y-\bar t_\ren)}
 \widetilde V_{\eta_\north}(Y)\widetilde V_{\eta_\south}(\bar Y).
\end{equation}
Note that the factors $e^{\eta_\north(Y-t_\ren)}, e^{\eta_\south(\bar Y-\bar t_\ren)}$ in the integrand arose from the first term in~(\ref{LWt}), and they correspond to the effect of vortex defects explained in~(\ref{tfl}). By rewriting the defect correlator using~(\ref{ggr}) one finds
\begin{align}
&\langle V_{\eta_\north}V_{\eta_\south}\rangle
 = \sum_{s\in\frac12(\eta_\north-\eta_\south+\mathbb Z)}
 \int_{\mathbb R}\frac{da}{2\pi}\,
 e^{-it_\ren(a+is)-\eta_\north t_\ren-i\bar t_\ren(a-is)-\eta_\south \bar t_\ren}
 \frac{\Gamma(\kappa_\north(\eta_\north)-\eta_\north+s+q-ia)}
      {\Gamma(-\kappa_\south(\eta_\south)+\eta_\south+s+1-q+ia)}
 \nonumber \\ &\hskip8mm=
 \sum_{s\in\frac12(\eta_\north-\eta_\south+\mathbb Z)}
 \int_{\mathbb R}\frac{da}{2\pi}
 \int\frac{d^2Y}{\pi}
  e^{-q(Y+\bar Y)}
  e^{i\ell\widetilde W+i\ell\overline{\widetilde W}
   +(\eta_\north-\lceil\eta_\north\rceil)Y
   +(\eta_\south-\lceil\eta_\south\rceil)\bar Y
   -\eta_\north t_\ren-\eta_\south\bar t_\ren}.
\label{vtxcr}
\end{align}
From this one can identify
\begin{equation}
 \widetilde V_{\eta_\north}(Y)= e^{-\kappa_\north(\eta_\north)Y},\quad
 \widetilde V_{\eta_\south}(\bar Y)= e^{-\kappa_\south(\eta_\south)\bar Y}.
\label{VetaY}
\end{equation}

One can generalize the above result to $U(1)$ SQEDs with several charged chiral multiplets. As in~(\ref{dc1}), let $(e_i,q_i,M_i)$ be the electric charge, R-charge and the mass of the $i$-th chiral multiplet. Then the twisted superpotential for the mirror LG theory on the squashed sphere is given by
\begin{eqnarray}
 \widetilde W &=& \sum_i \left\{\Big(e_i\Sigma+M_i \Big)Y_i
+\frac i\ell e^{-Y_i}\right\}-t_\ren\Sigma,\nonumber \\
 \overline{\widetilde W} &=& \sum_i \left\{\Big(e_i\bar\Sigma+M_i \Big)\bar Y_i-\frac i\ell e^{-\bar Y_i}\right\}-\bar t_\ren\bar\Sigma.
\end{eqnarray}
The defect correlator~(\ref{dc1}) can be rewritten as
\begin{eqnarray}
\langle V_{\eta_\north}V_{\eta_\south}\rangle
 &=& \sum_{s\in\frac12(\eta_\north-\eta_\south+\mathbb Z)}
 \int\frac{da}{2\pi}\prod_i\frac{d^2Y_i}{\pi}\;  e^{-\sum_i q_i(Y_i+\bar Y_i)} e^{i\ell\widetilde W+i\ell\overline{\widetilde W}}
\nonumber\\
&& \hskip 35mm
\times  e^{\eta_\north(\sum_ie_iY_i-t_\ren)+\eta_\south(\sum_ie_i\bar Y_i-\bar t_\ren)}\widetilde V_{\eta_\north}\widetilde V_{\eta_\south},
\label{gvm}
\end{eqnarray}
where
\begin{equation}\label{V-tilde-def}
 \widetilde V_{\eta_\north}=\prod_i e^{-\kappa_{i,\north}(e_i\eta_\north)\, Y_i},\quad
 \widetilde V_{\eta_\south}=\prod_i e^{-\kappa_{i,\south}(e_i\eta_\south)\,\bar Y_i}.
\end{equation}
The functions $\kappa_{i,\north}$ and $\kappa_{i,\south}$ are determined according to the boundary condition on the $i$-th chiral multiplet at the north and south poles. Note that, when defining the mirror of the vortex defects, we separated the part which arise from the classical delta-functional values of $D, F_{12}$ (that is, $e^{\eta_\north(\sum_ie_iY_i-t_\ren)}$ or $e^{\eta_\south(\sum_ie_i\bar Y_i-\bar t_\ren)}$ in the above) from the rest ($\widetilde V_{\eta_\north}$ and $\widetilde V_{\eta_\south}$).  The contributions from the smeared defect \cite{Okuda:2015yra} are in the first part, whose logarithm $\propto \partial_{\Sigma} \widetilde{W}$ is trivial in the twisted chiral ring. For flavor vortex defects, the operators in the mirror are given by the product of these two pieces. However, for the gauge vortex defects one can drop, if desired, the first part because the twisted superpotential is linear in $\Sigma$ and therefore the first piece can be formally absorbed into the redefinition of $a$ and $s$. Thus the mirror of the gauge vortex operators are given by (\ref{V-tilde-def}).

One can check that the mirror vortex operators satisfy the shift relation,
\begin{equation}
 \widetilde V_{\eta_\north+n}
 = (e^{-\sum_i e_iY_i})^n\,\widetilde V_{\eta_\north}
 = e^{-nt_\ren}\,\widetilde V_{\eta_\north}\,,
\end{equation}
where the twisted F-term condition is used at the second equality. Note that this holds irrespective of the choice of boundary conditions. Similar equality holds also for $\widetilde V_{\eta_\south}$. It is interesting to note that, though the twisted chiral ring relation in the GLSM side~(\ref{nss}) had subtlety when taking products, here the ring relation can be multiplied simply. This is not a contradiction: here the twisted chiral ring is described by the fields $\Sigma$ and $Y_i$, whereas the one for the GLSM can be obtained from it by integrating out $Y_i$.

\subsection{${\cal N}=2$ minimal model and its orbifold}
\label{sec:cal-n=2-minimal}

As a simple application of the above results,  let us consider ${\cal N}=2$ minimal model of level $k\in\mathbb Z_+$, which is the theory of a single chiral multiplet $\Phi$ of R-charge $2q=2/h~(h\equiv k+2)$ with superpotential
\begin{equation}
W={\rm g}_0 \cdot \Phi^h,
\end{equation}
where ${\rm g}_0$ is a coupling constant. This theory is known to give a CFT with the central charge $c=3(h-2)/h$. Also, it has a discrete flavor symmetry which phase-rotates $\Phi$ while keeping the superpotential invariant.

The minimal model of level $k$ is known to be mirror to its $\mathbb Z_h$ orbifold \cite{Vafa:1989xc}. So let us take the orbifold of the above theory of $\Phi$. We are interested in the $p$-th twisted sector, where $\Phi$ is quasi-periodic: $\Phi(\varphi+2\pi) = e^{2\pi i p/h} \Phi(\varphi)$.  Alternatively, one can work in the frame where $\Phi$ is periodic but there is a non-zero background gauge field, $A_\varphi=-p/h$ mod $\mathbb Z$ with $p=1,2,\cdots,h-1$.  We thus consider a pair of vortex defects $V_{\eta_{\north}}, V_{\eta_\south}$ with $\eta_\north=\eta_\south=-p/h$.
We require the chiral multiplet $\Phi$  to satisfy the normal boundary condition at both defects, so that the operators have positive dimensions
\begin{equation}
 [V_{\eta_\north}] = [V_{\eta_\south}] = \frac ph >0
\end{equation}
as read off from (\ref{flavor-renormalization}).
Their renormalized correlation function on the squashed sphere of size $\ell$ reads
\begin{equation}
 \left\langle V^\ren_{\eta_\north}V^\ren_{\eta_\south}\right\rangle
 = \ell^{-\frac{2p}h}\frac1h\frac{\Gamma(\frac{1+p}h)}{\Gamma(1-\frac{1+p}h)}
 = \ell^{-\frac{2p}h}\frac{\Gamma(\frac{1+p}h)^2}{h\pi}\sin\frac{(1+p)\pi}h\,.
\end{equation}
The factor $1/h$ arises because the correlator is given by the average of $h$ twisted sector contributions. Using the formula~(\ref{ggr}) one can also rewrite%
\footnote{%
Essentially the same expressions were obtained in equation~(5.27) of~\cite{Hori:2000ck} for the $tt^*$ correlation functions.
}
\begin{eqnarray}
 \frac1h\frac{\Gamma(\frac{1+p}h)}{\Gamma(1-\frac{1+p}h)} &=&
 \frac1{\pi h^2}\int_{\mathbb R}d{\text{Re}Y}\int_{-\pi h}^{\pi h}d\text{Im}Y
 e^{-\frac{1+p}h(Y+\bar Y)+e^{\bar Y}-e^{-Y}}
 \nonumber \\ &=& \frac1\pi\int d^2Z \, Z^p\bar Z^p e^{\bar Z^h-Z^h}.
\end{eqnarray}
This shows that the vortex defect correlators  in the orbifold theory reproduce the correlator of basic monomial operators in the mirror minimal model rather precisely.

A defect with $\eta=-p/h$ and the normal boundary condition imposed on $\Phi$, as considered above, can be given a regularization of the short distance singularity by smearing. 
To demonstrate this, let us work in flat space.
We enlarge the discrete symmetry $\mathbb{Z}_h$ to continuous $U(1)$ and promote the coupling constant ${\rm g}_0$ to the bottom component ${\rm g}$ of a non-dynamical chiral multiplet. With respect to the $U(1)$, $\Phi$ has charge +1 and ${\rm g}$ has charge $-h$.  This is the symmetry used in the previous paragraph to remove the quasiperiodicity of $\Phi$. Therefore, in the new gauge where $\Phi$ is periodic, the coupling becomes position-dependent: ${\rm g}={\rm g}_0\cdot e^{+ip\varphi}$.  We replace $A_{\varphi}=-p/h$ by the usual smeared defect configuration~(\ref{smearing-background-flat}).
We also need to modify ${\rm g}$ to a smooth background field. It has to approach ${\rm g}_0 \cdot e^{+i p\varphi}$ away from the defect.  To preserve SUSY,  it also needs to be annihilated by $D_{\bar  z}=\frac{e^{i\varphi}}{2}(\partial_r + \frac{i}{r} (\partial_\varphi +i h A_\varphi))$.  The function ${\rm g}_0 \cdot e^{+i p\varphi}$ is indeed annihilated by  $D_{\bar  z}$ for $r\gg\epsilon$.
It then follows that for $r\ll\epsilon$, ${\rm g}\sim r^p e^{+i p\varphi}=z^p$.
This is  regular because $p>0$.
We also know from Section~\ref{sec:eigenv-probl-flat} that the bulk modes for $\Phi$ obey the normal boundary condition because $\eta<0$, and that there are no localized modes because $-1<\eta$.
Thus we have succeeded in regularizing the defect with the normal boundary condition.

What about the flipped boundary condition?
A defect defined by the flipped boundary condition would, according to the flipped version of~(\ref{flavor-renormalization}), have a negative dimension and should be unphysical.
We can also see that such a defect cannot be regularized by smearing as follows.
To obtain a flipped boundary condition without localized modes, we need that $0<\eta<1$.
Thus we take $\eta=1-p/h$.
In the gauge where $A_\varphi=\eta=1-p/h$, the coupling constant depends on $\varphi$ as ${\rm g}={\rm g}_0 e^{i(p-h)\varphi} $ for $r\gg \epsilon$.
Solving $D_{\bar z}{\rm g}=0$, we find that ${\rm g}\sim z^{p-h}$ for $r\ll\epsilon$.
This is singular, so smearing fails to provide a UV regularization of the defect.

\section{Vortex defects at conical singularities}\label{sec:vcns}

Here we study the vortex defect $V_\eta$ inserted at a $\mathbb Z_K$-orbifold fixed point or, more generally, at a conical singularity. We first notice that, for suitable choice of $\eta$ and $K$, the vortex defects at conical singularities are related to orbifold twisted sectors. The correlator of such defects can be evaluated from partition functions by a simple orbifold projection, and one does not need a careful examination of boundary conditions of matter wave functions. In~\cite{Hosomichi:2015pia} a formula for defect correlators was proposed based on this approach. However, there is a subtlety in extrapolating the result in $\eta$ and $K$: contrarily to the assumption made in~\cite{Hosomichi:2015pia}, a detailed analysis of matter wave functions shows that the defect correlators depend non-trivially on $\eta$ as well as (generically non-integer) $K$.

\subsection{Orbifold projection}

Let us begin by considering the $\mathbb Z_K$ orbifold identification of a charged scalar $\phi$ on the complex plane,
\begin{equation}
 \phi(ze^{\frac{2\pi i}K})=e^{-\frac{2\pi ip}K}\phi(z).
 \quad(p,K\in\mathbb Z).
\end{equation}
The field $\phi$ is in the $p$-th twisted sector of the $\mathbb Z_K$ orbifold. This can be understood as the effect of a $U(1)$ vortex defect with $\eta=p/K$ inserted at the orbifold fixed point. By taking a similar $\mathbb Z_K$ orbifold of the squashed sphere, one can introduce vortex defects of opposite vorticity at the north and the south poles of the (topological) sphere with conical singularities. The SUSY on the squashed sphere can be preserved by combining this $\mathbb Z_K$ rotation with a suitable vector R- and local Lorentz rotations, so that the Killing spinors are invariant. To find out the condition for SUSY-preserving orbifold, we move to a frame in which the spin connection has no Dirac string singularity at the north pole,
\begin{equation} 
 ds^2=f(\theta)^2d\theta^2+\ell^2\sin^2\theta d\varphi^2,\quad
\omega^{12}=\left(1-\frac\ell f\cos\theta\right)d\varphi.
\label{orbmet}
\end{equation}
The Killing spinors in this frame take the form
\begin{equation}
 \xi
 =\left(\begin{array}{r} \xi^+ \\ \xi^- \end{array}\right)
 = \left(\begin{array}{r}
 ie^{-i\varphi}\sin\frac\theta2 \\ \cos\frac\theta2 \end{array}\right),
 \quad
 \bar\xi
 =\left(\begin{array}{r} \bar\xi^+ \\ \bar\xi^- \end{array}\right)
 = \left(\begin{array}{r}
 \cos\frac\theta2 \\ ie^{i\varphi}\sin\frac\theta2 \end{array}\right).
\label{orbks}
\end{equation}
Their components transform under the $\mathbb Z_K$ rotation as follows,
\begin{alignat}{2}
 \xi^+(\theta,\varphi+\frac{2\pi}K)
 &=\xi^+(\theta,\varphi)e^{-\frac{2\pi i}K},&\quad
 \bar\xi^+(\theta,\varphi+\frac{2\pi}K)
 &=\bar\xi^+(\theta,\varphi),\nonumber\\
 \xi^-(\theta,\varphi+\frac{2\pi}K)
 &=\xi^-(\theta,\varphi),&\quad
 \bar\xi^-(\theta,\varphi+\frac{2\pi}K)
 &=\bar\xi^-(\theta,\varphi)e^{\frac{2\pi i}K}.
\end{alignat}
The $\mathbb Z_K$ orbifold action is required to preserve their form. If we define the spin $J_3$ in such a way that the spinor components with $\gamma^3=\pm1$ have $J_3=\pm1/2$, the orbifold projection requires the field $X$ of spin $J_3$, R-charge $\RV$ and electric charge $e$ to satisfy
\begin{equation}
 X(\theta,\varphi+\frac{2\pi}K)
 =\exp\frac{2\pi i}K\left\{(K-1)(J_3+\frac12\RV)-p\hskip0.2mm e\right\}
 X(\theta,\varphi).
\label{zktw}
\end{equation}
Having this $\mathbb Z_K$ identification twisted by the symmetries $J_3,\RV$ and $\text{e}$ is equivalent to turning on the corresponding gauge fields as follows,
\begin{equation}
 \omega^{12} = (K-1)d\varphi,\quad
 V = -\frac12(K-1)d\varphi,\quad
 A \simeq p d\varphi
\label{bggf}
\end{equation}
and requiring the simple $\mathbb Z_K$ orbifold invariance. The latter definition of $\mathbb Z_K$ orbifold can be easily analytically continued to general real positive $K$.

Let us now use the standard orbifold technique to compute the correlator of the vortex defects with $\eta_\north=\eta_\south=p/K$ inserted at the two $\mathbb Z_K$ fixed points. For the evaluation of $Z_\text{1-loop}$, we only need to compute the determinants of $\susy^2$ on the sphere {\it without} defects but restricted to the $\mathbb Z_K$-invariant subspace of $\text{ker} J^+$ and $\text{ker} J^-$. The determinant $\text{det}(\susy^2)$ restricted to $\text{ker}J^+\subset {\cal H}$ is given by
\begin{eqnarray}
 \text{det}(\susy^2)|_{\text{ker}J^+}
 &=& \prod_{m_\north\ge0,~ m_\north=q(K-1)-p\text{ mod }K}
 \frac i\ell(m_\north+q-ia+s),\quad(s\ge0)
 \nonumber \\
 \text{det}(\susy^2)|_{\text{ker}J^+}
 &=& \prod_{m_\south\ge0,~ m_\south=q(K-1)-p\text{ mod }K}
 \frac i\ell(m_\south+q-ia-s),\quad(s\le0)
\end{eqnarray}
whereas $\text{det}(\susy^2)$ restricted to $\text{ker}J^-\subset{\cal H}'$ is
\begin{eqnarray}
 \text{det}(\susy^2)|_{\text{ker}J^-}
 &=& \prod_{m_\south\le0,~ m_\south=(q-1)(K-1)-p\text{ mod }K}
 \frac i\ell(m_\south+q-1-ia-s),\quad(s\ge0)
 \nonumber \\
 \text{det}(\susy^2)|_{\text{ker}J^-}
 &=& \prod_{m_\north\le0,~ m_\north=(q-1)(K-1)-p\text{ mod }K}
 \frac i\ell(m_\north+q-1-ia+s).\quad(s\le0)
\end{eqnarray}
We rewrite these formulae using the rescaled variables $\tilde a=a/K,\,\tilde s=s/K$ and $\eta=p/K$. Note that $s$ here is the magnetic flux on the sphere before orbifold, so it is $\tilde s$ which obeys the usual flux quantization law. One then obtains
\begin{eqnarray}
 \text{det}(\susy^2)|_{\text{ker}(J^+)}
 &=& \prod_{n\ge\lceil\eta-q(1-\frac1K)\rceil}
 \frac {iK}\ell(n+q-\eta-i\tilde a+|\tilde s|),
 \nonumber \\
 \text{det}(\susy^2)|_{\text{ker}(J^-)}
 &=& \prod_{n\ge-\lfloor\eta-(q-1)(1-\frac1K)\rfloor}
 \frac {-iK}\ell(n+1-q+\eta+i\tilde a+|\tilde s|).
\end{eqnarray}
Using the simple fact that $\lfloor x+1-\frac1K\rfloor=\lceil x\rceil$ for $x\in\frac1K\mathbb Z$, one can write the final formula for $Z_\text{1-loop}$ in two different ways.
\begin{eqnarray} \label{eq:formula-orbifold}
 Z_\text{1-loop} &=&
 \frac{\Gamma(\lceil\eta-q(1-\frac1K)\rceil+q-\eta-i\tilde a+\tilde s)}
      {\Gamma(-\lceil\eta-q(1-\frac1K)\rceil+1-q+\eta+i\tilde a+\tilde s)}
 \nonumber \\ &=&
 \frac{\Gamma(\lfloor\eta-(q-1)(1-\frac1K)\rfloor+q-\eta-i\tilde a+\tilde s)}
      {\Gamma(-\lfloor\eta-(q-1)(1-\frac1K)\rfloor+1-q+\eta+i\tilde a+\tilde s)}.
\label{lvo}
\end{eqnarray}

\subsection{Normal and flipped boundary conditions for conical singularities}
\label{sec:norm-flipp-bound}

The two expressions in (\ref{eq:formula-orbifold}) can be naturally generalized in two different ways for non-integer $K>0$. Consider a squashed sphere with conical singularities parametrized by $K>0$ at the north and south poles, and also introduce vorticities $\eta_\north,\eta_\south$ on these poles. The matter one-loop determinant on this background is
\begin{equation}
 Z_\text{1-loop} = \frac{\Gamma(\kappa_\north(\eta_\north)+q-\eta_\north-ia+s)}
 {\Gamma(-\kappa_\south(\eta_\south)+1-q+\eta_\south+ia+s)},
\label{lvog}
\end{equation}
where $\kappa_\north\,(\kappa_\south)$ depends on $\eta_\north\,(\eta_\south)$ as well as $q,K$ as follows.
\begin{alignat}{2}
 \kappa_{\north}(\eta_{\north}) &\equiv\lceil\eta_{\north}-q(1-\tfrac1K)\rceil
&\quad&\text{(normal b.c. at NP)}
\nonumber \\
 \kappa_{\north}(\eta_{\north}) &\equiv\lfloor\eta_{\north}-(q-1)(1-\tfrac1K)\rfloor
&\quad&\text{(flipped b.c. at NP)}
\end{alignat}
The above formula can be derived for general $K>0$ by studying the normalizable zeromodes of $J^\pm$ as in Section~\ref{sec:vcs} on the squashed sphere with conical singularities%
\footnote{%
Compared with (\ref{orbmet}), we redefined quantities as $\varphi_\text{new}=K \varphi_\text{old}$, $\sin\theta_\text{new}=K^{-1}\sin\theta_\text{old}$, $f_\text{new}d\theta_\text{new}= f_\text{old} d\theta_\text{old}$.
}
\begin{equation} \label{singular-metric}
 ds^2 = f^2(\theta)d\theta^2+\ell^2\sin^2\theta d\varphi^2,\qquad
 f(0)=f(\pi)= K\ell, \qquad \varphi \sim \varphi+2\pi.
\end{equation}
The zeromodes $X\in\text{ker}J^+$ behave near the north and south poles as
\begin{alignat}{2}
\text{(north)}&&\quad
 X &\sim e^{im_\north\varphi}\cdot
 (\sin\theta)^{K(m_\north-\eta_\north)+q(K-1)},
 \nonumber \\
\text{(south)}&&\quad
 X &\sim e^{im_\south\varphi}\cdot
 (\sin\theta)^{K(m_\south-\eta_\south)+q(K-1)}.
\end{alignat}
Therefore, depending on the choice of boundary conditions $m_{\north,\south}$ have to satisfy
\begin{alignat}{2}
\text{(normal)}&&\quad
 m_{\north,\south} &\ge\lceil\eta_{\north,\south}-q(1-\tfrac1K)\rceil,
\nonumber \\
\text{(flipped)}&&\quad
 m_{\north,\south} &\ge\lfloor\eta_{\north,\south}-(q-1)(1-\tfrac1K)\rfloor.
\end{alignat}
The other zeromodes $\Xi\in\ker J^-$ can be studied in the same way, which leads to the formula~(\ref{lvog}). Note the non-trivial $K$-dependence of the defect correlator, which is somewhat contrary to the known squashing independence of the partition function~\cite{Gomis:2012wy}. The $K$-dependence remains even after sending $\eta\to 0$ in the above formula.

\subsection{Resolution of the conical singularities and the localized modes}
\label{sec:resol-conic-sing}

Let us introduce a small parameter $\epsilon>0$ and consider resolving the conical singularities in (\ref{singular-metric}) as
\begin{equation}
  ds^2 =    f_\epsilon^2(\theta)d\theta^2+\ell^2\sin^2\theta d\varphi^2 ,
\end{equation}
where the function $f_\epsilon(\theta)$ ($0\leq \theta\leq \pi$) behaves as
\begin{equation}
  f_\epsilon(\theta) \sim \left \{
    \begin{array}{cll}
\ell      & \text{for} & \sin\theta \ll \epsilon , \\
K\ell & \text{for} & \epsilon \ll \sin\theta \ll 1 .
    \end{array}
\right.
\end{equation}
At the same time, we smooth the gauge field singularity as in (\ref{spsm}) and (\ref{S-g-epsilon}), with $f$ replaced by~$f_\epsilon$.
The resolution of the metric singularity was studied in~\cite{Mori:2015bro} in the case $K^{-1}$ is an integer.

The one-loop determinant in the resolved background is given by  $(\ref{one-loop-smear})$, {\it i.e.}, 
\begin{equation} \label{one-loop-resolved}
 Z_\text{1-loop} = \frac{\Gamma(q-\eta_\north-ia+s)}
 {\Gamma(1-q+\eta_\south+ia+s)},
\end{equation}
because the background we are considering is a special case of the set-up in Section~\ref{sec:smear-sphere}.
This observation is consistent with the results of~\cite{Mori:2015bro}.
In particular, the one-loop determinant as well as and the whole partition function are independent of the parameter $K>0$.

We expect that the difference between (\ref{one-loop-resolved}) and (\ref{lvog}) is accounted for, as in Section~\ref{sec:smear-sphere}, by the localized modes and the missing frozen bulk modes.
Here we focus on the localized modes.

A mode $\Psi= \hat\Psi e^{i m_\north \varphi}$ ($m_\north\in\mathbb Z$)) annihilated by $J^+$ behaves near the north pole as
\begin{equation}
\hat\Psi\sim \left \{
    \begin{array}{cll}
\theta^{m_\north}    & \text{for} & \theta \ll \epsilon , \\
\theta^{K(m_\north-\eta)+(K-1)q}& \text{for} & \epsilon \ll \theta \ll 1 .
    \end{array}
\right.
\end{equation}
Regularity requires that $m_\north \geq 0$, and the mode is localized near the north pole when $K(m_\north-\eta)+(K-1)q<-1$.
When $\eta_\north -(q-1)(1-K^{-1}) >0$, the number of values of such $m_\north$ is $\lfloor \eta_\north -(q-1)(1-K^{-1})\rfloor$, which coincides with $\kappa_\north(\eta_\north)$ for the flipped boundary condition.
From the experience in Section~\ref{sec:eigenv-probl-flat}, we expect that the non-localized bulk modes obey the flipped boundary condition at the north pole.

Similarly, a mode $\Psi= \hat\Psi e^{i m_\south \varphi}$ annihilated by $J^-$ behaves near the south pole as 
\begin{equation}
\hat\Psi\sim \left \{
    \begin{array}{cll}
(\pi-\theta)^{-m_\south}    & \text{for} &\pi- \theta \ll \epsilon , \\
(\pi-\theta)^{K(\eta_\south - m_\south )+ (1-K)(q-1)}& \text{for} & \epsilon \ll \pi-\theta \ll 1 .
    \end{array}
\right.
\end{equation}
A regular mode localized near the south pole has $m_\south \leq 0$ and $K(\eta_\south-m_\south)+(1-K)(q-1)<-1$.
When $\eta_\south -q(1-K^{-1}) <0$, the number of values of such $m_\south$ is $-\lceil \eta_\south -q(1-K^{-1})\rceil$, which coincides with $\kappa_\south(\eta_\south)$ for the normal boundary condition.
We expect that the non-localized bulk modes obey the normal boundary condition at the south pole.

\section{Vortex defects in non-Abelian theories}
\label{sec:vnat}

Here we briefly discuss vortex defects in non-Abelian theories. The vorticity parameter $\eta$ is taken to be in Cartan subalgebra, and is further subject to identification by Weyl reflections. We would like to study the SUSY path integrals on the squashed sphere in the presence of defects with vorticity $\eta_\north,\eta_\south$ at the two poles.

General supersymmetric saddle point configurations are given by
\begin{alignat}{2}
 \sigma&=\frac a\ell, &\qquad
 D &= -\frac a{f\ell}
 +2\pi i\eta_\north\delta^2_{(\text{NP})}
 +2\pi i\eta_\south\delta^2_{(\text{SP})},
 \nonumber \\
 \rho &= -\frac s\ell,&\qquad
 A &= \left\{\begin{array}{ll}
   s(\cos\theta-1)d\varphi+\eta_\north d\varphi & (\text{north}) \\
   s(\cos\theta+1)d\varphi+\eta_\south d\varphi & (\text{south})
	    \end{array} \right.,
\end{alignat}
with Cartan subalgebra-valued parameters $a,s,\eta_\north,\eta_\south$. Weyl-reflecting any of these parameters gives rise to a new saddle point, but the simultaneous reflection of all of them is a gauge transformation. The localized path integral therefore involves summing over $s$, integrating over $a$ as well as summing over Weyl images of $\eta_\north,\eta_\south$. Note also that the flux quantization requires that $2s+\eta_\south-\eta_\north$ have integer inner product with any weight vectors of the gauge group.

The fluctuation of fields around saddle point configurations gives rise to one-loop determinants. For a chiral multiplet with $\RV=2q$ in representation $\Lambda$, the determinant is given by a product over the weight vectors $w$,
\begin{equation}
 Z_{\rch,\Lambda} = \prod_{w\in\Lambda}
 \frac{\Gamma(q+w\cdot(s-\eta_\north-ia)+\kappa_\north(w\cdot\eta_\north))}{\Gamma(1-q+w\cdot(s+\eta_\south+ia)-\kappa_\south(w\cdot\eta_\south))}.
\end{equation}
Here $\kappa_\north,\kappa_\south$ are ceiling or floor functions depending on the choice of boundary condition at the poles.

For non-Abelian theories, vector multiplet also gives rise to a non-trivial determinant. It can be most easily evaluated by introducing cohomological variables similar to~(\ref{dfch}). We reorganize the fields $(A_m,\rho,\sigma,\lambda,\bar\lambda,D)$ into five Grassmann-even and four Grassmann-odd variables,
\begin{alignat}{4}
 & X^+ &&\equiv \bar\xi\gamma^m\bar\xi A_m+\bar\xi\gamma^3\bar\xi\rho,
 \quad &
 & \susy X^+ && =-\tfrac12\bar\xi\bar\lambda,
 \nonumber \\
 & X^0 &&\equiv \bar\xi\gamma^m\xi A_m+\bar\xi\gamma^3\xi\rho,
 \quad &
 & \susy X^0 && =+\tfrac12\bar\xi\lambda-\tfrac12\xi\bar\lambda,
 \nonumber \\
 & X^- &&\equiv \xi\gamma^m\xi A_m+\xi\gamma^3\xi\rho,
 \quad &
 & \susy X^- && =+\tfrac12\xi\lambda,
 \nonumber \\
 & \Xi &&\equiv \tfrac12\bar\xi\lambda+\tfrac12\xi\bar\lambda,
 \quad &
 & \susy\Xi &&= D+\tfrac1f\sigma+i\bar\xi\gamma^3\xi(F_{12}-\tfrac\rho f),
 \nonumber \\
 & \hat\Sigma &&\equiv \bar\xi\xi\sigma+i\bar\xi\gamma^3\xi\rho+i\bar\xi\gamma^m\xi A_m, \quad &
 &~\;(\susy\hat\Sigma &&=0)
\label{cvv}
\end{alignat}
which are all Lorentz scalars. In addition, for the gauge fixing we need to introduce the ghosts $c,\bar c,B$ and the BRST symmetry. The BRST charge $\brs$ acts on all the physical fields in the standard way, namely as $\text{Gauge}(c)$. For the ghost fields, we define the action of $\susy$ and $\brs$ on the saddle point labeled by $s,a$ as
\begin{alignat}{4}
 &\brs c &&= cc,
 \qquad &&
 \susy c &&= \langle\hat\Sigma\rangle-\hat\Sigma,
 \nonumber \\
 &\brs \bar c &&= B,
 \qquad &&
 \susy \bar c &&=0,
 \nonumber \\
 &\brs B &&=0,
 \qquad &&
 \susy B &&= -\bar\xi\gamma^m\xi\partial_m\bar c+[\langle\hat\Sigma\rangle,\bar c],
 \qquad(\langle\hat\Sigma\rangle\equiv a/\ell)
\end{alignat}
so that the total supercharge $\hat\susy\equiv \susy+\brs$ squares to
\begin{equation}
 \hat\susy^2 = \frac1\ell\left\{\partial_\varphi+\frac i2\RV+\text{Gauge}(\hat a)\right\},\quad
 \hat a\equiv\left\{\begin{array}{ll}
 a+is-i\eta_\north &
 \text{(north patch)} \\
 a-is-i\eta_\south &
 \text{(south patch)} \end{array}\right.
\end{equation}
on all the fields. This is clearly similar to the property~(\ref{q2wd}) of $\susy$ for charged matter fields coupled to background vector multiplet, so the rest of the computation of determinant is the same as before.

After moving to the cohomological variables, one can regard the vector multiplet as made of three Grassmann-even scalars $X^{+,-,0}$, three Grassmann-odd scalars $\Xi,\bar c,c$ and their $\hat\susy$-superpartners. The one-loop determinant for vector multiplet is thus equal to that for an adjoint chiral multiplet with $q=1$. For $\eta_\north=\eta_\south=0$, the determinant is given by a product over the roots $\alpha\in\Delta$ or positive roots $\alpha\in\Delta_+$,
\begin{equation}
 Z_\text{vec} = \prod_{\alpha\in\Delta}
 \frac{\Gamma(1+\alpha\cdot(s-ia))}
 {\Gamma(\alpha\cdot(s+ia))}
 = (-1)^{4\rho\cdot{\boldsymbol s}}\prod_{\alpha\in\Delta_+}
 |\alpha\cdot(s+ia)|^2.
\label{zv0}
\end{equation}
Here $\rho$ is the Weyl vector.

To evaluate the determinant $Z_\text{vec}$ on defect backgrounds, one first needs to determine the boundary condition on the cohomological variables (\ref{cvv}) near the defects. The fields $X^{+,-,0}$ behave near the north pole $\theta=0$ as (with $z\equiv \theta e^{i\varphi}$)
\begin{alignat}{2}
 X^+ \;&\simeq\; +\frac1\ell A_{\bar z} &&\;\simeq\; \frac{i\eta_\north}{2\ell\bar z}+\text{(fluctuation)},\nonumber \\
 X^- \;&\simeq\; -\frac1\ell A_z &&\;\simeq\; \frac{i\eta_\north}{2\ell z}+\text{(fluctuation)},\nonumber \\
 X^0 \;&\simeq\; -\frac1\ell A_\varphi-\rho &&\;\simeq\;
 \frac{s-\eta_\north}\ell+\text{(fluctuation)}.
\end{alignat}
Recalling the boundary condition for the fields in vector multiplet discussed in Section~\ref{sec:BPS-defects}, one finds that the fluctuation of $X^\pm$ is allowed to diverge mildly as $\theta^\gamma\,(\gamma>-1)$, whereas that of $X^0$ has to be finite at the defect. Similarly, the fields $\Xi, c, \bar c$ can be shown to obey the same boundary condition as for the parameter of gauge transformation, so they have to be finite at the defect. These are identified with the behavior of an adjoint chiral multiplet (with $q=1$) satisfying the flipped boundary condition.
The determinant $Z_\text{vec}$ is thus given by
\begin{eqnarray}
 Z_\text{vec} &=& \prod_{\alpha\in\Delta}
 \frac{\Gamma(1+\alpha\!\cdot\!(s-\eta_\north-ia)+\lfloor\alpha\!\cdot\!\eta_\north\rfloor)}
{\Gamma(\alpha\!\cdot\!(s+\eta_\south+ia)-\lfloor\alpha\!\cdot\!\eta_\south\rfloor)}
 \nonumber \\ &=&
 (-1)^{2\rho\cdot(2s+\eta_\south-\eta_\north)}\cdot F(\eta_\north,s-ia)
 F(\eta_\south,s+ia)\,,
\end{eqnarray}
where
\begin{equation}
 F(\eta,s)~\equiv~
 \prod_{\alpha\in\Delta_+}(-1)^{\lceil\alpha\cdot\eta\rceil}
 \cdot
 \prod_{\alpha\in\Delta_+,\,\alpha\cdot\eta\in\mathbb Z}
 \alpha\!\cdot\!s.
\end{equation}
The roots satisfying $\alpha\cdot\eta_\north\in\mathbb Z$ forms the root system for the Lie algebra ${\mathbb L}_\north$ which is preserved by the defect at the north pole. In particular, for $\eta_\north=\eta_\south$ the determinant $Z_\text{vec}$ reduces to the one without defects~(\ref{zv0}) for the unbroken gauge symmetry ${\mathbb L}_\north={\mathbb L}_\south$. Note also that $F(\eta,s)$ is odd under the Weyl group acting simultaneously on both arguments:
\begin{equation}
 F(\pi(\eta),\pi(s))
=(-1)^{|\pi|}F(\eta,s).\qquad
\big(\pi\in W\big)
\end{equation}

As an illustrative example, consider $U(N)$ gauge group, in which case $a,s,\eta_\north,\eta_\south$ are $N\times N$ diagonal matrices. Let us assume for simplicity that the diagonal elements of $\eta_{\north,\south}$ all satisfy $0\le\eta^a_{\north,\south}<1$. The one-loop determinant for vector multiplet then takes the form
\begin{equation}
 Z_\text{vec} ~=~
 (-1)^{(N-1)\text{Tr}(2s+\eta_\south-\eta_\north)}\cdot F(\eta_\north,s-ia)
 F(\eta_\south,s+ia).
\end{equation}
The first sign factor can be absorbed by redefining the $\theta$-angle, and the function $F$ is given by
\begin{eqnarray}
 F(\eta,s)
 &\equiv& (-1)^{\sigma(\eta)}\cdot\prod_{a<b,\,\eta_a=\eta_b}(s_a-s_b).
 \nonumber \\
 \sigma(\eta) &\equiv&
 \text{number of pairs $(a,b)$ such that $\eta_a>\eta_b$}.
\end{eqnarray}
As a special case, if the matrix $\eta$ is proportional to identity, then $F$ becomes the usual Vandermonde determinant. Another special case is when $\eta$ has no degenerate eigenvalues, for which $F$ equals the parity of the permutation which brings $\{\eta^a\}$ to the ascending order.  Vortex defect correlators  in some $U(N)$ SQCDs have been studied using this formula in~\cite{Hosomichi:2015pia}.
It would be nice to perform non-trivial checks of these results.

\section{Discussion}
\label{sec:discussion}

In this paper we discussed the definition and correlators of vortex defects in two dimensions. Our analysis focused mostly on Abelian theories, where we found that the vortex defects are equivalent to local functionals of twisted chiral fields. It is not fully clear whether the vortex defects in non-Abelian theories lead to new observables or have interesting applications.

In contrast, in 4d or higher, the codimension two vortex defects are essentially new; for example the introduction of surface operators in 4d Omega background leads to an interesting generalization of instanton partition functions. The defects in 4d were studied in~\cite{Gomis:2016ljm} by realizing them as coupled (0d-)2d-4d systems. See also~\cite{Gorsky:2017hro}. It would be nice to study them using the approach taken in this paper.

As we mentioned at the end of Section~\ref{sec:eigenv-probl-flat}, the vortex defects defined by the normal and the flipped boundary conditions can be related to smeared defect configurations for some values of vorticity $\eta$. There is, however, a restriction on the combination of the type of boundary condition and the sign of $\eta$ for such a relation to hold. For a single chiral multiplet of charge $+1$, the flipped boundary condition with $\eta>0$ and the normal boundary condition with $\eta<0$ are related to smeared vortex defects in this way.  In this paper we also obtained a couple of results which seem to suggest the importance of whether a given vortex defect can be smeared or not. An example is the pair of chiral multiplets of opposite $U(1)$ charge discussed in page \pageref{pg:mads}: the lifting by superpotential is complete only when the choice of matter boundary condition is such that the defect can be smeared for some $\eta$. As another example, we saw in Section~\ref{sec:cal-n=2-minimal} that only the vortex defects that can be smeared leads to consistent twist fields in the orbifolded minimal model. The consistent smearing in this case involves an extra condition that the coupling constant can be promoted to a smooth configuration of a non-dynamical chiral multiplet.

It would be interesting to see if our results for vortex defects without dynamical vector multiplets can be used to compute the {\it orbifold} Gromov-Witten invariants, which have been
 studied only relatively recently~\cite{Chen2001}.
Such a connection can be motivated by the fact that the mathematical formulas for the invariants proposed in~\cite{Cheong2015} involve the ceiling and the floor functions (also known as the round-up and the round-down). It would be nice to develop field theoretical techniques for computing the Gromov-Witten invariants for both toric and non-toric orbifolds, which may lead to physical understanding of the so-called crepant resolution conjecture~\cite{MR2234886}.

\section*{Acknowledgements}

We would like to thank Costas Bachas, Jaume Gomis, Kentaro Hori, Tatsuma Nishioka, Sara Pasquetti, David Tong,  and Itamar Yaakov for useful discussions.
The research of S.L. is supported in part by the National Research Foundation of Korea (NRF) Grant NRF-2017R1C1B1011440.
The research of T.O. is supported in part by  JSPS Grants-in-Aid for Scientific Research No.~25287049 and No.~16K05312.
T.O. thanks the Galileo Galilei Institute for Theoretical Physics (GGI) for the hospitality and INFN for partial support during the completion of this work, within the program “New Developments in AdS3/CFT2 Holography”.

\appendix

\section{The $\epsilon\rightarrow 0$ limit of the bulk modes}
\label{sec:dominant-terms-bulk}

Here we study the behavior of the general non-zeromode wave function satisfying~(\ref{eigenvalue-equation-in-r}).
A related analysis was done in~\cite{Kapustin:2012iw}, where the vorticity was restricted to $-1/2<\eta<1/2$; here we consider the more general case  $\eta\in\mathbb{R}-\mathbb{Z}$. Let us begin by rewriting it using $u\equiv r/\epsilon$ and $\mu\equiv \epsilon\lambda^{1/2}$.
\begin{equation} \label{eigenvalue-equation-in-u}
  \left[(u\partial_u)^2+\eta u\partial_u g-(m-\eta\cdot g)^2+\mu^2 u^2 \right]\hat\Psi(u)=0 \,.
\end{equation}
For $u\gg1$ the solution can be expressed as in~(\ref{Psi-hat-J}) using two Bessel functions. The main aim here is to show that the coefficients $\alpha_\pm$ satisfy the relations~(\ref{alpha-inequalities}), which we reproduce here
\begin{equation} \nonumber
 \text{ (\ref{alpha-inequalities})}: \qquad
 \frac{\alpha_+}{\alpha_-} ~\sim~ \left\{
 \begin{array}{ll}
  \epsilon^{-2|m-\eta|+\text{non-positive}} \quad & (m<0) \\
  \epsilon^{-2|m-\eta|+2} \quad & (0\le m<\eta) \\
  \epsilon^{-2|m-\eta|-2} \quad & (\text{max}\{0,\eta\}<m)
 \end{array}
 \right.\,.
\end{equation}

\paragraph{Crude analysis.}

In the limit of small $\epsilon$ with $\lambda$ kept fixed, $\mu$ is very small.
This motivates us to consider dropping the last term in the LHS of~(\ref{eigenvalue-equation-in-u}). Let us denote the regular solution of the resulting equation by $h(u)$ to distinguish it from $\hat{\Psi}$.
Up to overall normalization it behaves as
\begin{equation} \label{h-u-behavior}
h(u) = \left\{
  \begin{array}{cll}
    u^{|m|} &  \text{ for } & 0\leq u\ll 1 \,, \\
    b_+ u^{+|m-\eta|} + b_- u^{-|m-\eta|}   & \text{ for } &1\ll u\,.
  \end{array}
\right.
\end{equation}
Since~(\ref{eigenvalue-equation-in-u}) becomes independent of $\mu$ once the last term is dropped, the coefficients $b_\pm$ are also independent of it. A comparison of~(\ref{Psi-hat-J}) and~(\ref{h-u-behavior}) gives
\begin{equation} \label{eq:b-plus-nonzero}
 \frac{\alpha_+}{\alpha_-} \sim \mu^{-2|m-\eta|}\cdot \frac{b_+}{b_-}\,.
\end{equation}
For generic $\eta$ and $m$, we will soon see that $b_+$ is non-zero, and hence, including the case $b_-=0$, we have $ |\alpha_+| \gg |\alpha_-|$.

Let us find when $b_+$ vanishes. For $b_+ = 0$ one can show that $h$ satisfies a first order equation $(\partial_x-m+\eta g) h=0$, where $x\equiv \log u$, as follows.
\begin{equation}
 0 = -\int_{-\infty}^\infty dx h^\ast(\partial_x+m-\eta g)(\partial_x-m+\eta g) h = \int_{-\infty}^\infty dx|(\partial_x-m+\eta g)h|^2.
\end{equation}
In the first equality we used the differential equation for $h$ written in $x$, and then integrated by parts using that $h$ or $\partial_x h$ decays in either direction $x\to\pm\infty$. The solution coincides with~(\ref{h-u-behavior}) with $b_+=0$ if and only if $0\leq m <\eta$.

Function $h(u)$ provides an approximation of $\hat\Psi(u)$ for $u\ll \mu^{-1}$ because the term we dropped is small. This combined with~(\ref{eq:b-plus-nonzero}) does not, however,  imply that $|\alpha_-| \gg |\alpha_+|$ for $0\le m<\eta$ because we need to take into account the effect of the term we dropped. If we write
\begin{equation}
 \hat\Psi(u)\simeq \tilde b_+ u^{+|m-\eta|}+\tilde b_- u^{-|m-\eta|}\quad\text{for}\quad 1\ll u\ll \mu^{-1},
\label{tbpm}
\end{equation}
then the coefficients $\tilde b_\pm$, as we will see, satisfy
\begin{equation}\label{tbrl}
 \frac{\tilde b_+}{\tilde b_-} ~\sim~ \left\{
 \begin{array}{ll}
  \mu^\text{non-positive}   \quad & (m<0)\\
  \mu^2 \quad & (0\le m<\eta)\\
  \mu^{-2} \quad & (\text{max}\{0,\eta\}<m)
 \end{array}
 \right.\,.
\end{equation}
The relation~(\ref{alpha-inequalities}) follows immediately once we accept this. The second line of this relation implies the following. For $0\le m<\eta$, the second term of~(\ref{tbpm}) is indeed dominant at $u\sim 1$, but as one increases $u$ generically the first term starts dominating much earlier than $u$ becomes of order $\mu^{-1}$. The only exception is when $m=\lfloor \eta\rfloor$ and the decay of the second term is very slow, or in other words it is diverging mildly towards the defect.

\paragraph{Smearing by a step function.}

For the special choice $g(u)\equiv\Theta(u-1)$ one can find out the ratio of $\alpha_\pm$ explicitly, by putting
\begin{equation}
 \hat\Psi(u)=\left\{
\begin{array}{ll}
J_{|m|}(\mu u)\quad &(0\le u\le 1) \\
\alpha_+ J_{|m-\eta|}(\mu u)+\alpha_-J_{-|m-\eta|}(\mu u)\quad&(1\le u)
\end{array}
\right.
\end{equation}
and by solving the junction condition at $u=1$,
\begin{equation}
 \hat\Psi(1_+)=\hat\Psi(1_-),\quad
 \partial_u\hat\Psi(1_+)=\partial_u\hat\Psi(1_-)-\eta\hat\Psi(1_-).
\end{equation}
Here $1_+,1_-$ means approaching $u=1$ from above or below. One finds
\begin{equation}
\frac{\tilde b_+}{\tilde b_-} ~\sim~
 \frac{\alpha_+J_{+|m-\eta|}(\mu)}{\alpha_-J_{-|m-\eta|}(\mu)}
~\sim~ \left\{
 \begin{array}{ll}
   1 \quad & (m<0)\\
  \mu^2 \quad & (0\le m<\eta)\\
  \mu^{-2} \quad & (\text{max}\{0,\eta\}<m)
 \end{array}
 \right.\,.
\end{equation}
This is a special case of~(\ref{tbrl}).

\paragraph{Perturbative analysis for general smearing.}

Finally, let us derive~(\ref{alpha-inequalities}) for the general choice of $g(u)$. We begin by constructing a perturbative solution to~(\ref{eigenvalue-equation-in-u}). Here we use $x=\log u$ and introduce
\begin{equation}
 \hat\Phi\equiv (\partial_x-m+\eta g)\hat\Psi.
\end{equation}
The differential equation can then be cast into coupled first order equations, for which one can easily construct the solution in the form of a path-ordered exponential. The solution for given initial values $\hat\Psi(x_0),\hat\Phi(x_0)$ is
\begin{equation} \label{series-path-order}
\begin{pmatrix}\hat\Psi(x)\\\hat\Phi(x)\end{pmatrix}=\text{P}\exp\int_{x_0}^x dy
\begin{pmatrix}m-\eta g(y) & 1 \\  -\mu^2 e^{2y} & -m+\eta g(y)\end{pmatrix}
\begin{pmatrix}\hat\Psi(x_0)\\\hat\Phi(x_0)\end{pmatrix}.
\end{equation}
One can rewrite this further as a series organized by the number of appearances of off-diagonal matrix elements. The resulting formal series expression for $\hat\Psi(x)$ reads
\begin{eqnarray}
&& ~\hskip-5mm \hat\Psi(x) = U(x,x_0)\hat\Psi(x_0)
 +\int dx_1 U(x,x_1)U^{-1}(x_1,x_0)\hat\Phi(x_0)
 \nonumber \\ &&
  +\int dx_1dx_2 U(x,x_1)U^{-1}(x_1,x_2)(-\mu^2e^{2x_2})U(x_2,x_0)\hat\Psi(x_0)
 \nonumber \\ &&
 +\int dx_1dx_2dx_3 U(x,x_1)U^{-1}(x_1,x_2)(-\mu^2e^{2x_2})U(x_2,x_3)U^{-1}(x_3,x_0)\hat\Phi(x_0)+\cdots,
\nonumber \\ && \hskip25mm\text{with }\quad
U(x_1,x_2)\equiv \exp\int_{x_2}^{x_1} dy(m-\eta g(y))\,.
\end{eqnarray}
The integration variables here are understood to satisfy $x\ge x_1\ge x_2\ge\cdots\ge x_0$.

Let us focus on the first two terms in this series. For an $x_0$ negatively large, the solution is given in terms of Bessel functions and can be expanded in $\mu$ as
\begin{eqnarray}
 \hat\Psi(x_0)&=& e^{|m|x_0}\left(1-\frac{\mu^2e^{2x_0}}{4(|m|+1)}+{\cal O}(\mu^4)\right),\nonumber \\[1mm]
 \hat\Phi(x_0) &= & \left\{\begin{array}{ll}
  \displaystyle -2m e^{|m|x_0} +\mathcal{O}(\mu^2) & (m<0) \\[1mm]
  \displaystyle  -\frac{\mu^2e^{(|m|+2)x_0}}{2(m+1)} +\mathcal{O}(\mu^4)\quad& (m\ge0)
 \end{array}
 \right.\,.
\end{eqnarray}
Substituting this into~(\ref{series-path-order}) as the initial condition, we obtain an expression for $\hat\Psi(x)$ as an expansion in $\mu$.

We wish to compare it with the $\mu$-expansion of the second line of~(\ref{Psi-hat-J}), {\it i.e.},
\begin{equation} \label{Psi-hat-J-x}
   \hat\Psi(x) =  \alpha_+  J_{+|m-\eta|}(\mu\, e^x) + \alpha_-  J_{-|m-\eta|}(\mu\, e^x) \,.
  \end{equation}
We emphasize that the coefficients $\alpha_\pm$ depend on $\mu$.
As functions of $x$ we have
\begin{eqnarray}
 U(x,x_0) & = & (\text{non-zero})\cdot e^{(m-\eta)x},
\nonumber \\
 \int_{x_0}^x dx_1 U(x,x_1)U^{-1}(x_1,x_0) & = &  (\text{const})\cdot e^{(m-\eta)x} + (\text{non-zero}) \cdot e^{-(m-\eta)x} \,.
\end{eqnarray}
Here (const) is a not necessarily non-zero constant, while (non-zero) is a non-zero constant.
We thus find
\begin{equation}
 \hat\Psi(x)
~=~\left\{
 \begin{array}{ll}
  (\text{const})\cdot e^{(m-\eta)x}+(\text{non-zero})\cdot e^{-(m-\eta)x} + \mathcal{O}(\mu^2)\quad
 & (m<0)\\[2mm]
   \begin{array}{l}
\hspace{-0.5mm}
 (\text{non-zero})\cdot e^{(m-\eta)x} + (\text{non-zero})\cdot \mu^2\ e^{-(m-\eta)x} \\[2mm]
\hspace{30mm}
+ (\text{const})\cdot  \mu^2 \ e^{(m-\eta)x}+ \mathcal{O}(\mu^4)
   \end{array}
\quad
 & (m\ge0)
 \end{array}
\right.\,,
\end{equation}
where all the constants are independent of $\mu$.
For $m\geq 0$, by comparing this with~(\ref{Psi-hat-J-x}) we obtain~(\ref{tbrl}).
For $m<0$, if the coefficient of $e^{(m-\eta)x}$ is non-zero we find that $\tilde{b}_+/\tilde{b}_- \sim 1$.
If the coefficient vanishes  $\tilde{b}_+/\tilde{b}_-$ must be $\mathcal{O}(\mu^\text{positive})$ to be consistent with the inequality $ |\alpha_+| \gg |\alpha_-|$ found by the crude analysis.
Thus we obtain~(\ref{tbrl}), which implies~(\ref{alpha-inequalities}).

\section{A review of mirror symmetry}\label{sec:quintic}

Here we summarize some basic facts about sigma models on Calabi-Yau manifolds, their moduli spaces and the simplest mirror pair of quintic hypersurfaces. For more details see~\cite{Cox:2000vi} and~\cite{Jockers:2012dk}.

\paragraph{Local special K\"ahler manifolds and prepotential.}

The moduli spaces of complex structures ${\cal M}_\text{C}$ and complexified K\"ahler structures ${\cal M}_\text{K}$ of a Calabi-Yau three-fold $M$ are both local special K\"ahler manifolds, and they have complex dimensions $h^{2,1}(M)$ and $h^{1,1}(M)$, respectively. A local special K\"ahler manifold of dimension $n$ is a K\"ahler manifold with the K\"ahler potential
\begin{equation}
 K = -\log i(\bar X^I{\cal F}_I-X^I\bar{\cal F}_I), \qquad(I=0,1,\cdots,n)
\end{equation}
where $X^I,{\cal F}_I$ are local functions of the holomorphic coordinates $\{z_i\}_{i=1}^n$. Rescaling them by a common holomorphic function of $z_i$ amounts to a K\"ahler transformation, so they are regarded as a kind of homogeneous coordinates. A set of holomorphic coordinates can be defined, for example, by $z_i \equiv X^i/X^0$. Moreover, special geometry requires that ${\cal F}_I$, if expressed as functions of $X^J$, satisfy $\partial{\cal F}_I /\partial X^J= \partial{\cal F}_J/\partial X^I$. This implies there is a local holomorphic function ${\cal F}(X)$ of scaling weight 2 satisfying ${\cal F}_I = \partial{\cal F}/\partial X^I$. Such an ${\cal F}$ is called the prepotential.

For ${\cal M}_\text{C}$, one can identify $X^I,{\cal F}_I$ with the period integrals of the holomorphic 3-form $\Omega$
\begin{equation}
 X^I = \int_{A^I}\Omega,\quad
 {\cal F}_I = \int_{B_I}\Omega,\quad
 K = -\log i\int \Omega\wedge\bar\Omega,
\end{equation}
over the symplectic basis of 3-cycles $A^I,B_I$ satisfying
\begin{equation}
 \langle A^I,B_J\rangle = \delta^I_J,\quad
 \langle A^I,A^J\rangle = \langle B_I,B_J\rangle = 0.
\end{equation}

For ${\cal M}_\text{K}$, $X^I,{\cal F}_I$ are identified with the complexified volumes of even-dimensional cycles,
\begin{equation}
 X^0 = 1,\quad
 X^i = \int_{C^i} J,\quad
 {\cal F}_i = \frac12\int_{\hat C_i}J\star J,\quad
 {\cal F}_0 = -\frac16\int_{\hat M}J\star J\star J,
\end{equation}
where $J=B+i\omega$ is the complexified K\"ahler form and $C^i$ are the basis 2-cycles and $\hat C_i$ are the dual basis 4-cycles satisfying $\#(C^i\cap\hat C_j)=\delta^i_j$. Note the product $\star$ is that of quantum cohomology and differs from the classical wedge product of differential forms by instanton corrections. The K\"ahler potential can be written as
\begin{equation}
 K = -\log\left[-i\int_M \exp_\star(J)\wedge \exp_\star(-\bar J)\right].
\end{equation}
If we use $\tau^i\equiv X^i/X^0$ as the coordinates, ${\cal F}_I$ are related to the prepotential ${\cal F}(X) = (X^0)^2F(\tau^i)$ as
\begin{equation}
 {\cal F}_I = X^0\frac{\partial F}{\partial\tau^i},\qquad
 {\cal F}_0 = X^0\left(2F - \tau^i\frac{\partial F}{\partial\tau^i}\right).
\end{equation}
The prepotential $F(\tau)$ is known to consist of the polynomial part and the instanton part,
\begin{equation}
 F(\tau) = \frac1{3!}\sum_{\ell,m,n}\kappa_{\ell mn}\tau^\ell\tau^m\tau^n
 +\frac12\sum_{\ell,m}a_{\ell m}\tau^\ell\tau^m
 +\sum_\ell b_\ell\tau^\ell -\frac {ic}{16\pi^3}+F_\text{inst}(\tau),
\end{equation}
where $\kappa_{\ell mn}, a_{\ell m}, b_\ell, c$ are topological invariants of $M$, in particular
\begin{equation}
 \kappa_{\ell mn}=\#(\hat C_\ell\cap\hat C_m\cap\hat C_n),\quad
 c=\chi(M)\zeta(3).
\end{equation}
The instanton part $F_\text{inst}$ is the sum over contributions of worldsheet instantons wrapping various 2-cycles $\eta=\sum_i\eta_iC^i\in H_2(M,\mathbb Z)$,
\begin{equation}
 F_\text{inst}(\tau) = \frac1{(2\pi i)^3}\sum_{\eta\ne0}
 N_\eta\,\text{Li}_3\big(e^{2\pi i\sum_i\eta_i\tau^i}\big),\quad
 \text{Li}_k(x)\equiv \sum_{n\ge1}\frac{x^n}{n^k}.
\end{equation}
The coefficients $N_\eta$ are called Gromov-Witten invariants. Using these properties of the prepotential one can show that the K\"ahler potential takes the form
\begin{eqnarray}
 e^{-K} &=&
 -\frac i6\kappa_{\ell mn}(\tau-\bar\tau)^\ell(\tau-\bar\tau)^m(\tau-\bar\tau)^n
 +\frac c{4\pi^3}
 \nonumber \\ &&
 + 2i(F_\text{inst}-\bar F_\text{inst})
 -i(\tau-\bar\tau)^\ell\left(
  \frac{\partial F_\text{inst}}{\partial\tau^\ell}
 +\frac{\partial\bar F_\text{inst}}{\partial\bar\tau^\ell}
 \right).
\label{emk}
\end{eqnarray}

\paragraph{Sphere partition function for quintic.}

For the GLSM for quintic Calabi-Yau, the sphere partition function reads
\begin{eqnarray}
 Z_{S^2}
 &=& \sum_{s\in\frac12\mathbb Z}
 \int_{\mathbb R}\frac{da}{2\pi}e^{-it(a+is)-i\bar t(a-is)}
 \left[
 \frac{\Gamma(s+q-ia)}
      {\Gamma(s+1-q+ia)}
 \right]^5
 \frac{\Gamma(1-5s-5q+5ia)}
      {\Gamma(-5s+5q-5ia)}.
\label{zqi}
\end{eqnarray}
The FI parameter $r=\text{Re}(t)$ controls the size of the $\mathbb C\mathbb P^4$ and therefore the size of the Calabi-Yau. For large positive $r$ the model is said to be in the geometric phase, and for large negative $r$ it is in the orbifold phase. In correspondence with this, one can close the $a$-integration contour of~(\ref{zqi}) in the lower or upper half planes and write the integral as different residue-sums. In order to ensure that the poles from matter fields  with positive (negative) charges are all below (resp. above) the real axis, we require
\begin{equation}
 0<q,\quad 0<2-10q,
\end{equation}
where $q$ is half the R-charge of the five chiral multiplets of charge $+1$.

Let us study the residue integral in the geometric phase in detail. The poles of the determinant of the five chiral fields with charge $+1$ are labeled by a pair of non-negative integers $k,\bar k$. The values of $a,s$ are determined from them by the relations
\begin{equation}
 k=ia-s-q, \quad
 \bar k=ia+s-q.
\end{equation}
Rewriting~(\ref{zqi}) as a sum of the residue integrals with respect to $\epsilon\equiv ia-q-\frac12(k+\bar k)$ around the origin, one obtains the following form
\begin{equation}
 Z_{S^2} = -e^{-q(t+\bar t)}
 \oint_0\frac{d\epsilon}{2\pi i}\frac{5}{\epsilon^4}
 \frac{\Gamma(1+5\epsilon)}{\Gamma(1-5\epsilon)}
 \frac{\Gamma(1-\epsilon)^5}{\Gamma(1+\epsilon)^5}
 w(z,\epsilon)w(\bar z,\epsilon),\quad
 z\equiv -5^5e^{-t},
\end{equation}
where $w(z,\epsilon)$ is defined by
\begin{equation}
 w(z,\epsilon)
 \equiv \sum_{j\ge0}\epsilon^jw_j(z)
 \equiv \sum_{k\ge0}z^{k+\epsilon}
 \frac{\prod_{j=1}^{5k}(j+5\epsilon)}{5^{5k}\prod_{j=1}^k(j+\epsilon)^5}.
\end{equation}
It is clear that the sphere partition function can be written as a bilinear of $\{w_0,\cdots,w_3\}$. Also, since the Picard-Fuchs differential~(\ref{pfd}) acts on the above $w(z,\epsilon)$ as
\begin{equation}
\left\{(z\frac{d}{dz})^4-z(z\frac{d}{dz}+\frac45)(z\frac{d}{dz}+\frac35)(z\frac{d}{dz}+\frac25)(z\frac{d}{dz}+\frac15)\right\}w(z,\epsilon)=z^\epsilon \epsilon^4,
\end{equation}
the functions $\{w_0,\cdots,w_3\}$ are the four independent solutions of the PF equation, and $Z_{S^2}$ also satisfies the PF equation.

The GLSM is known to be singular at $t=5\ln5+i\pi$ or $z=1$ due to the non-compact Coulomb branch. This implies that the PF system also has a singularity at $z=1$. we will later confirm this on the mirror side.

Using the formula for the polygamma functions
\begin{equation}
 \psi^{(n)}(1)=(-1)^{n+1}n!\,\zeta(n+1),\quad
 \left(\psi^{(n)}(x)\equiv\frac{d^{n+1}}{dx^{n+1}}\ln\Gamma(x)\right)
\end{equation}
one easily finds
\begin{eqnarray}
 Z_{S^2} &=&
 -5^{-10q}(z\bar z)^q
 \oint_0\frac{d\epsilon}{2\pi i}
 \left(\frac{5}{\epsilon^4}-\frac{400\zeta(3)}{\epsilon}\right)
 w(z,\epsilon)w(\bar z,\epsilon).
\end{eqnarray}
Neglecting instanton corrections and substituting $w(z,\epsilon)\simeq e^{-\epsilon t}$, one finds
\begin{eqnarray}
 Z_{S^2} &\simeq&
 5^{-10q}e^{-q(t+\bar t)}
 \left(\frac56(t+\bar t)^3+400\zeta(3)+\cdots\right),
\end{eqnarray}
which agrees with the perturbative part of~(\ref{emk}) under the identification $-t=2\pi i\tau$. Indeed it was found in~\cite{Jockers:2012dk} that the log of sphere partition function agrees with the K\"ahler potential of the moduli space of K\"ahler structures.

\paragraph{Mirror quintic and Picard-Fuchs equation.}
The mirror of the quintic, denoted by $\tilde M$, is defined by the hypersurface
\begin{equation}
 X_1^5+X_2^5+X_3^5+X_4^5+X_5^5-5\psi X_1X_2X_3X_4X_5=0
\label{mtd}
\end{equation}
in the orbifold $\mathbb C\mathbb P^4/\Gamma$, where $\Gamma\equiv(\mathbb Z_5)^3$ is the group of discrete phase rotations of $X_i$,
\begin{equation}
 (X_1,\cdots,X_5) \mapsto (\omega^{a_1}X_1,\cdots,\omega^{a_5}X_5)\quad
\Big(
 \omega\equiv e^{\frac{2\pi i}5},~~ a_i\in\mathbb Z,~~
 \sum_ia_i=0~\text{mod }5
\Big)
\end{equation}
modulo identifications $(a_1,\cdots,a_5)\sim (a_1+1,\cdots,a_5+1)$. The complex structure of $\tilde M$ is parametrized by $\psi$, but the hypersurface equation with the modulus $\psi$ and $\omega\psi$ are equivalent as they are related by $X_1\to \omega X_1$. Therefore, a good coordinate on the moduli space is $\psi^5$. Note also that the mirror quintic becomes singular when $\psi^5=1$.

The periods are the integrals of the holomorphic 3-form $\Omega$ over the basis 3-cycles of $\tilde M$. For the hypersurface~(\ref{mtd}), one can express $\Omega$ using the inhomogeneous coordinates $x_i\equiv X_i/X_5$ as
\begin{equation}
\Omega = \frac{dx_1\wedge dx_2\wedge dx_3}{\partial p/\partial x_4},\quad
 p(x_i) \equiv x_1^5+x_2^5+x_3^5+x_4^5+1-5\psi x_1x_2x_3x_4,
\end{equation}
where $x_4$ should be eliminated by solving $p(x_i)=0$. Alternatively, one may express $\Omega$ as an integral of a 4-form in $\mathbb P^4/\Gamma$ along a small loop that goes around the hypersurface~(\ref{mtd}). For later convenience, let us generalize the hypersurface equation~(\ref{mtd}) as follows,
\begin{equation}
 \hat P(X) \equiv
 b_1X_1^5+b_2X_2^5+b_3X_3^5+b_4X_4^5+b_5X_5^5+b_6X_1X_2X_3X_4X_5,
\end{equation}
and define the 4-form by
\begin{equation}
 \hat\Omega \equiv
 \Big\{ X_1 dX_2dX_3dX_4dX_5+X_2 dX_3dX_4dX_5dX_1+(\text{3 more terms})\Big\}\Big/\hat P(X).
\end{equation}
The periods (generalized by the parameters $b_1,\cdots,b_6$) can now be expressed as certain four-dimensional integrals of $\hat\Omega$.

These generalized periods are annihilated by the differential operators,
\begin{equation}
 \partial_{b_1}\partial_{b_2}\partial_{b_3}\partial_{b_4}\partial_{b_5}-\partial_{b_6}^5,\quad
 \sum_{i=1}^6b_i\partial_{b_i}+1,\quad
 b_i\partial_{b_i}-b_5\partial_{b_5}~(i=1,\cdots,4).
\end{equation}
The first two simply annihilate $1/\hat P(X)$. The third one generates the rescaling of $b_i$ and $b_5$ which can be absorbed by the rescaling of $X_i$ and $X_5$, so it annihilates the periods. It follows from the last two differential equations that the periods all have to be $1/b_6$ times some functions of
\begin{equation}
z\equiv -5^5b_1b_2b_3b_4b_5/(b_6)^5 = \psi^{-5}.
\end{equation}
The first equation can then be rewritten into the Picard-Fuchs form.
\begin{eqnarray}
 0 &=& b_1b_2b_3b_4b_5b_6
 \Big\{\partial_{b_1}\partial_{b_2}\partial_{b_3}\partial_{b_4}\partial_{b_5}-\partial_{b_6}^5\Big\}\frac1{b_6}\omega(z)
 \nonumber \\
 &=& \left\{
 (z\partial_z)^5-z(z\partial_z+1)(z\partial_z+\frac45)(z\partial_z+\frac35)(z\partial_z+\frac25)(z\partial_z+\frac15) \right\}\omega(z)
 \nonumber \\
 &=& z\partial_z\left\{
 (z\partial_z)^4-z(z\partial_z+\frac45)(z\partial_z+\frac35)(z\partial_z+\frac25)(z\partial_z+\frac15) \right\}\omega(z).
\end{eqnarray}
Thus we see that $\psi^5=1~(z=1)$ is a singularity of the PF system.

\section{Interpretation of $\kappa(\eta)$}
\label{sec:interpret-kappa}
Here we point out that the exponent $\kappa(\eta)$ ($=\kappa_\north(\eta)$ or $\kappa_\south(\eta)$)  appears in the solutions of the BPS vortex equation in the Higgs phase in the presence of defects.  To discuss this point, we use the simplest SQED with a single chiral field $\Phi$.

The phase of $\Phi$ and $\text{Im}Y$ are T-dual of each other under mirror symmetry. In particular the term $e^{-Y}$ in the twisted superpotential is known to correspond to the BPS vortex instanton with unit instanton number,
\begin{equation}
 n\equiv \frac1{2\pi}\int d^2x F_{12} = 1.
\end{equation}
Classically, the vortex instantons are described by the solutions to the equations
\begin{equation}
 F_{12}= e^2(r-\bar\phi\phi),\quad
 D_{\bar z}\phi=0,
\label{bpsv}
\end{equation}
where we assume $r>0$ and use $z\equiv x^1+ix^2$. For $n$-vortex instanton solutions, the scalar $\phi$ has the absolute value $|\phi|\simeq\sqrt{r}$ at infinity, and its phase winds $n$ times as one goes around the large circle at infinity counterclockwise once.

It is thus natural to compare $\kappa(\eta)$ with the winding number of the phase of $\phi$ in the vortex defect background. Let us put the defect $V_\eta$ at $z=0$ and solve the equation~(\ref{bpsv}) around it. For simplicity we take the rotation-symmetric ansatz,
\begin{equation}
 A = \{n-H(\rho)\} d\varphi,\quad
 \phi=\sqrt{r}\Phi(\rho)e^{in\varphi}\qquad
(z\equiv \sqrt{\rho}e^{i\varphi})
\end{equation}
with real functions $H(\rho),\Phi(\rho)$. See \cite{Hori:2000kt} for a related analysis, which we generalize here. The BPS vortex equation then becomes
\begin{equation}
 -2H' = e^2r(1-\Phi^2),\quad
 2\rho\Phi' = H\Phi.
\label{bpsv2}
\end{equation}
We are interested in the solution satisfying $\Phi(\infty)=1,\,H(\infty)=0$ and $H(0)=n-\eta$. One can eliminate $\Phi$ and find the asymptotic form of $H$ for $e^2r |z|^2\gg 1$,
\begin{equation}
 \rho H''=HH'+\frac{e^2r}{2}H,\quad
 H =\text{const}\cdot \sqrt{m|z|}e^{-m|z|}+\cdots.\quad
 (m\equiv \sqrt{2e^2r})
\end{equation}
It follows from this asymptotics and the equation~(\ref{bpsv2}) that the radial evolution of $H$ and $\Phi$ are both monotonic, and there are three possible cases.
\begin{alignat}{5}
\text{(i)}&~~~& H'&<0,~&\Phi'&>0 &~~~~ \Longrightarrow~~~~ n&>\eta,&~~ \Phi&\sim |z|^{n-\eta} \text{ vanishes at the origin.} \nonumber\\
\text{(ii)}&~~~& H'&\equiv0,~&\Phi'&\equiv0 &~~~~ \Longrightarrow~~~~ n&=\eta,&~~ \Phi& \text{ is constant.} \nonumber\\
\text{(iii)}&~~~& H'&>0,~&\Phi'&<0 &~~~~ \Longrightarrow~~~~ n&<\eta,&~~ \Phi&\sim |z|^{n-\eta} \text{ diverges at the origin.}
\end{alignat}

What is the minimum allowed value for the winding number $n$? For the chiral multiplet obeying the normal boundary condition at the defect, the allowed behavior for the scalar is (i) or (ii), so that $n\ge\lceil\eta\rceil=\kappa(\eta)$. For the chiral multiplet obeying the flipped boundary condition, the behavior (iii) is also allowed as long as $n-\eta>-1$, so that $n\ge\lfloor\eta\rfloor=\kappa(\eta)$. Thus $\kappa(\eta)$ agrees with the minimum winding number for the BPS vortex solutions on the defect background.

We also note that $|\kappa(\eta)|$ is the number of localized modes when the defect with the normal or the flipped boundary condition arises as a limit of the smeared defect.

\bibliography{references}

\end{document}